\documentclass[letterpaper, inpress]{jds} 

\jdsmonth{October}                 
\jdsyear{2022}                  
\jdsvolume{xx}                  
\jdsissue{xx}                   
\jdsdoi{xx.xxxx/xxxxxxxxx}      
\jdsreceived{October, 2022}       
\jdsaccepted{MM, YYYY}         

\usepackage{longtable}
\usepackage{calc}
\usepackage{natbib}
\usepackage{graphicx}
\usepackage{float}
\PassOptionsToPackage{hyphens}{url}\usepackage{url}
\usepackage{color}
\usepackage[dvipsnames]{xcolor}
\usepackage{amsmath}
\usepackage{amsfonts} 
\usepackage[ruled,vlined]{algorithm2e}
\usepackage{placeins}
\usepackage{makecell}
\usepackage{hyperref}
\usepackage{bm}
\usepackage{bbm}
\usepackage{booktabs}
\usepackage{multirow}
\usepackage{arydshln}
\usepackage{enumitem}
\usepackage{eucal}
\usepackage{subcaption}
\graphicspath{ {figures/} }

\newcommand{\bs}{{\boldsymbol{s}}}

\newcommand{\team}{\texttt}

\usepackage{amsfonts,amsmath,amssymb,amsthm}
\usepackage{booktabs}

\usepackage{lipsum}

\title[The Second Competition on Spatial Statistics for Large Datasets] {The Second Competition on Spatial Statistics for Large Datasets}

\author[1]{Sameh Abdulah\thanks{Corresponding author. Email: sameh.abdulah@kaust.edu.sa}}
\author[2]{Faten Alamri}
\author[3]{Pratik Nag}
\author[1,3]{Ying Sun}
\author[1]{Hatem Ltaief}
\author[1]{David E. Keyes}
\author[1,3]{Marc G. Genton}
\affil[1]{Extreme Computing Research Center, King Abdullah University of Science and Technology, Thuwal 23955-6900, Saudi Arabia}
\affil[2]{Mathematical Science Department, Princess Nourah bint Abdulrahman University, Riyadh 84428, Saudi Arabia}
\affil[3]{Statistics Program, King Abdullah University of Science and Technology, Thuwal 23955-6900, Saudi Arabia}

\begin{document}

\nocite{nesi2022multi}
\nocite{cao2022accelerating}
\nocite{leandro2021exploiting}
\nocite{abdulah2018exageostat}
\nocite{abdulah2018parallel}
\nocite{abdulah2021accelerating}
\nocite{hong2021efficiency}
\nocite{salvana2021high}
\nocite{abdulah2021accelerating}
\nocite{abdulah2019geostatistical}
\maketitle

\begin{abstract}

In the last few decades, the size of spatial and spatio-temporal datasets in many research areas has rapidly increased with the development of data collection technologies. As a result, classical statistical methods in spatial statistics are facing computational challenges. For example, the kriging predictor in geostatistics becomes prohibitive on traditional hardware architectures for large datasets as it requires high computing power and memory footprint when dealing with large dense matrix operations. Over the years, various approximation methods have been proposed to address such computational issues, however, the community lacks a holistic process to assess their approximation efficiency. To provide a fair assessment, in 2021, we organized the first competition on spatial statistics for large datasets, generated by our {\em ExaGeoStat} software, and asked participants to report the results of estimation and prediction. Thanks to its widely acknowledged success and at the request of many participants, we organized the second competition in 2022 focusing on predictions for more complex spatial and spatio-temporal processes, including univariate nonstationary spatial processes, univariate stationary space-time processes, and bivariate stationary spatial processes. In this paper, we describe in detail the data generation procedure and make the valuable datasets publicly available for a wider adoption. Then, we review the submitted methods from fourteen teams worldwide, analyze the competition outcomes, and assess the performance of each team.

\end{abstract}


\section{Introduction}%
\label{sec:intro}

With the explosion of spatial and spatio-temporal data coming from different sources such as sensors and satellites, it has become crucial to handle these data in a robust manner. In the last few decades, studies focusing on spatial statistics have followed one main direction to deal with large geospatial data: adopting an approximation approach to reduce the modeling and prediction process complexity. However, these methods have been mostly benchmarked using small and medium-sized datasets because of the prohibitive computation of the exact solution for comparison purposes.

~\cite{abdulah2018exageostat} proposed the {\em ExaGeoStat} software to perform large-scale statistical modeling and prediction for geospatial data on leading-edge parallel hardware architectures. {\em ExaGeoStat} is able to deal with millions of spatial locations and complete the statistical estimation and prediction in exact and approximate formats \citep{abdulah2018parallel,abdulah2021accelerating}. It is also supported by a data generation tool that is able to generate synthetic spatial datasets from different realizations with millions of locations. As mentioned by~\cite{vu2021discussion}, {\em ExaGeoStat} can be described as the gold standard for spatial statistical data modeling tools since it can generate and analyze large geospatial data with exact computations for millions of spatial locations. Thus, in 2021, we organized the first KAUST spatial statistics competition for large synthetic datasets for the spatial statistics community to assess existing geospatial modeling methods using {\em ExaGeoStat}. The competition involved a set of synthetic datasets generated from univariate spatial processes with up to 1M locations. Out of twenty-nine research teams worldwide who registered to participate in the competition, twenty-one teams successfully submitted their results; see \cite{huang2021competition} for the analysis of the submissions. Over the past year, {\em ExaGeoStat} has been further developed to support richer classes of models for spatial and spatio-temporal processes. Therefore, we organized a second competition in 2022 by providing geospatial datasets with new features.

This work follows several studies that have aimed to assess the efficiency of existing tools and methods to perform kriging for geospatial datasets. For instance, in~\cite{englund1990variance},
a collection of spatial datasets were sent to twelve investigators to analyze them and perform spatial inference for missing locations. The ``Walker Lake'' dataset \citep{srivastava1987non} was the core source of the study datasets. The author illustrated the clear variability in the submitted results, motivating and encouraging the statistics community to develop a set of performance-based guidelines to assess the quality of the statistical analysis with different models and methods. The work in~\cite{weber1992evaluation} is an extension of Englund's work, where the relative accuracy of fifteen inference methods was assessed for analyzing the fifty-four datasets of the ``Walker Lake'' data. For example, the spatial predictions based on the inverse distance were compared to those obtained by the kriging method. The study explained in detail the pros and cons of each method for these datasets. 
More recently, a study by~\cite{heaton2017methods} assessed existing kriging tools with simulated and real datasets by introducing one competition to a pre-selected set of research groups. Twelve methods were compared in the competition, with the code and details of each method made available to the statistics community. \cite{wikle2017common} proposed the ``secret sauce'' to build a benchmarking framework for objective comparison of spatial prediction methods, which has been called a Common Task Framework (CTF). The article argues that any fair comparison between spatial prediction methods should follow this framework by including a set of publicly available spatial datasets, predefined prediction assessing rules, and an automatic scoring referee. The datasets should be complete and well described, and the scoring referee should be unique for all the submissions. In the literature, other studies for the assessment of geospatial methods exist but with different perspectives and performed on other datasets, such as the work in~\cite{li2011review, li2014spatial}, \cite{shahbeik2014comparison}, and \cite{bradley2016comparison}.

Following the success of the 2021 competition, this year we organized another competition by providing synthetic datasets generated by {\em ExaGeoStat}. These datasets include three types of data: spatial nonstationary datasets, space-time and bivariate spatial stationary datasets. The evaluation criteria focuses on point prediction performance only. Specifically, we organized six sub-competitions, two sizes for each type of data: one medium-sized and the other large-sized. The competition was launched on March 1st, 2022, with twenty teams registered to participate in one or more sub-competitions. This year, the competition was hosted on the Kaggle platform, which allows fast and accurate assessment of the teams' submissions. The competition ended on May 1st, 2022, and fourteen teams successfully submitted their results to the platform. 

The rest of the paper is organized as follows. In Section 2, we provide an overview of the {\em ExaGeoStat} software. In Section 3, we describe in detail the models used for the generation of the competition datasets. In Section 4, we highlight the competition results and list the ranks of different teams based on their performance in different sub-competitions. In Section 5, we summarize the methods used by different teams in this competition. Finally, in Section 6, we conclude with observations and discussions.

\section{The ExaGeoStat Software}%
\label{sec:Overview}
Gaussian processes (GPs) are widely used to model geospatial data, for which the covariance function plays an important role. The modeling process relies on  likelihood-based methods that require constructing a covariance matrix whose entries represent the covariance between any two observations via a predefined covariance function. Suppose the covariance function has a parametric form $C({\boldsymbol s}_i,{\boldsymbol s}_j; \boldsymbol{\theta})$, where ${\boldsymbol s}_i,{\boldsymbol s}_j \in \boldsymbol{R}^2$ ($i,j=1,\ldots,n$) are the spatial locations and $\boldsymbol{\theta}$ is a parameter vector of interest. The Gaussian log-likelihood function is:
\begin{equation} \label{eqn:loglikelihood_function}
l(\boldsymbol{\theta}) = -\frac{n}{2} \log (2 \pi) - \frac{1}{2} \log |\boldsymbol{\Sigma}(\boldsymbol{\theta})| - \frac{1}{2} \mathbf{Z}^{\top} \boldsymbol{\Sigma}(\boldsymbol{\theta})^{-1} \mathbf{Z},
\end{equation} and needs to be maximized with respect to $\boldsymbol{\theta}$ to obtain its maximum likelihood estimator (MLE). Here $\boldsymbol{\Sigma}(\boldsymbol{\theta}) = \{ C({\bm s}_i,{\bm s}_j; \boldsymbol{\theta}) \}_{i,j = 1}^n$ is the $n \times n$ covariance matrix (symmetric and positive definite) of the $n$-dimensional data vector $\mathbf{Z}$, $n$ represents the number of spatial locations, and $|\boldsymbol{\Sigma}(\boldsymbol{\theta})|$ is 
the determinant of $\boldsymbol{\Sigma}(\boldsymbol{\theta})$.

The computation of the log-likelihood (\ref{eqn:loglikelihood_function}) is prohibitive for large sample size $n$ as the complexity of computing the inverse of the covariance matrix, $\boldsymbol{\Sigma}(\boldsymbol{\theta})^{-1}$, is $O(n^3)$, and requires $O(n^2)$ memory.
This computation becomes even more expensive when considering the multivariate and spatio-temporal case because the sample size becomes $pnm$, where $p$ is the number of spatial variables and $m$ is the number of time points. Thus, dealing with large geospatial datasets requires more advanced computational techniques, and integrating the capabilities of HPC with existing spatial statistics methods becomes mandatory with the vast increase in the sizes of geospatial datasets. ~\cite{abdulah2018exageostat} introduced the {\em ExaGeoStat} parallel software to help in scaling the geospatial operations using leading-edge parallel hardware architectures. {\em ExaGeoStat} relies on the state-of-the-art parallel linear algebra libraries to allow fast and efficient computation of the log-likelihood function.  {\em ExaGeoStat} also depends on modern runtime systems such as {\em StarPU} and {\em PaRSEC} to improve the portability of the code on the different parallel hardware architectures, including GPUs. {\em ExaGeoStat} has driven breakthroughs in the spatial statistics field by supporting rich classes of covariance models for large-scale spatial and spatio-temporal data; see, e.g., \cite{salvana2021high}, \cite{mondal2022parallel}, and \cite{salvana2022parallel}.
 \begin{figure}
\centering
\includegraphics[width=16cm]{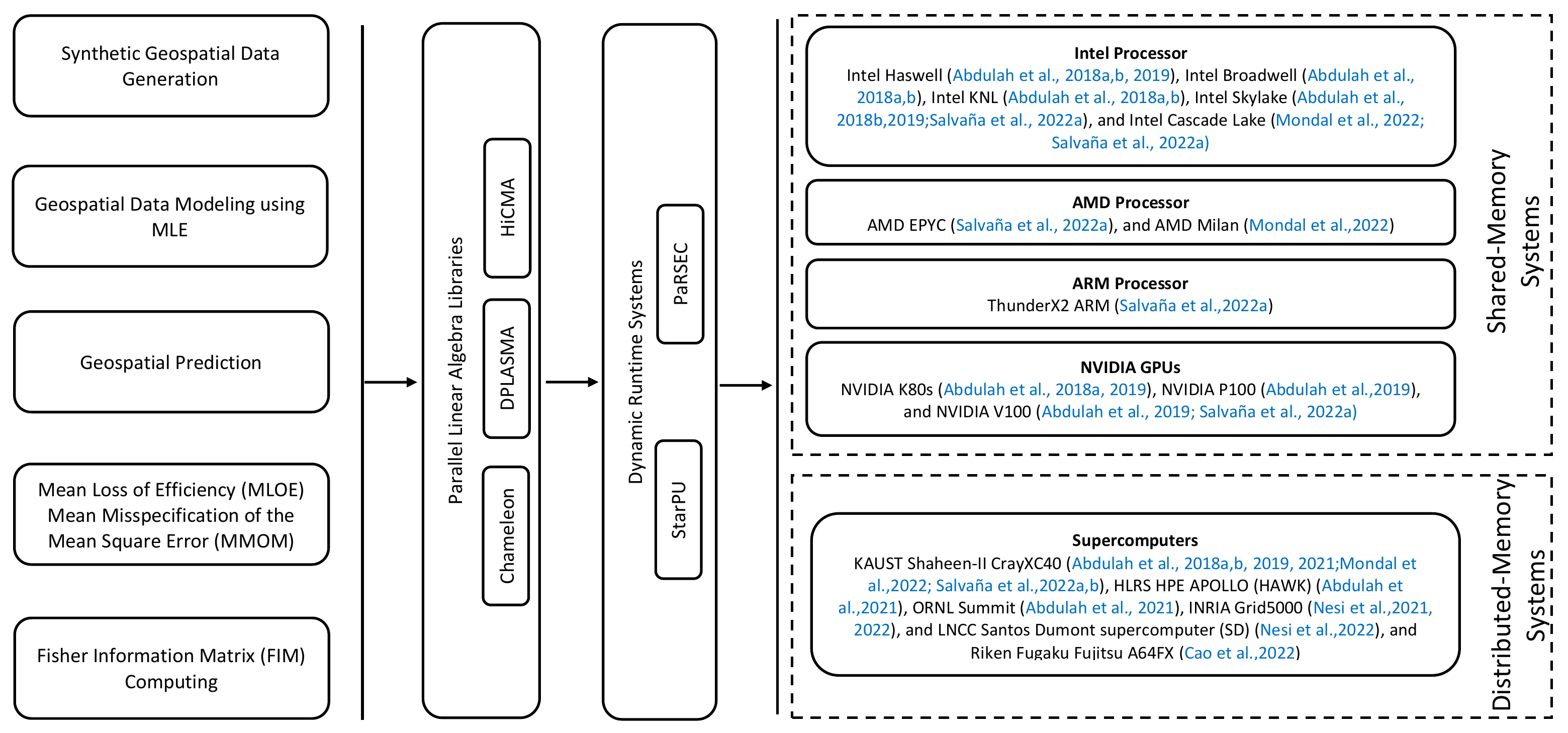}
\caption{Statistical tasks supported by {\em ExaGeoStat} with a visual literature review of the software portability on different hardware architectures.}
\label{fig:exageostat}
\end{figure}

Several studies have demonstrated the capabilities of {\em ExaGeoStat} in utilizing today's high-performance hardware for spatial statistics. Figure~\ref{fig:exageostat} explains in detail the statistical tasks supported by {\em ExaGeoStat} and a visual literature review of the software portability on different hardware architectures. The figure shows at a high-level the five main components currently included in the software, i.e., data generation, MLE modeling, geospatial prediction, mean loss of efficiency (MLOE) and mean misspecification of the mean square error (MMOM) tools, and the Fisher information matrix (FIM) computing. The MLOE/MMOM tools are used to assess the loss of the prediction efficiency by using the approximated or misspecified covariance models~\citep{hong2021efficiency}. The Fisher information matrix provides information about the importance of a single observation in estimating unknown statistical parameters, i.e., uncertainty quantification. Figure~\ref{fig:exageostat} also highlights the parallel linear algebra libraries on which {\em ExaGeoStat} relies to perform the required covariance matrix operations, i.e., Chameleon and DPLASMA for dense matrix computations and HiCMA for Tile Low-Rank (TLR) matrix approximations. In addition, as seen Figure~\ref{fig:exageostat}, {\em ExaGeoStat} relies on two different runtime systems, i.e., StarPu and PaRSEC.

\section{Detailed Description of the Competition Datasets}

In this competition, we generated three types of data: univariate nonstationary spatial data, univariate stationary space-time data, and bivariate stationary spatial data. We consider different  settings for each type. The following subsections give more details about the data generating models and the associated settings. A summary of the six sub-competitions can be found in Table~\ref{tbl:subcomp} at the end of this section.

\subsection{Univariate Nonstationary Spatial Model (Sub-competitions 1a and 1b)}\label{sec3.1}

We considered a univariate Gaussian random field (GRF), $Z(\bs)$, $\bs \in [0,1]^2$, modeled as:
\begin{equation}
Z(\bs) = m(\bs) + Y(\bs)+\epsilon(\bs),
\label{eq:general}
\end{equation}
where $m(\cdot)$ is the mean function, $Y(\cdot)$ is a spatially dependent and zero-mean GRF with covariance function $C(\cdot, \cdot)$, $\epsilon(\cdot) \sim {\cal N}(0, \tau^2)$ is an independent noise process with $\tau^2$ denoting the nugget effect, and $\epsilon(\cdot)$ is independent of $Y(\cdot)$. In Sub-competitions 1a and 1b, four nonstationary datasets were generated under the model in equation~(\ref{eq:general}).

For datasets 1a-1 and 1b-1, the nonstationarity is in the deterministic mean function as described in \cite{xiong2007non} and \cite{ba2012composite}. Data were generated by $Z(\bs) = m(\bs)+\epsilon(\bs)$, $\bs=(s_x, s_y)$ on a regular grid. For the dataset 1a-1 of size $100$K, $\tau=0.1$ and $m(\bs)$ is:
 \begin{equation}
 \begin{split}
    m(\bs)=&  5\sin\left\{30\left(\frac{s_x + s_y}{2} -0.9\right)^{3}\right\}\cos\left\{20\left(\frac{s_x + s_y}{2} -0.9\right)^4\right\} \\ &+\frac{1}{2}\exp\{\sin(30s_x) + \sin(13s_y)\}+\frac{1}{2} \left(\frac{s_x + s_y}{2} -0.2\right).  
     \end{split}
     \label{eq:e1}
 \end{equation}
For the dataset 1b-1 of size $1$M, $\tau=0.3$ and $m(\bs)$ takes the form: 
 \begin{equation}
 \begin{split}
    m(\bs)=&  3\sin\left\{20\left(\frac{s_x + s_y}{2} +1.9\right)\right\}\cos\left\{20\left(\frac{s_x + s_y}{2} -1.2\right)^6\right\} \\ &+\frac{3}{5}\exp\{\sin(25s_x) + \sin(13s_y)\}+\frac{1}{2} \left(\frac{s_x + s_y}{2} -0.2\right).  
     \end{split}
     \label{eq:e2}
 \end{equation}

Datasets 1a-2 ($100$K) and 1b-2 ($1$M) were generated at  irregular locations from a zero-mean Gaussian process $Z(\bs)$ with a nonstationary Mat\'ern covariance function \citep[and references therein]{li2019efficient}:
\begin{equation}
\begin{aligned}
  C^{NS}(\bm s_{i},\bm s_{j};\bm{\theta})=& \,\,\tau^2 \mathbbm{1} _{[i=j]}(\bm s_{i},\bm s_{j})+ \frac{\sigma(\bm s_{i})\sigma(\bm s_{j})}{\Gamma(\nu)2^{\nu-1}}|{\boldsymbol \Sigma}(\bm s_{i})|^{1/4}|{\boldsymbol \Sigma}(\bm s_{j})|^{1/4} \\
   &\times \left \vert\frac{{\boldsymbol \Sigma}(\bm s_{i})+{\boldsymbol \Sigma}(\bm s_{j})}{2}\right\vert^{-1/2}
   \left( 2\sqrt{\nu Q_{ij}} \right )^{\nu} {\cal K}_{\nu} \left( 2\sqrt{\nu Q_{ij}} \right ),
  \end{aligned}
  \label{eq:non_stat_cov_fun}
\end{equation}
where $\sigma(\bs_i)$ is the spatially varying standard deviation, ${\boldsymbol \Sigma}(\bs_i)$ is the kernel matrix at $\bs_i$, ${\cal K}_\nu$ is the modified Bessel function of the second kind of order $\nu>0$, $\nu$ is the smoothness parameter, and $Q_{ij}$ is the square Mahalanobis distance between $\bs_i$ and $\bs_j$.

The nonstationarity of the GRF is controlled by the spatially  varying parameters $\theta(\bm s_i) \in \{ {\boldsymbol \Sigma}(\bs_i),\sigma(\bs_i)\}$. The kernel matrices are obtained through a spectral decomposition:
\begin{equation*}
\begin{aligned}
{\boldsymbol \Sigma}(\bs_i)= 
\begin{bmatrix}
\cos{(\phi) \& -\sin(\phi)} \\
\sin{(\phi) \& \cos(\phi)} \\
\end{bmatrix}
\begin{bmatrix}
\lambda_{1}(\bs_i) & 0 \\
0 & \lambda_{2}(\bs_i) \\
\end{bmatrix}
\begin{bmatrix}
\cos{(\phi) \& \sin(\phi)} \\
-\sin{(\phi) \& \cos(\phi)} \\
\end{bmatrix},
  \end{aligned}
\end{equation*}

\noindent
where $\lambda_{1}(\bs_i)>0$ and $\lambda_{2}(\bs_i) > 0 $ are eigenvalues that represent spatial ranges and $\phi \in [0,\pi/2]$ represents the angle of rotation.

The generation process depends on dividing the spatial region into a grid of subregions centered at reference locations $(\tilde{ \bm s}_k)^{M}_{k=1}$. A spatially varying parameter $\theta$ at location~$\bm s_{i}$ is defined as follows:
\begin{equation*}
\begin{aligned}
    \theta(\bm s_{i})=\sum^{M}_{k=1} w(\bm s_i, \tilde{\bm s}_k) \theta_{k}, \qquad w(\bm s_{i}, \tilde{\bm s}_{k})=\frac{K(\bm s_{i}, \tilde{\bm s}_{k})}{\sum^{M}_{k=1} K(\bm s_{i}, \tilde{\bm s}_{k})},
    \end{aligned}
\end{equation*}
where $M$ represents the number of subregions, $\theta_{k}$ is the parameter value at the reference location~$\tilde{\bm s}_{k}$ associated with the $k$-th subregion, $w(\bm s_{i},\tilde{\bm s}_{k})$ is a weight function, and $K(\cdot)$ denotes a bivariate kernel function. Herein, we chose a Gaussian kernel defined as $K(\bm s_{i},\tilde{\bm s}_{k})=\exp \{-\|\bm s_{i}-\tilde {\bm s}_{k}\|^{2}/(2h)\}$, where $h>0$ is the bandwidth parameter  fixed to $0.09$.

 We chose $M=4$ subregions with four reference locations with the coordinates $(0.25,0.25)$, $(0.25,0.75)$, $(0.75,0.25)$, $(0.75,0.75)$ in the unit square. Parameter settings at the 4 reference locations are
$\sigma= (3.5, 1.9, 1.8, 0.7)$, $\lambda_1=\lambda_2= (0.03, 0.07, 0.1, 0.3)$, with constant $\nu= 0.7$, $\tau^2=0.3$, and $\phi=\pi/2$. 

\begin{figure}[H]
\captionsetup[subfigure]{labelformat=empty}
\centering
   \begin{minipage}[t]{\linewidth}
    \vspace{3mm}
 \subcaption{Nonstationary deterministic mean function}
 \vspace{-2mm}
\centering
\captionsetup[subfigure]{labelformat=empty}
 \begin{subfigure}[b]{0.25\textwidth}
 \centering
 \includegraphics[width=\textwidth]{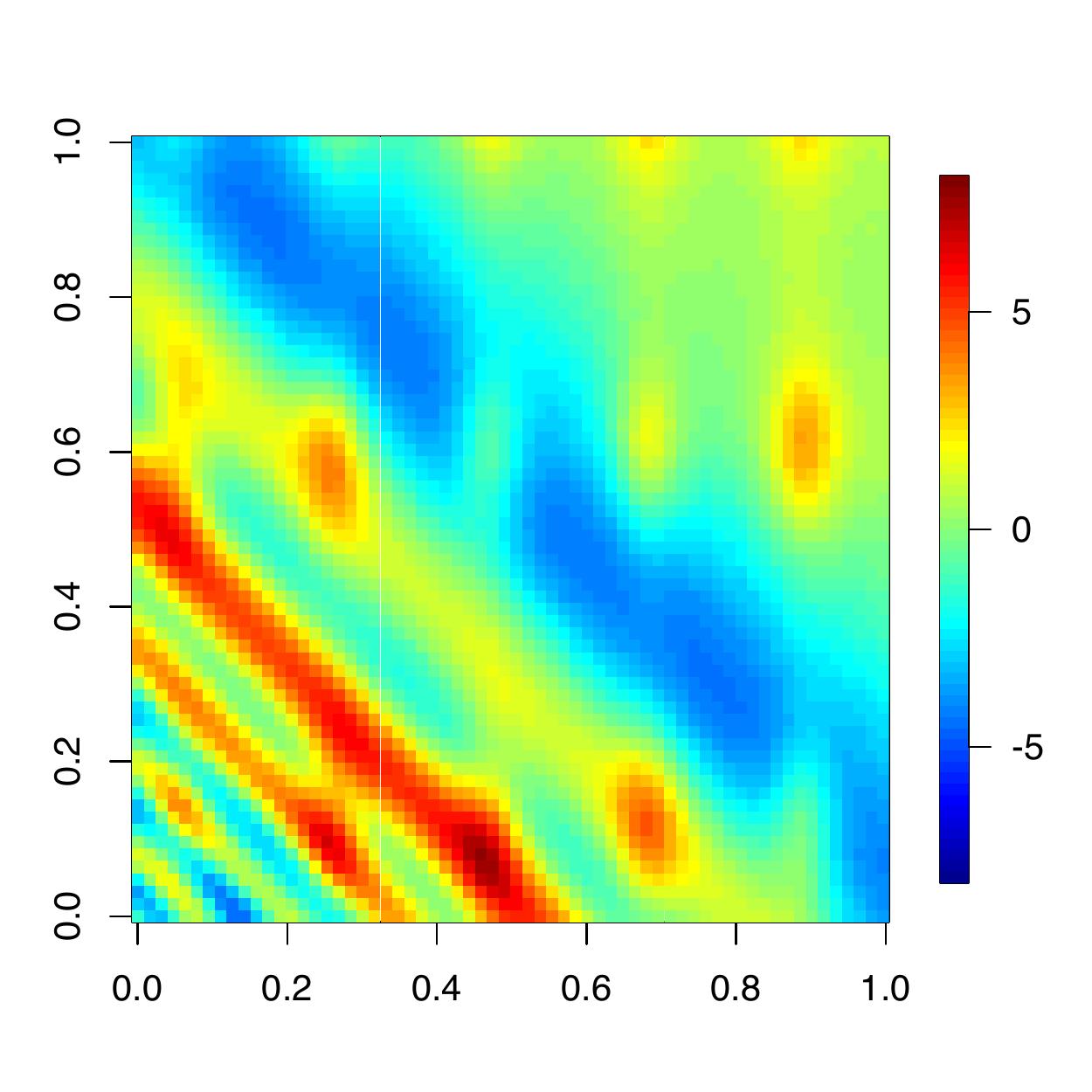}
 \caption{1a-1} 
 \label{fig:1a-1}
 \end{subfigure}
 \hspace{16mm}
 \begin{subfigure}[b]{0.25\textwidth}
 \centering
 \includegraphics[width=\textwidth]{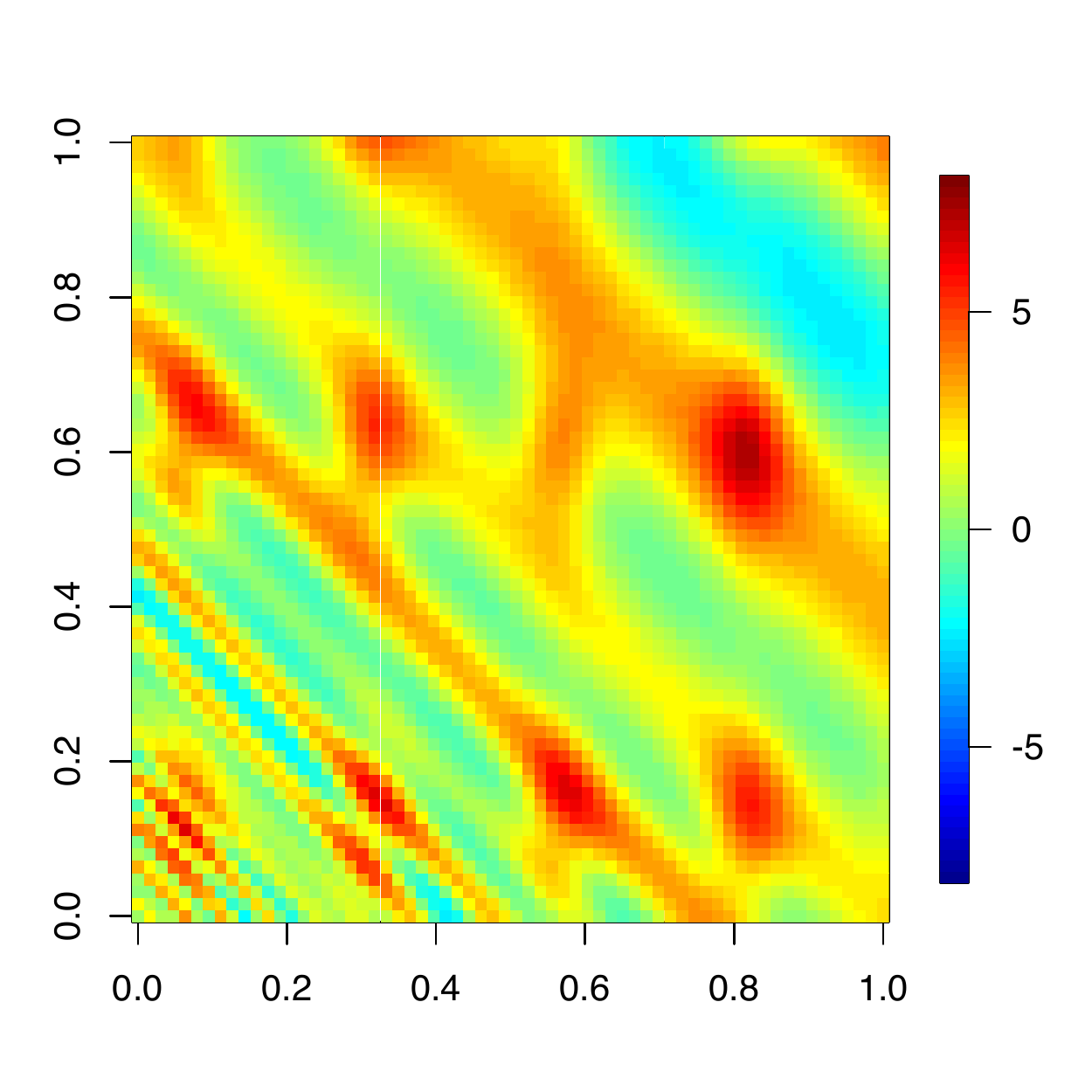}
 \caption{1b-1} 
  \label{fig:1b-1}
 \end{subfigure}
   \end{minipage}
 
    \begin{minipage}[t]{\linewidth}
    \centering
      \vspace{5mm}
 \subcaption{Nonstationary Mat\'ern covariance function}
 \vspace{-2mm}
 \begin{subfigure}[b]{0.25\textwidth}
 \centering
 \includegraphics[width=\textwidth]{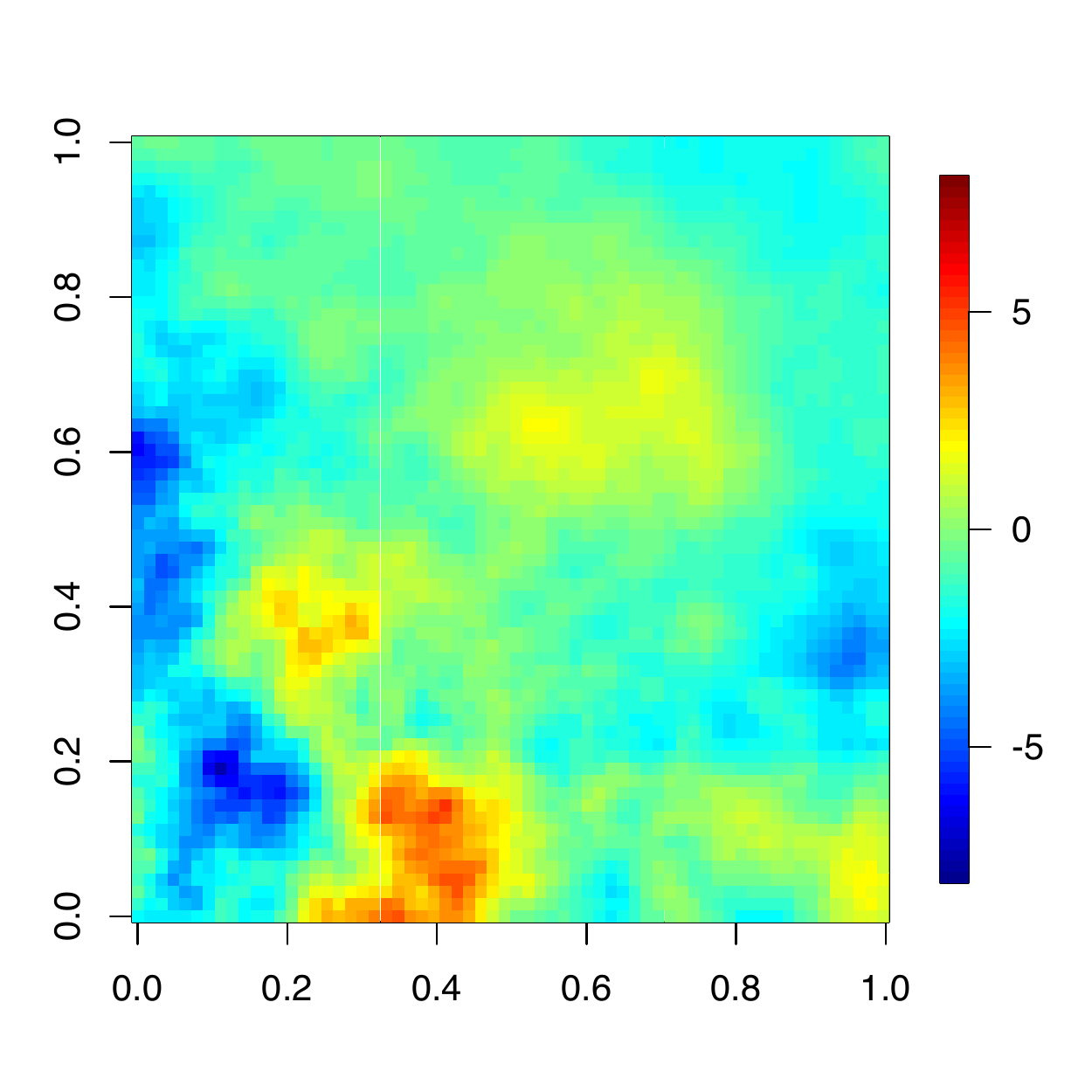}
 \caption{1a-2} 
  \label{fig:1a-2}
 \end{subfigure}
 \hspace{16mm}
 \begin{subfigure}[b]{0.25\textwidth}
 \centering
 \includegraphics[width=\textwidth]{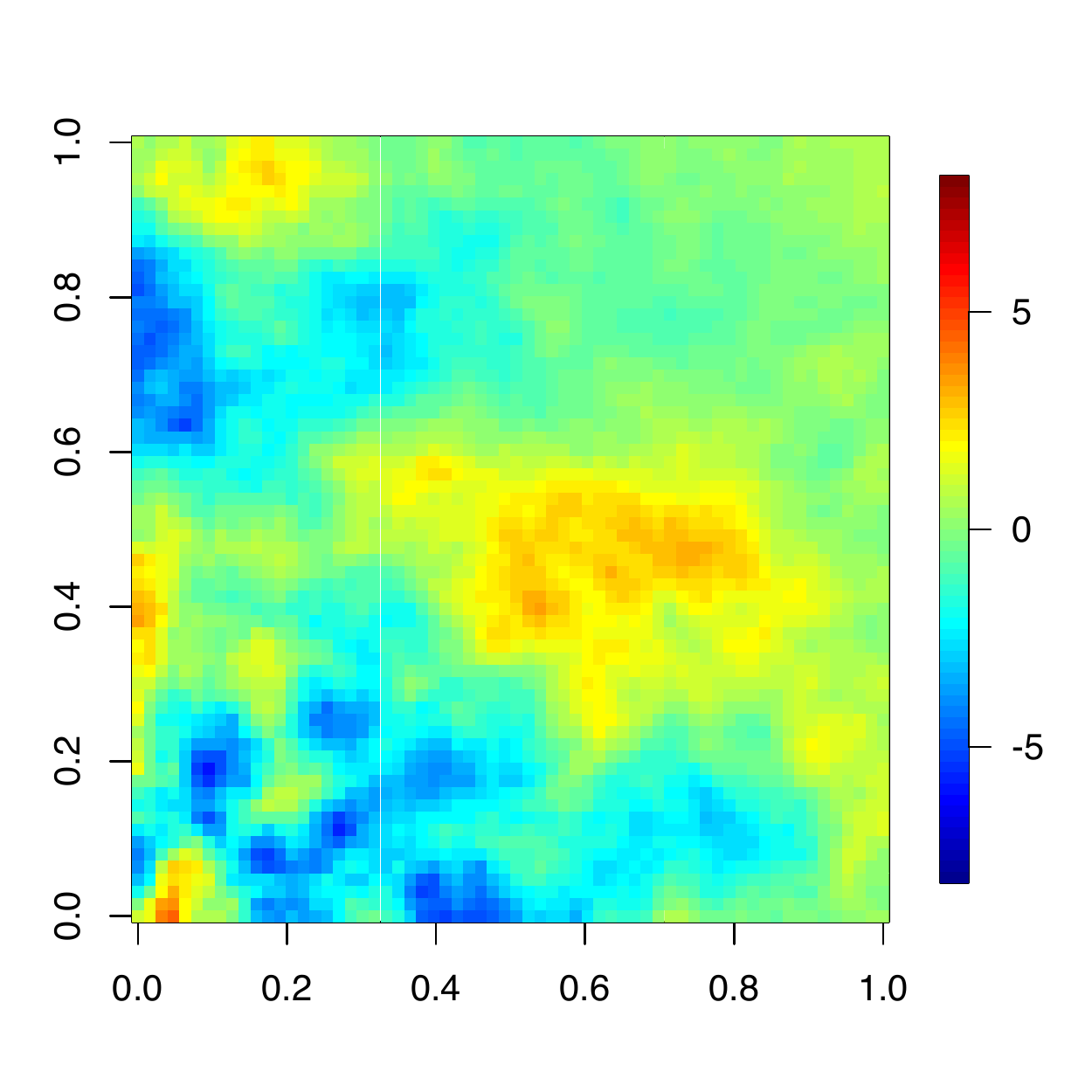}
 \caption{1b-2}
  \label{fig:1b-2}
 \end{subfigure}
   \end{minipage}
 \caption{Synthetic univariate nonstationary spatial datasets in Sub-competitions 1a and 1b: (1a-1) A $100$K nonstationary dataset generated using the determinsitic mean function in (\ref{eq:e1}); (1b-1) A $1$M nonstationary dataset generated using the deterministic mean function in (\ref{eq:e2}); (1a-2) A $100$K nonstationary dataset generated using the covariance function in (\ref{eq:non_stat_cov_fun}); and (1b-2) A $1$M nonstationary dataset generated using the covariance function in (\ref{eq:non_stat_cov_fun}).}
 \label{fig:non1}
 \end{figure}

For Sub-competitions 1a and 1b, 90\% of the data were randomly selected as the training datasets, and the remaining 10\% were used as testing datasets. Training datasets were given to the participants to perform prediction on the locations of the testing data. Figure~\ref{fig:non1}  shows the visual images of the four training datasets in Sub-competitions 1a and 1b that illustrate different features of nonstationarity.

\subsection{Univariate Stationary Space-Time Model (Sub-competitions 2a and 2b)}
\label{sec:spacetime}

The second part of the competition includes eighteen univariate space-time datasets generated from a zero-mean Gaussian process with a non-separable stationary space-time covariance function \citep{gneiting2002nonseparable} at space-time locations $(\bm s,t)\in [0,1]^{2}\times \mathbb{R}$ :
\begin{equation}
    C(\bm h,u;\boldsymbol \theta)= \frac{\sigma^2}{a_{t}|u|^{2\alpha}+1}
    {\cal{M_\mathcal{\nu}}} \left\{\frac{\|\bm h\|/a_{s} }{(a_{t}|u|^{2\alpha}+1)^{\beta/2} }\right\}, \label{eq:st-cov}
\end{equation}
where $\sigma^2> 0$ is the variance, $\nu >0 $ and $\alpha \in (0,1]$ are the smoothing parameters, $a_{s}$ and $a_{t}>0$ are the range parameters in space and time, respectively, $\beta \in (0,1]$ is the space-time interaction parameter, and ${\cal{M_\mathcal{\nu}}}$ is the univariate Mat\'ern correlation function:
\begin{equation*}
{\cal{M_\mathcal{\nu}}}(r)= \frac{1}{2^{\nu-1}\Gamma(\nu)}r^{\nu} {\cal{K_\mathcal{\nu}}} \left(r \right).
\end{equation*}

The sizes of the generated datasets were $1$K and $10$K locations, with $100$ time-slots. We considered different spatial and temporal dependencies: strong, moderate and weak for a non-separable model with $\beta =0.9$. The parameter settings for each case were as follows: weak-strong $(a_{\bm s},a_{t})$ = (0.02, 1), moderate-moderate $(a_{\bm s}, a_{t})$ = (0.08, 0.24), strong-strong $(a_{\bm s}, a_{t})$ = (0.4, 1), $\alpha = (0.08, 0.6)$, and $ (\sigma^2, \beta, \nu)= (0.9,  0.9, 1)$.  Table~\ref{tab:1} summarizes the parameter settings used for each dataset, as well as the time/space effective range (EffRange).

We considered three different scenarios of leaving out space-time points for prediction:

 \begin{itemize}
        \item[1.] Random spatial locations with all times left out (RS); 
        
        \item[2.] Random locations in space/time left out (RST);
        
        \item[3.] All spatial  locations are missing on the last 10 time points (T10).
    \end{itemize}

\begin{figure}[t!]
\captionsetup[subfigure]{labelformat=empty}
   \begin{minipage}[t]{\linewidth}
       \vspace{3mm}
   \subcaption{Space-time Mat\'ern ($a_s=0.02$ and $a_t =1$)}
 \vspace{-3mm}
 \begin{subfigure}[b]{0.46\textwidth}
\centering
\includegraphics[width=0.33\textwidth]{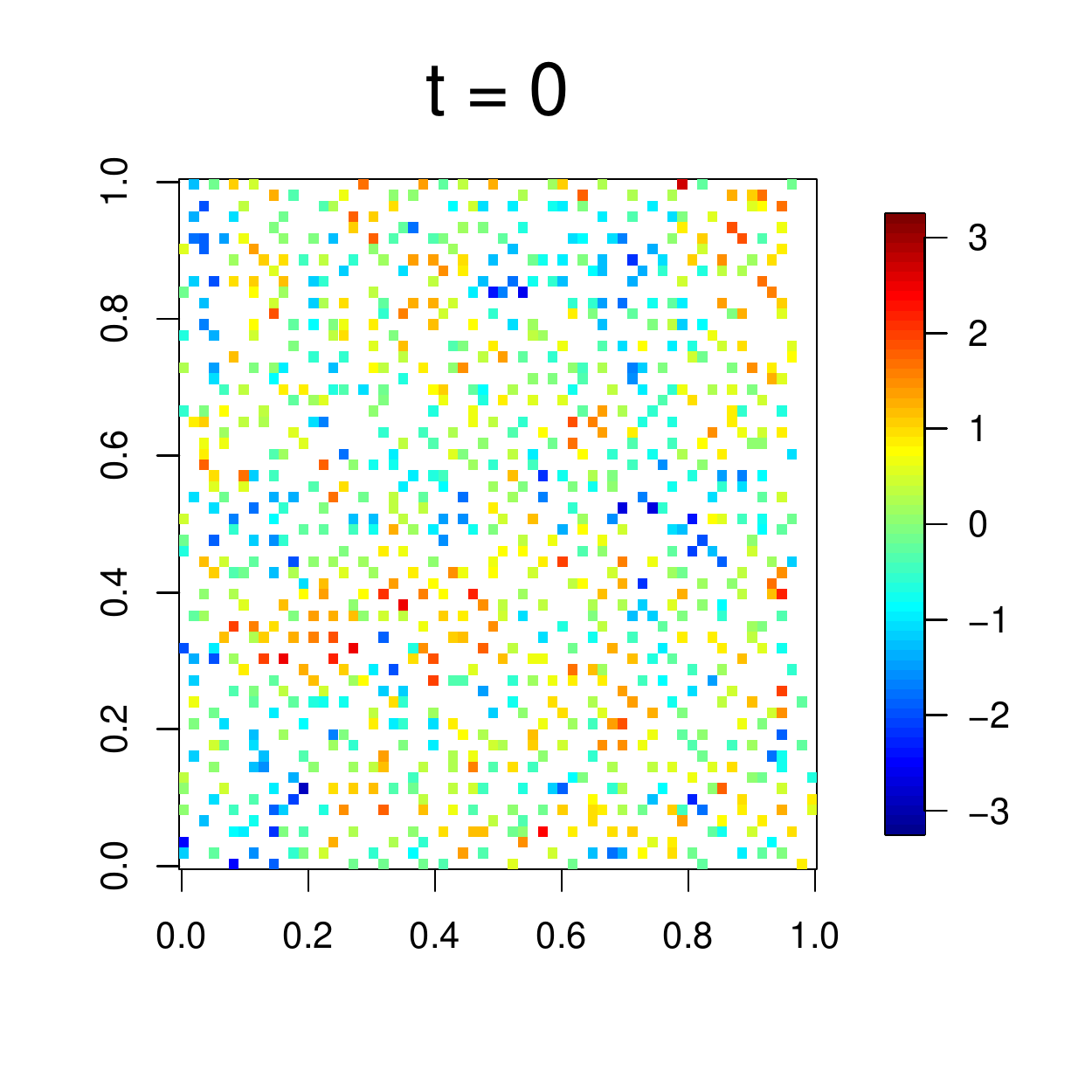}
\hspace{-2mm}
\includegraphics[width=0.33\textwidth]{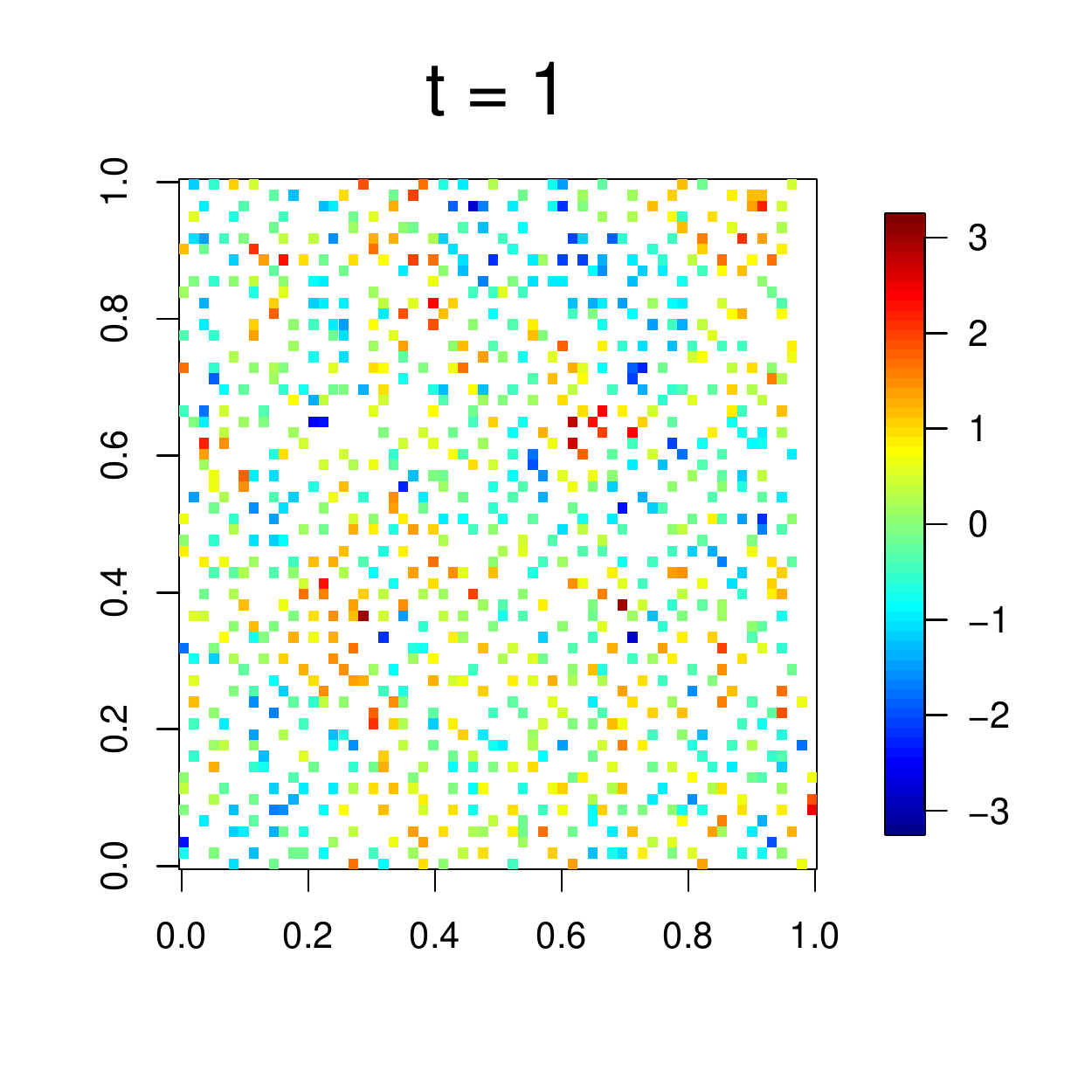}
\hspace{-2mm}
\includegraphics[width=0.33\textwidth]{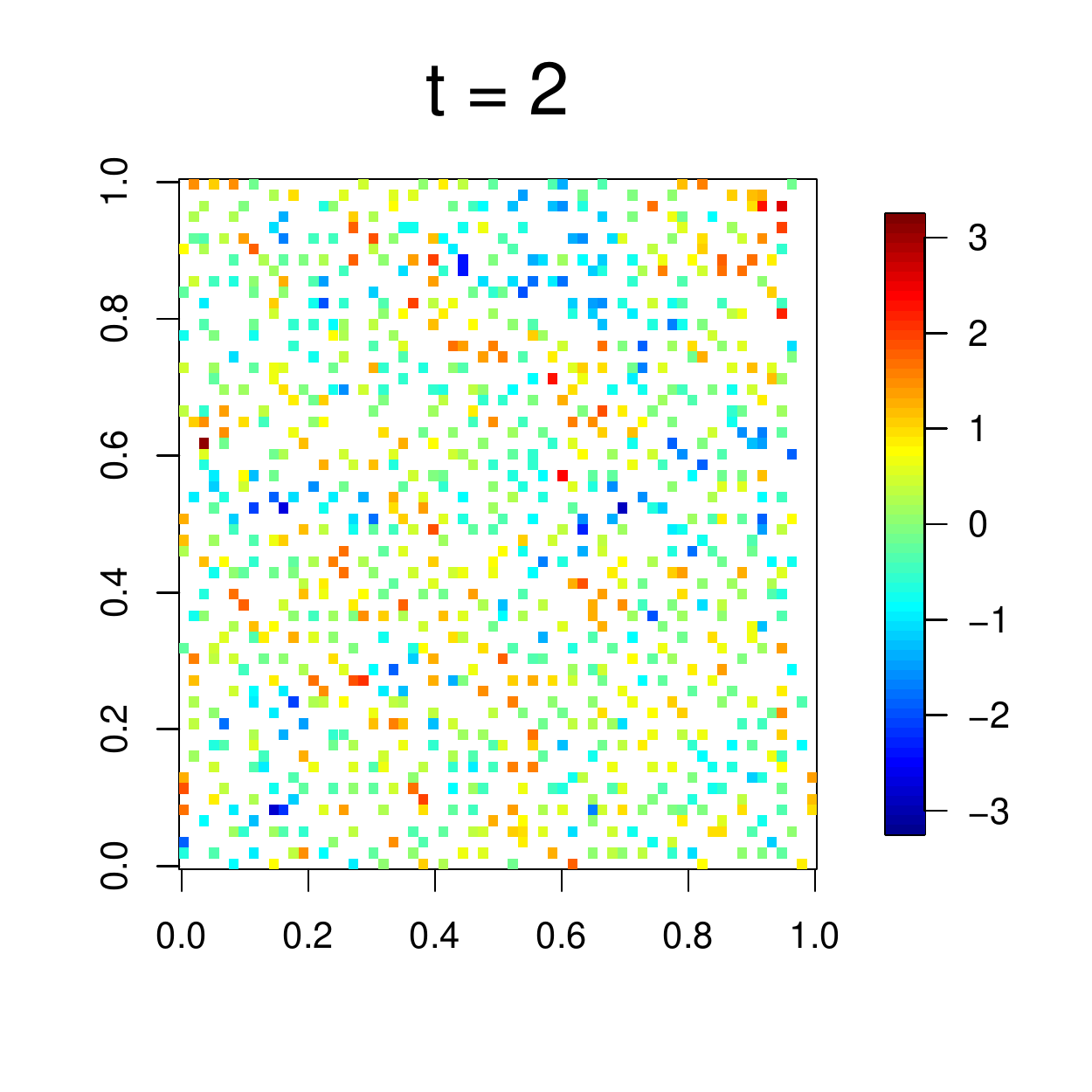}
\caption{2a-1} 
 \end{subfigure}
 \hspace{8mm}
\begin{subfigure}[b]{0.46\textwidth}
\centering
\includegraphics[width=0.33\textwidth]{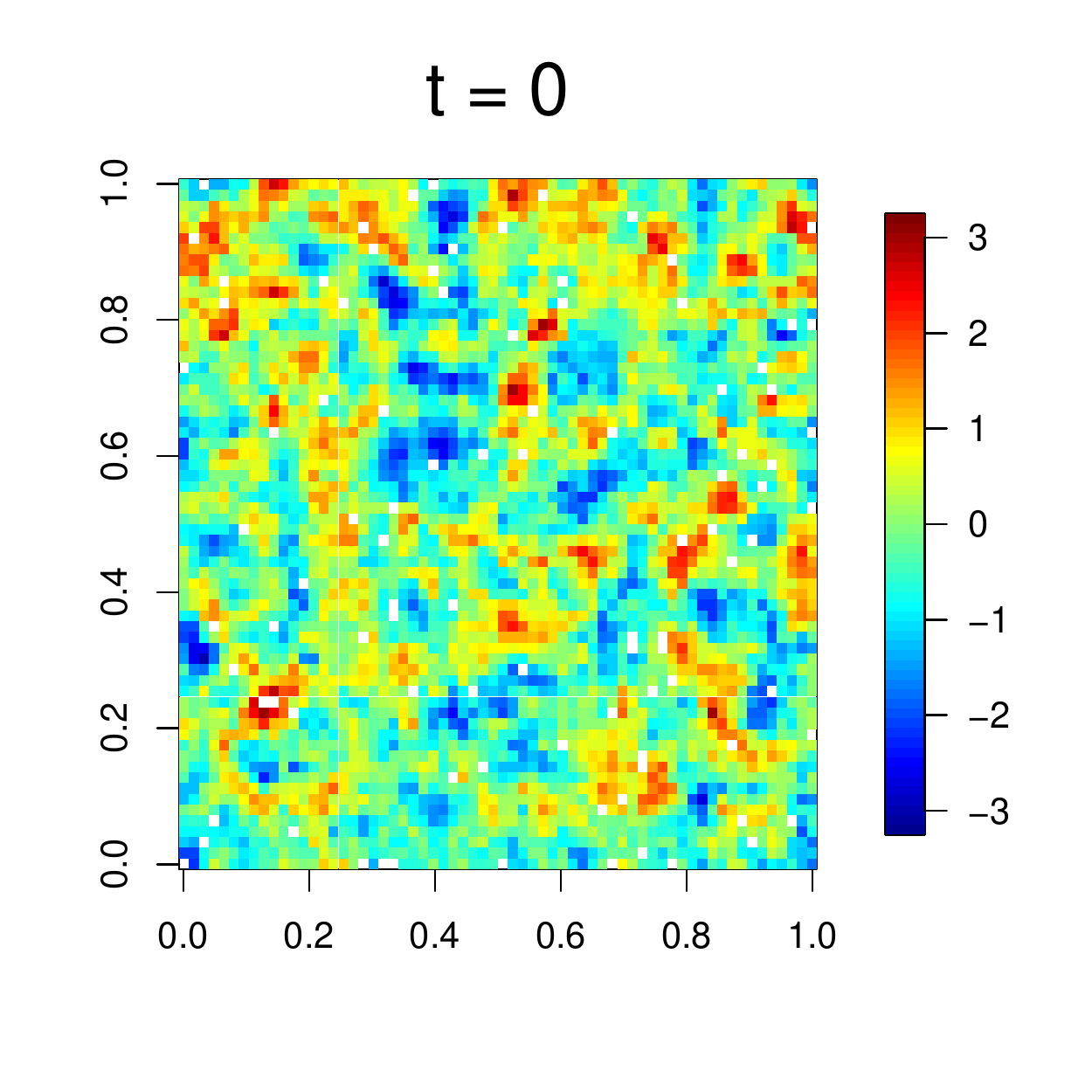}
\hspace{-2mm}
\includegraphics[width=0.33\textwidth]{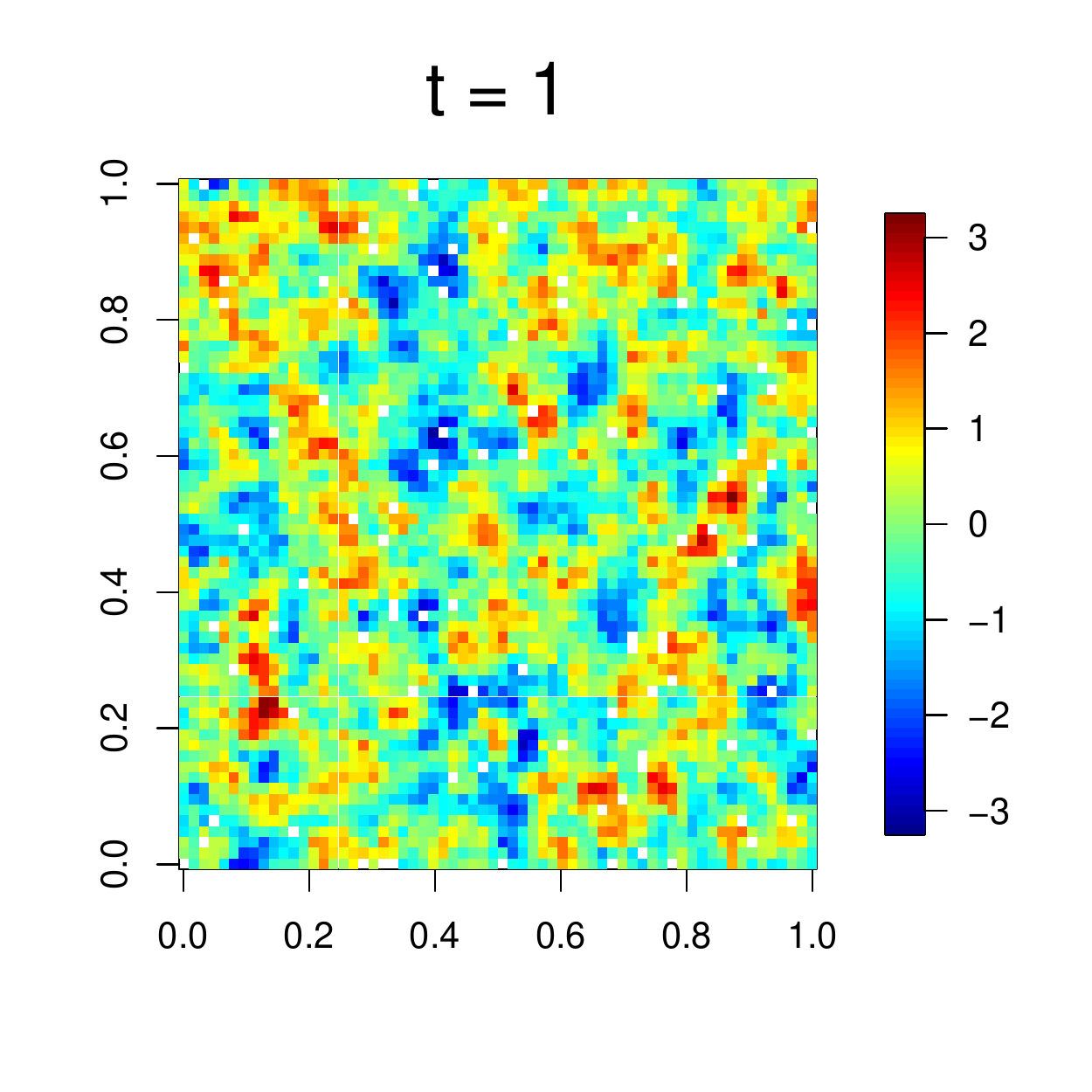}
\hspace{-2mm}
\includegraphics[width=0.33\textwidth]{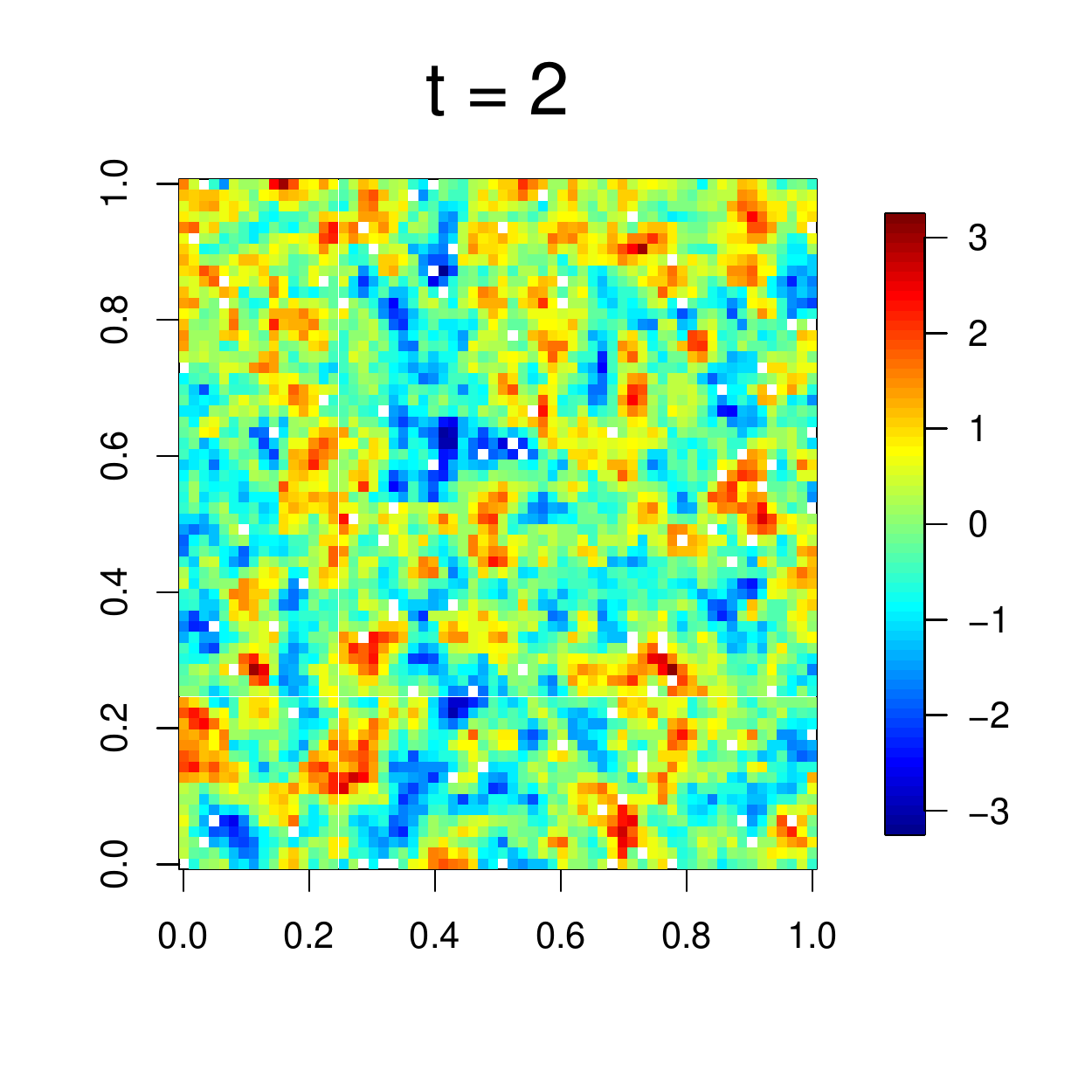}
\caption{2b-1} 
 \end{subfigure}

\end{minipage}
   \begin{minipage}[t]{\linewidth}
       \vspace{3mm}
   \subcaption{Space-time Mat\'ern ($a_s=0.08$ and $a_t =0.24$)}
 \vspace{-2mm}
\centering
\begin{subfigure}[b]{0.46\textwidth}
\centering
\includegraphics[width=0.33\textwidth]{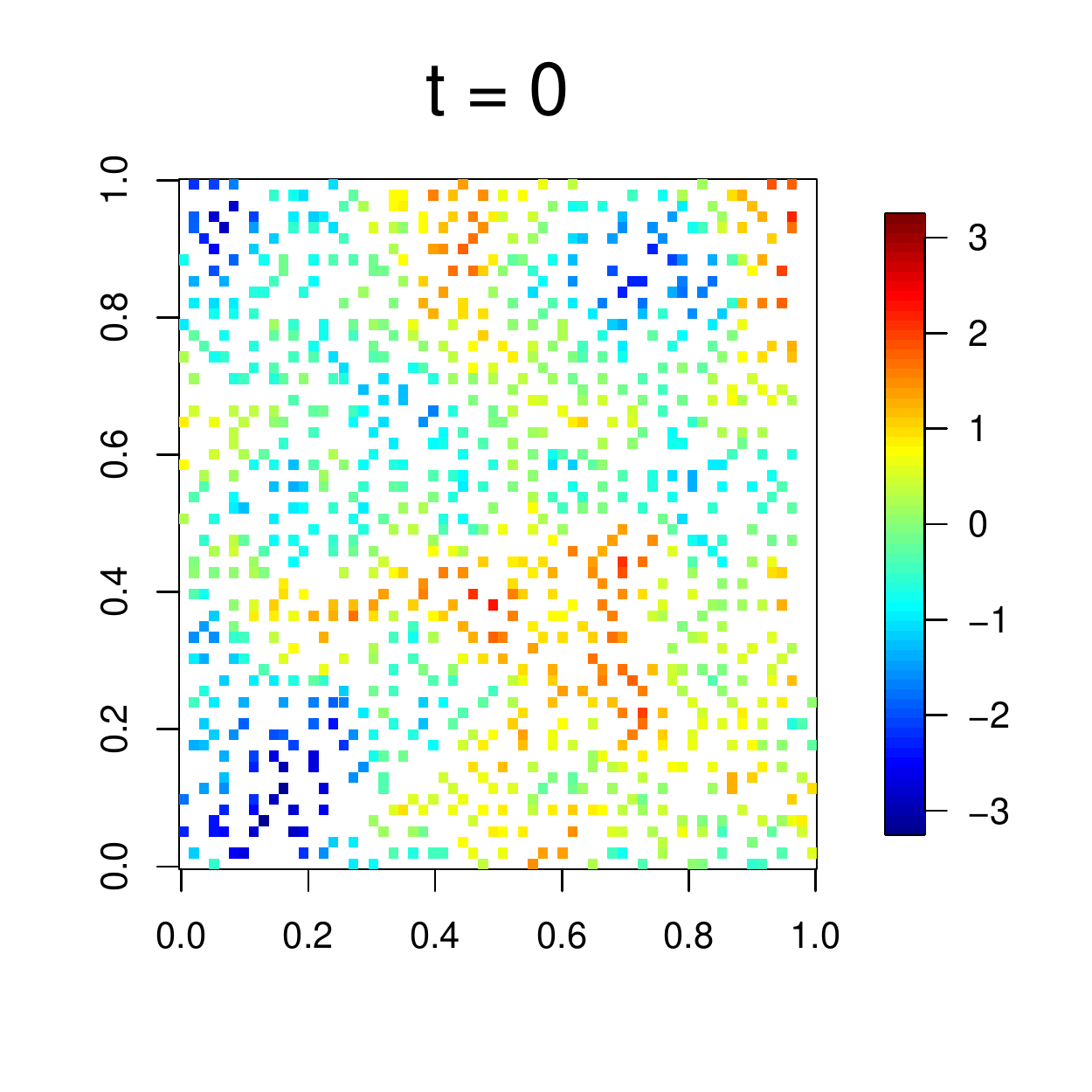}
\hspace{-2mm}
\includegraphics[width=0.33\textwidth]{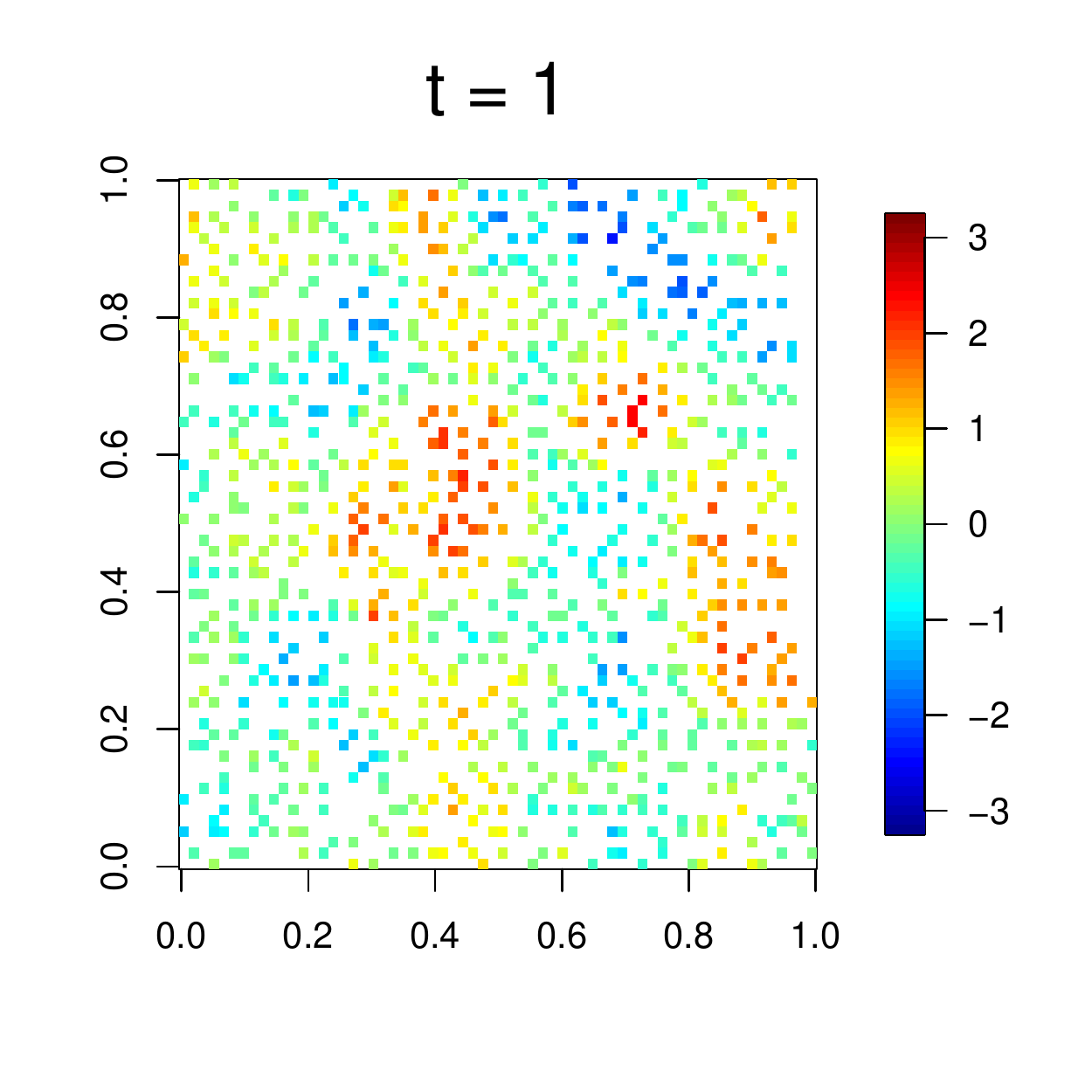}
\hspace{-2mm}
\includegraphics[width=0.33\textwidth]{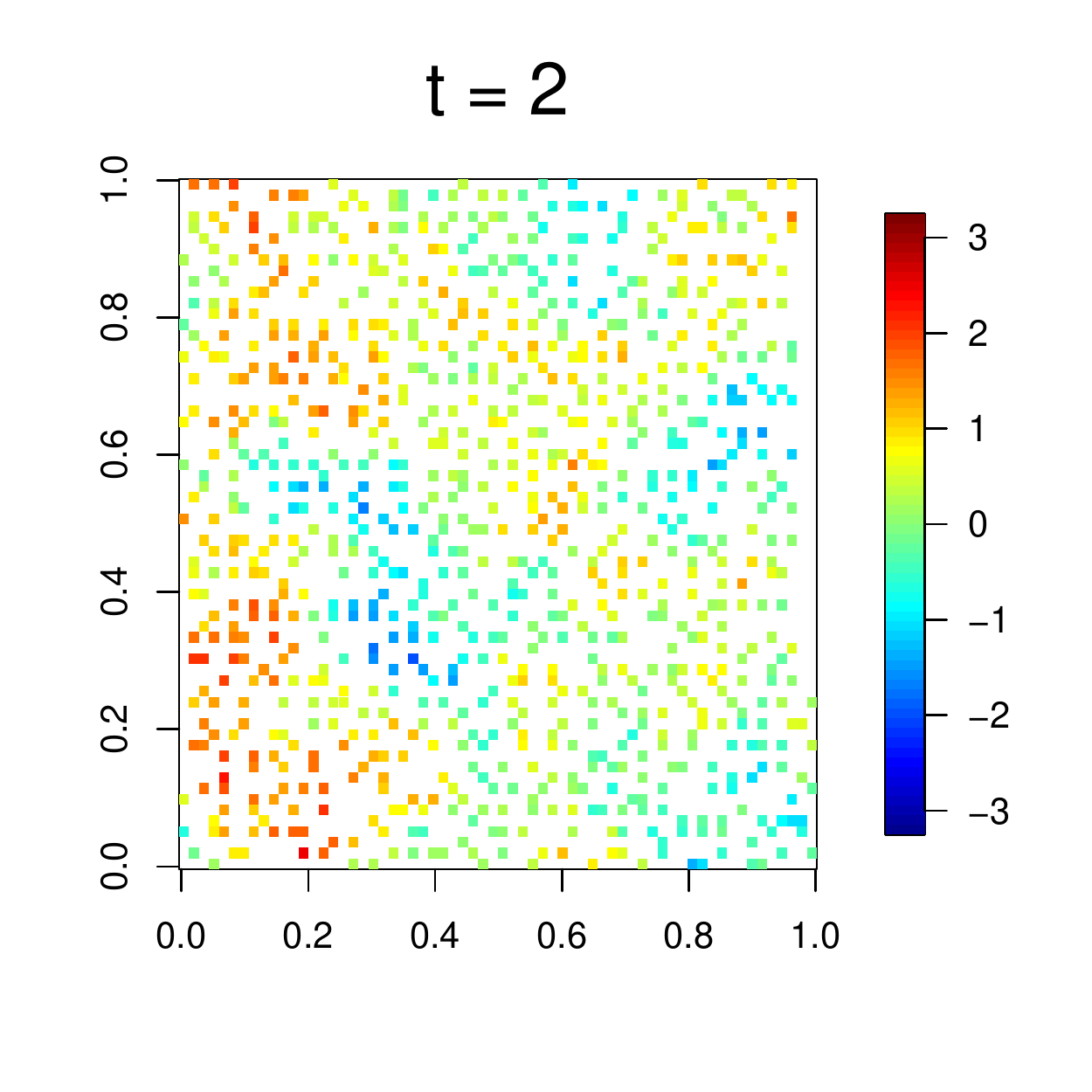}
\caption{2a-2} 
 \end{subfigure}
  \hspace{8mm}
\begin{subfigure}[b]{0.46\textwidth}
\centering
\includegraphics[width=0.33\textwidth]{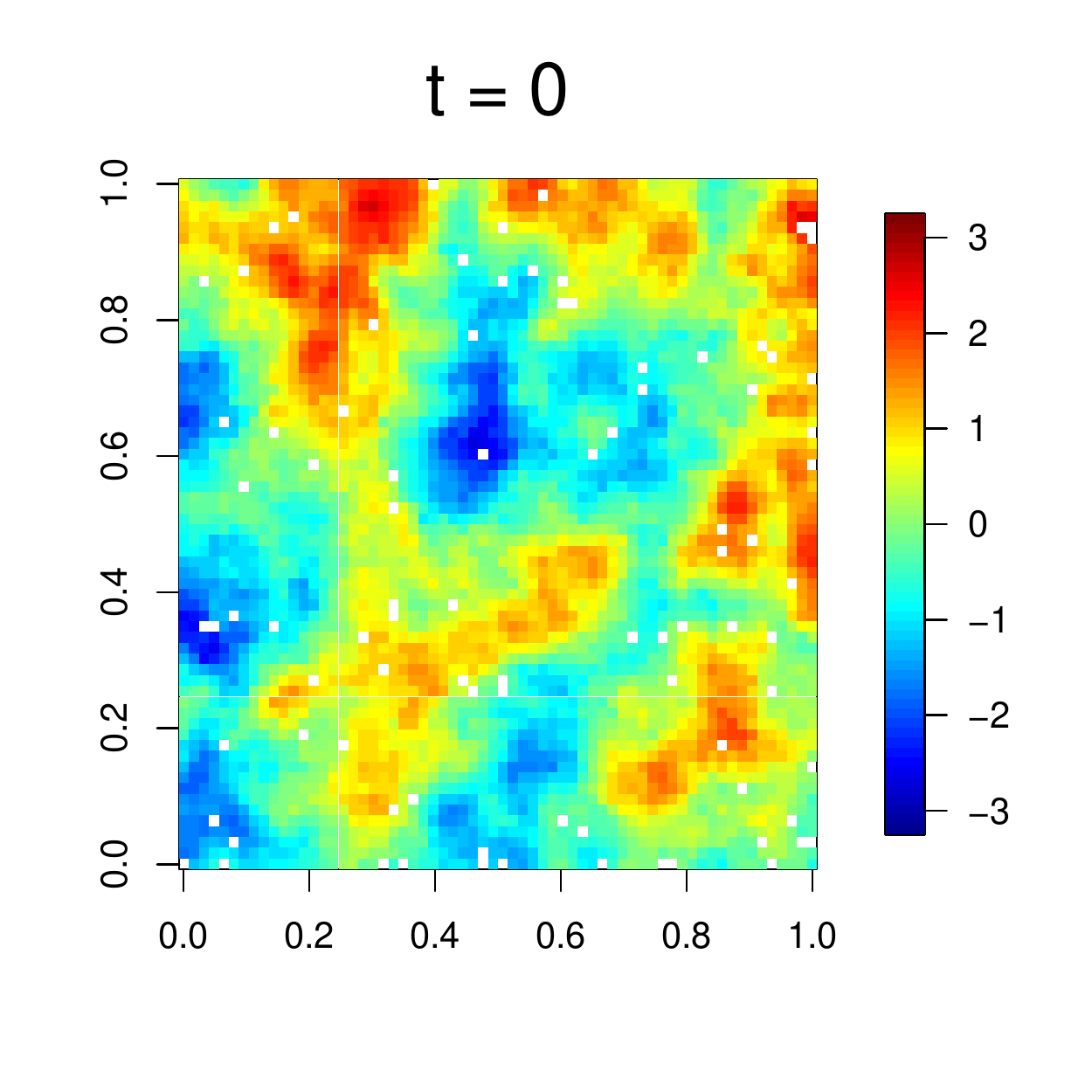}
\hspace{-2mm}
\includegraphics[width=0.33\textwidth]{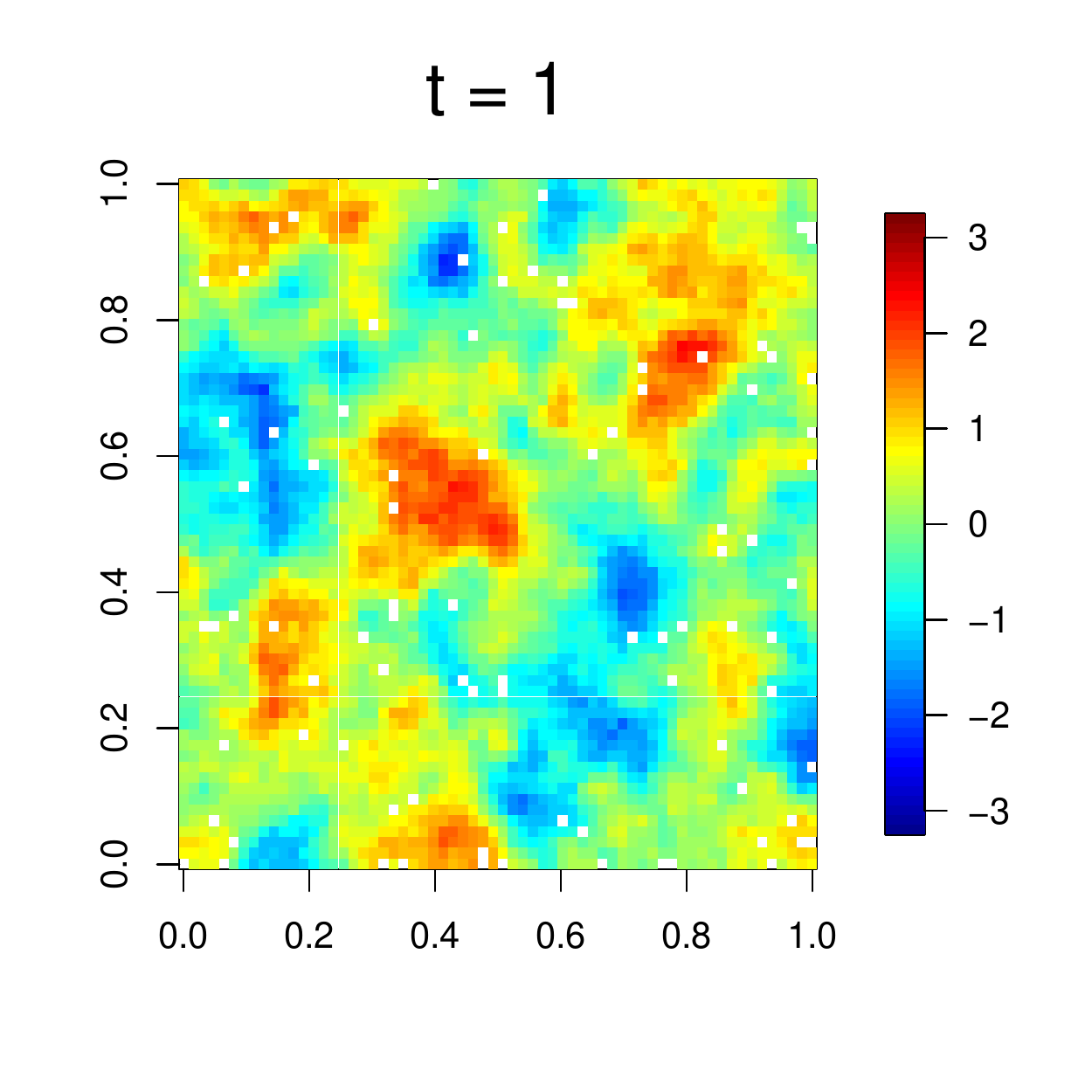}
\hspace{-2mm}
\includegraphics[width=0.33\textwidth]{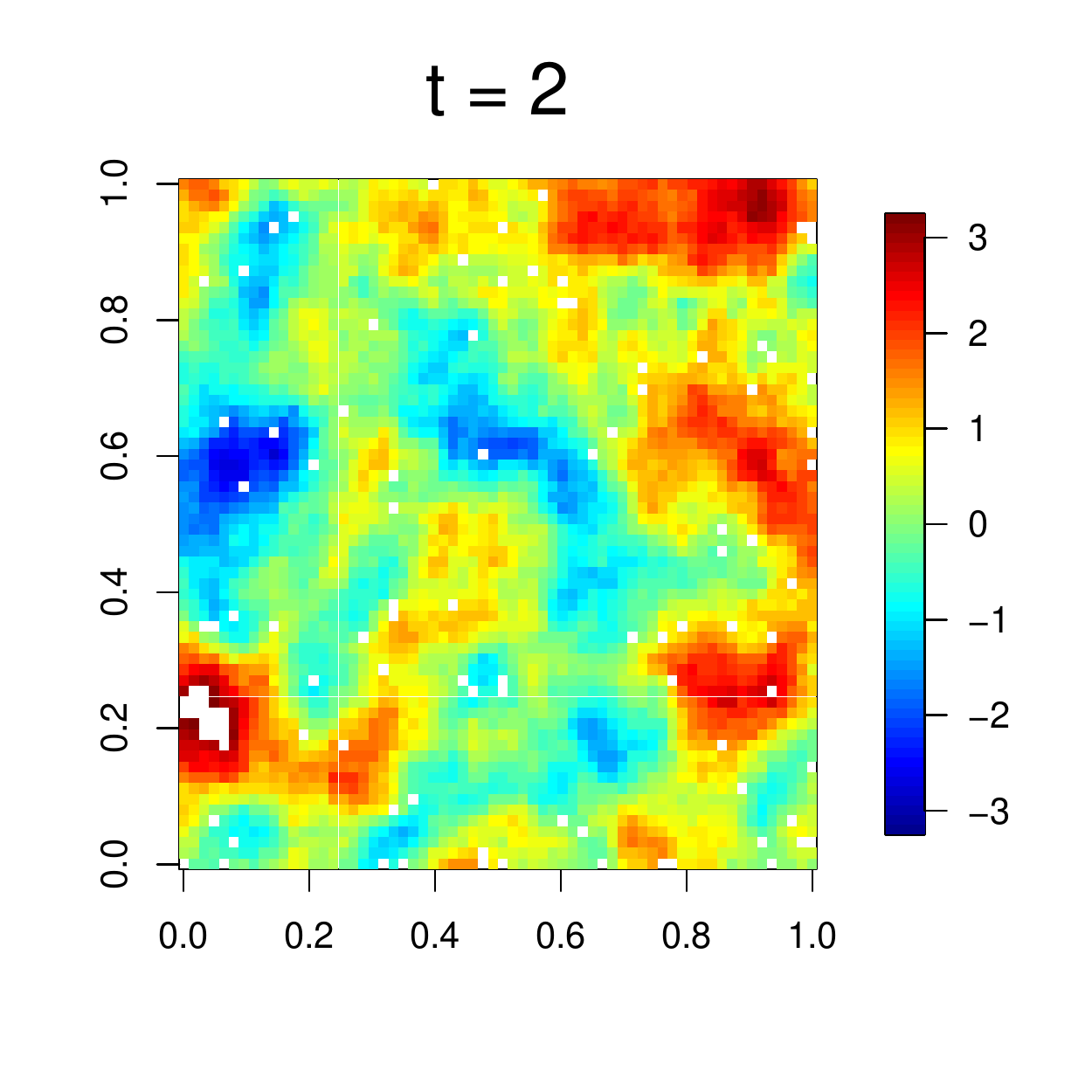}
\caption{2b-2} 
 \end{subfigure}

\end{minipage}

   \begin{minipage}[t]{\linewidth}
    \vspace{3mm}
   \subcaption{Space-time Mat\'ern ($a_s=0.4$ and $a_t =1$)}
 \vspace{-2mm}
\begin{subfigure}[b]{0.46\textwidth}
\centering
\includegraphics[width=0.33\textwidth]{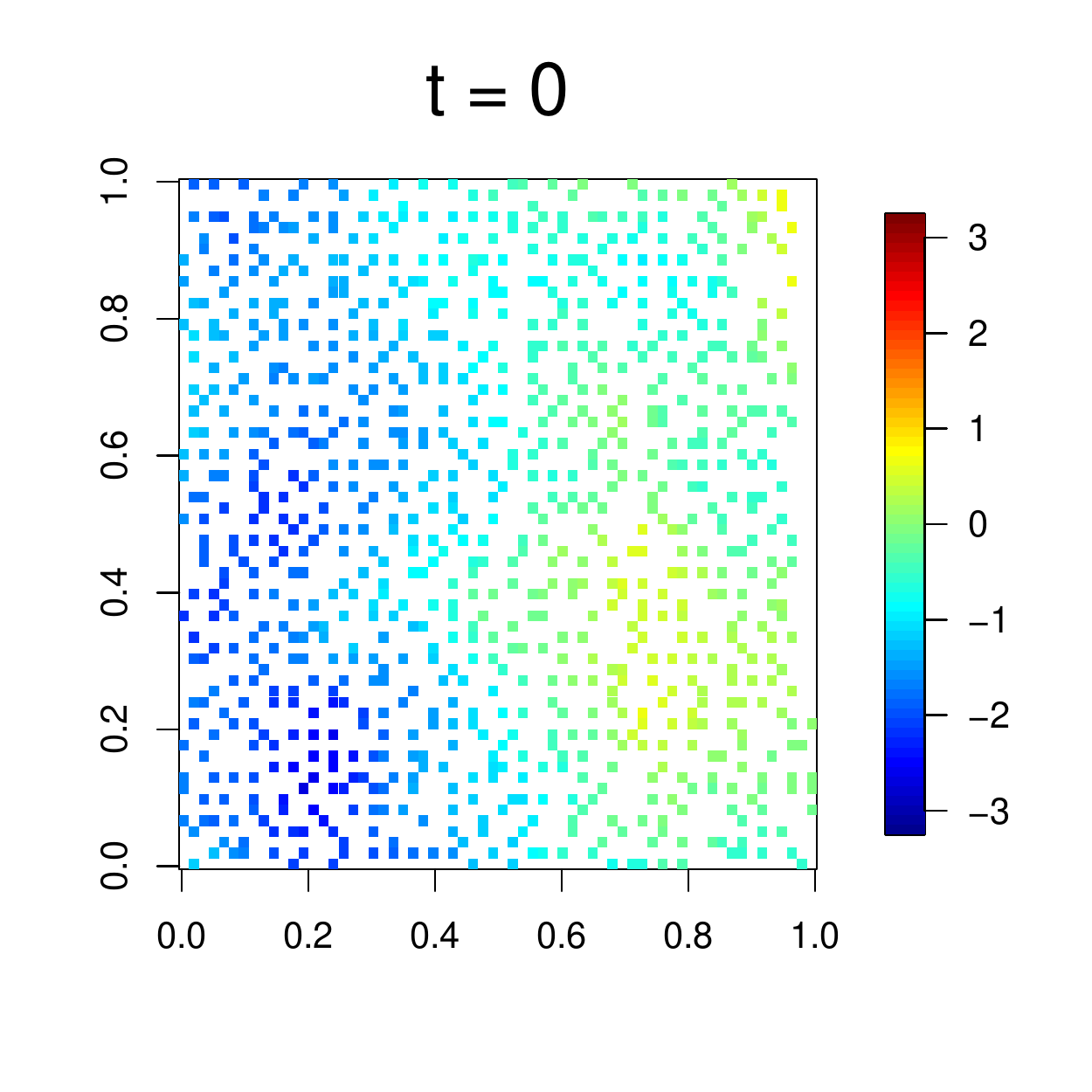}
\hspace{-2mm}
\includegraphics[width=0.33\textwidth]{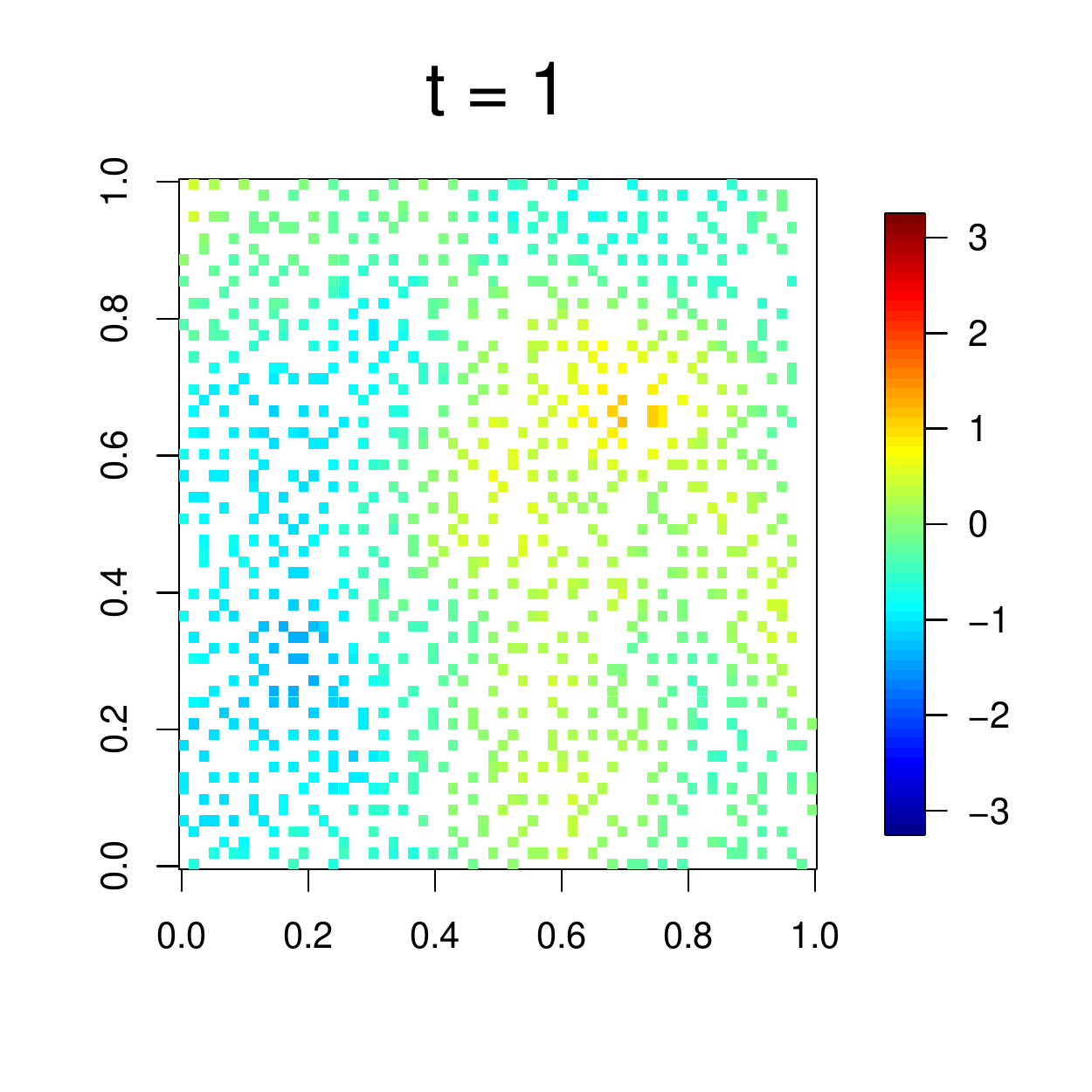}
\hspace{-2mm}
\includegraphics[width=0.33\textwidth]{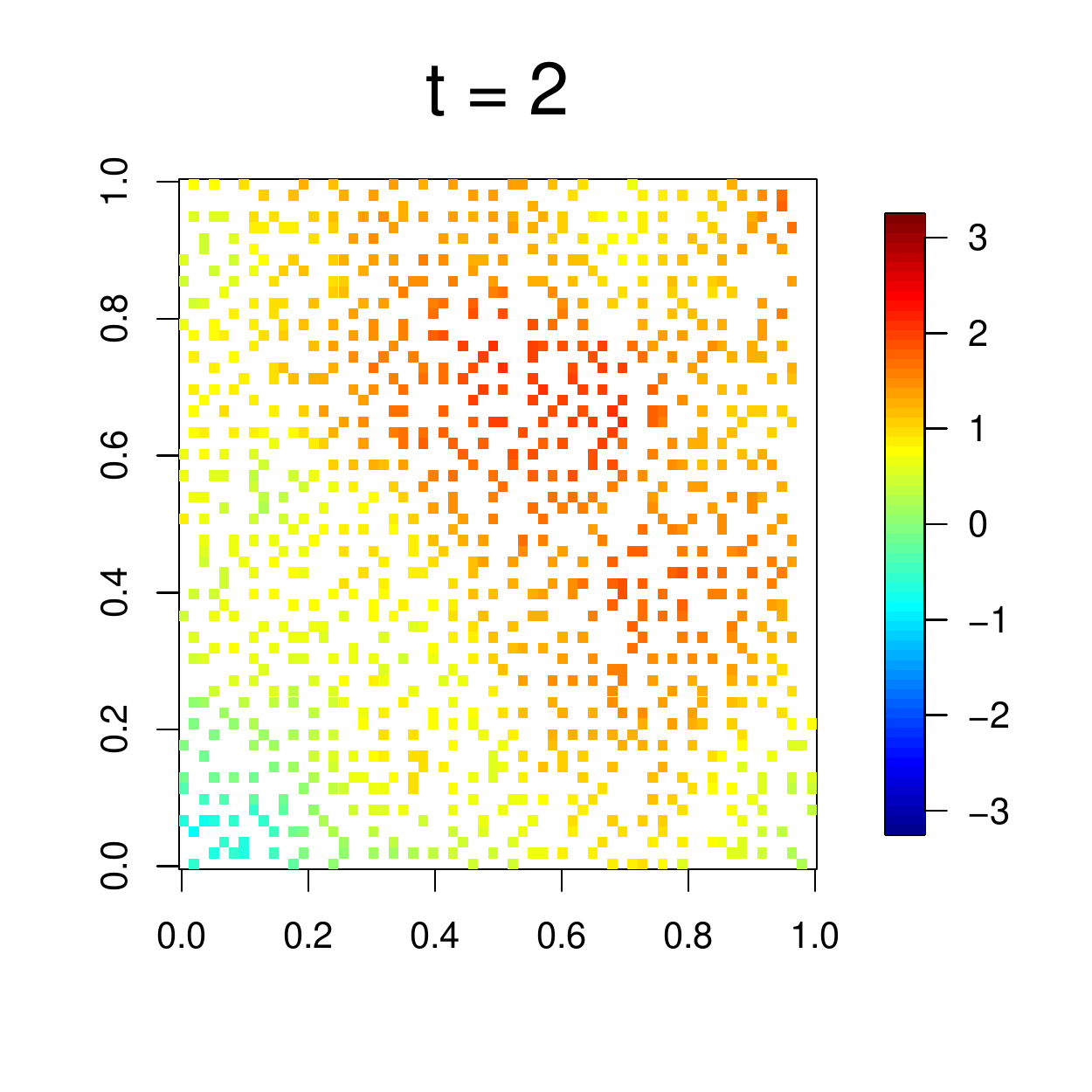}
\caption{2a-3} 
 \end{subfigure}
  \hspace{8mm}
\begin{subfigure}[b]{0.46\textwidth}
\centering
\includegraphics[width=0.33\textwidth]{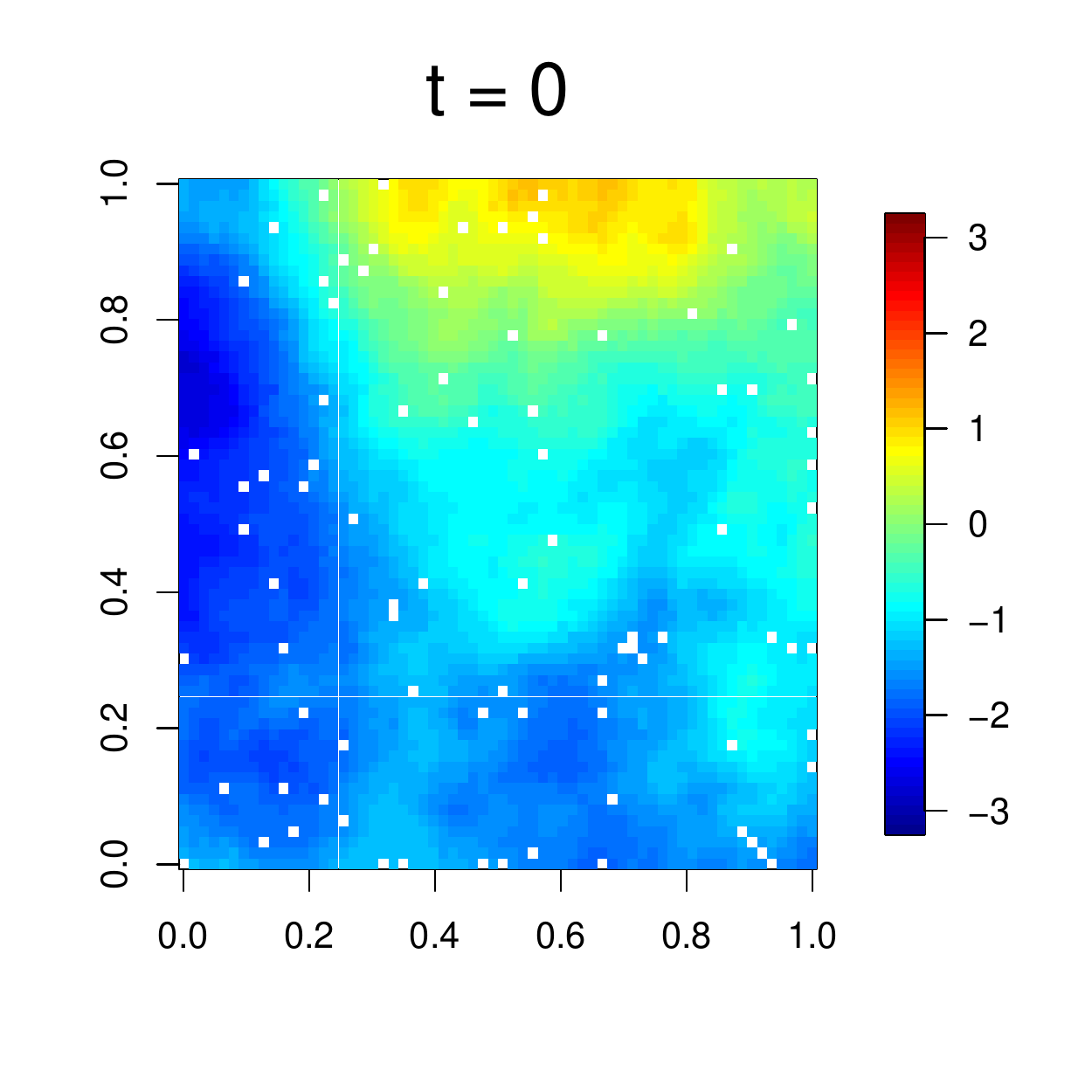}
\hspace{-2mm}
\includegraphics[width=0.33\textwidth]{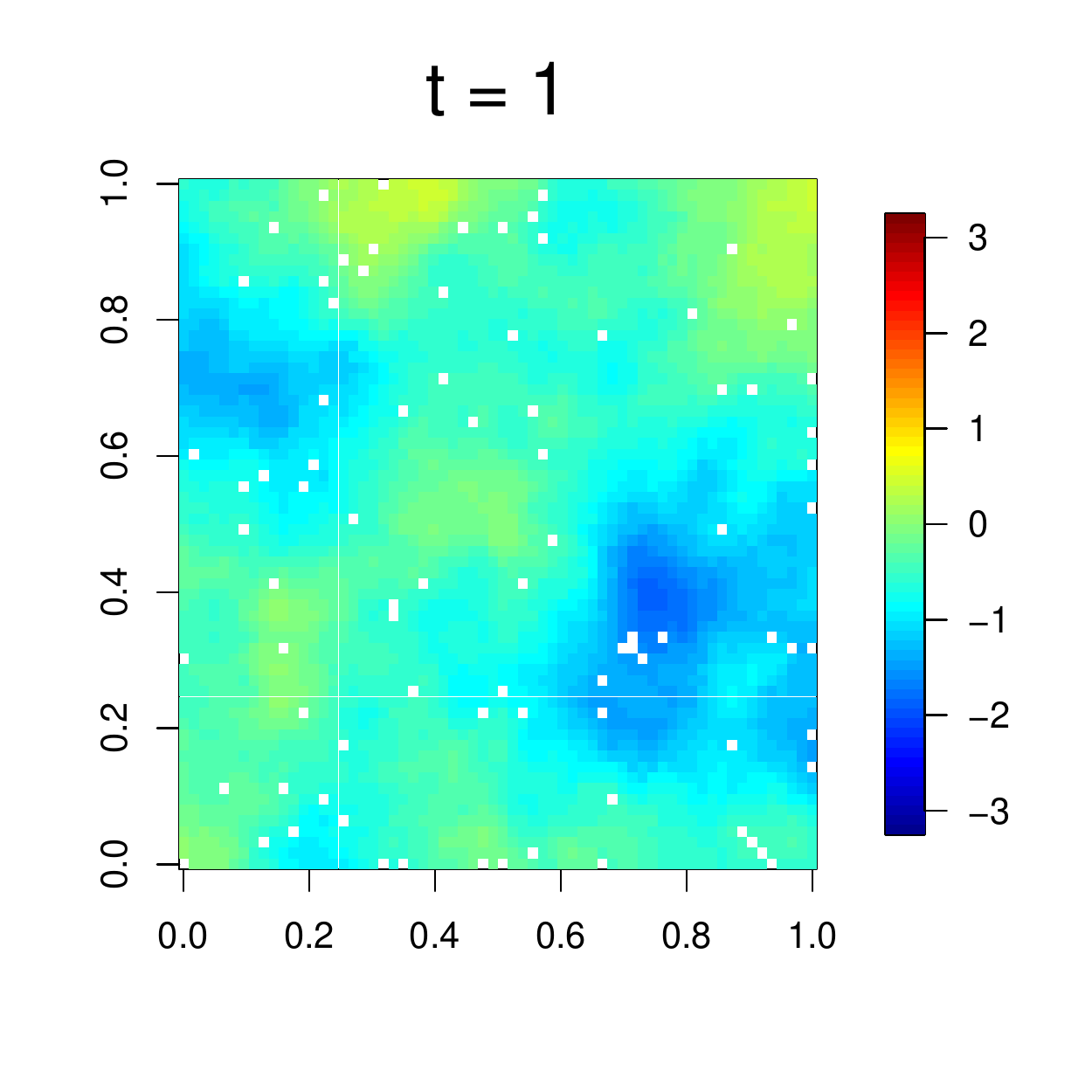}
\hspace{-2mm}
\includegraphics[width=0.33\textwidth]{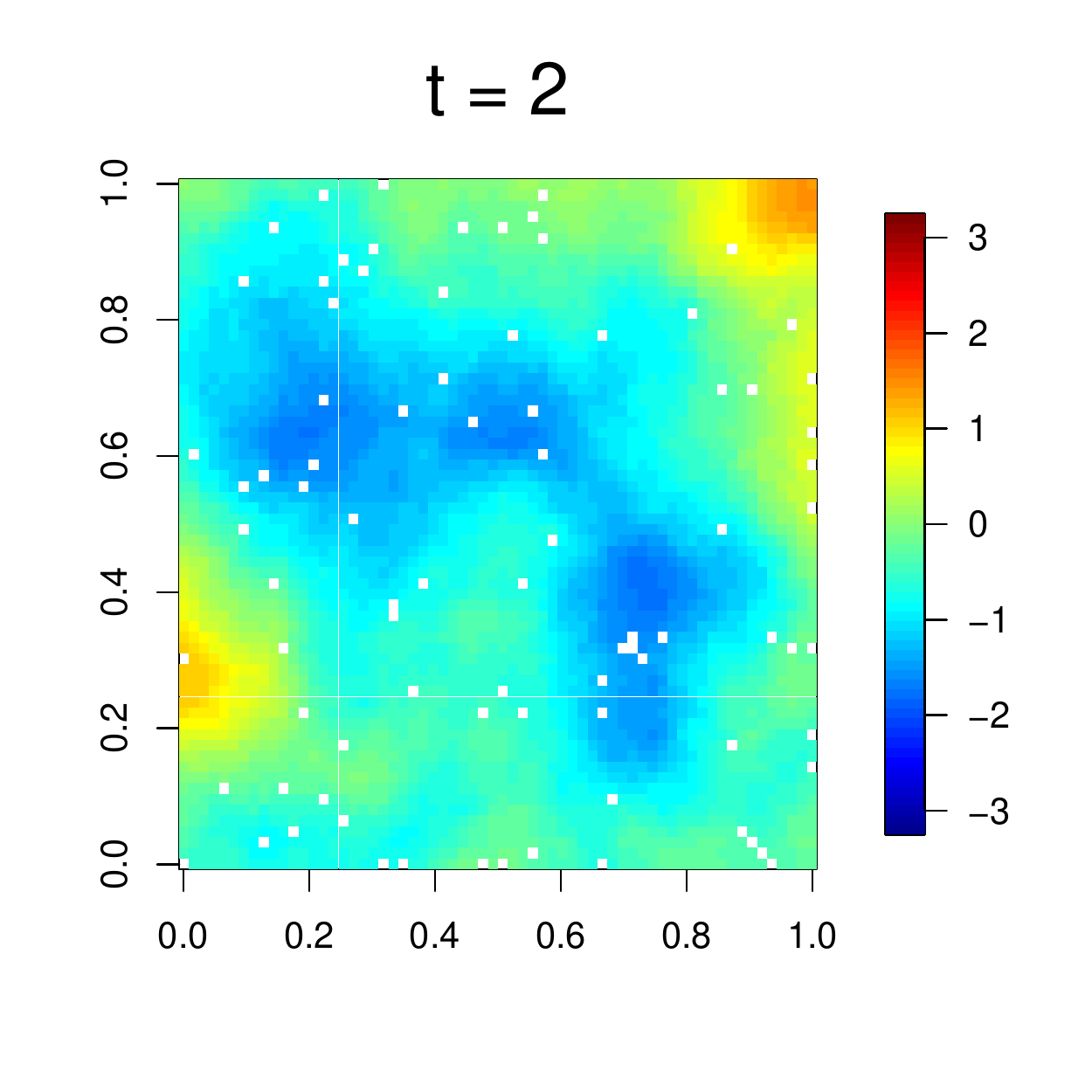}
\caption{2b-3} 
 \end{subfigure}
\end{minipage}
 \caption{(2a-1, 2a-2, 2a-3) Synthetic space-time datasets with $1$K locations and 100 time-slots generated from a non-separable space-time covariance function; (2b-1, 2b-2, 2b-3) Synthetic space-time datasets with 10K locations and 100 time-slots generated from a non-separable space-time covariance function. The figure visualizes the datasets at $t=0$, $t=1$, $t=2$ generated under the first three settings of each sub-competition in Table \ref{tab:1} with RS left out.}
\label{fig:1m-st}
\end{figure}

There were eighteen space-time datasets generated by considering different parameter settings and leaving out schemes. Figure~\ref{fig:1m-st} visualizes datasets at $t=0$, $t=1$, and $t=2$  generated under the first three settings of each sub-competition in Table \ref{tab:1} with RS left out. Other datasets (under RST and T10) were generated using the same parameter settings.

\begin{table}[h!]
  \centering
  \caption{The settings to generate datasets in Sub-competitions 2a and 2b using a non-separable stationary space-time Mat\'ern covariance model.}
  \scalebox{0.8}{
\begin{tabular}{cccccccccccc}
		\hline
		 Dataset  &  \makecell{\# spatial \\ locations}    & \makecell{\# time \\ slots}   &$\sigma^2$ & $\alpha$& $\beta$ & $\nu$& $a_{s}$ &$a_{t}$ & \makecell{Time \\ EffRange} & \makecell{Space \\ EffRange}  & \makecell{Prediction \\ setting }\\
		 	\hline
   ST1 (2a-1)   &   1K  & 100 & 0.9   & 0.6 & 0.9   & 1       &0.02  &1 & 11.63  &0.08 
   & RS\\
   
   ST2 (2a-2)  &  1K  & 100 & 0.9   & 0.6 & 0.9   & 1       &0.08 & 0.24 & 38.20&0.32 
   &RS\\
   
     ST3 (2a-3) &  1K  & 100 & 0.9   & 0.08 & 0.9   & 1     &0.4& 1 &  $\infty$& 1.6 
     &RS\\
     
    ST4 (2a-4) &  1K  & 100 & 0.9   & 0.6 & 0.9   & 1       &0.02 &1 &11.63& 0.08&RST\\
 ST5 (2a-5) &  1K  & 100 & 0.9   & 0.6 & 0.9   & 1       &0.08 & 0.24&38.20&0.32&RST\\
 ST6 (2a-6) &    1K  & 100 &   0.9   & 0.08 & 0.9   & 1     &0.4& 1 &$\infty$&1.6&RST\\
     ST7 (2a-7)&  1K  & 100 & 0.9   & 0.6 & 0.9   & 1       &0.02 &1  &11.63&0.08&T10\\
  ST8 (2a-8)&    1K  & 100 & 0.9   & 0.6 & 0.9   & 1       &0.08 & 0.24 &38.20&0.32&T10\\
    
  ST9 (2a-9) &  1K   & 100 &  0.9   & 0.08 & 0.9   & 1     &0.4& 1 &$\infty$&1.6&T10 \\
  \hline 
   ST10 (2b-1) & 10K   & 100 & 0.9   & 0.6 & 0.9   & 1       &0.02 &1&11.63&0.08&RS\\
  ST11 (2b-2) &  10K   & 100 & 0.9   & 0.6 & 0.9   & 1     &0.08 & 0.24&38.20&0.32&RS\\
   ST12 (2b-3)& 10K    & 100 & 0.9   & 0.08 & 0.9   & 1     &0.4& 1 &$\infty$&1.6&RS\\
  ST13 (2b-4)&  10K   & 100 & 0.9   & 0.6 & 0.9   & 1       &0.02 &1&11.63&0.08&RST\\
   ST14 (2b-5)& 10K    & 100  &0.9   & 0.6 & 0.9   & 1       &0.08 & 0.24&38.20&0.32&RST\\
  ST15 (2b-6) &  10K   & 100   &0.9   & 0.08 & 0.9   & 1     &0.4& 1 &$\infty$&1.6&RST\\
  ST16 (2b-7) &  10K   & 100  &0.9   & 0.6 & 0.9   & 1       &0.02 &1 &11.63&0.08&T10\\
  ST17 (2b-8) &  10K    & 100  &0.9   & 0.6 & 0.9   & 1       &0.08 & 0.24 &38.20&0.32&T10\\
  ST18 (2b-9) & 10K   & 100 &  0.9   & 0.08 & 0.9   & 1     &0.4& 1 &$\infty$&1.6&T10\\
    	\hline
    \end{tabular}%
    }
  \label{tab:1}%
\end{table}%


\subsection{Bivariate Stationary Spatial Model (Sub-competitions 3a and 3b)}

The third part of the competition involved generating datasets from bivariate zero-mean Gaussian random fields $(Z_1(\bm s),Z_2(\bm s))^\top$ using two different cross-covariance functions (without nugget): the parsimonious \citep{gneiting2010} and the flexible \citep{apanasovich2012valid} Mat\'ern. 

The parsimonious Mat\'ern cross-covariance function has the following form:
\begin{equation}
    C_{ij}(\bm h;\bm \theta)= \frac{\rho_{ij} \sigma_{ii}\sigma_{jj}}{2^{\nu_{ij}-1}\Gamma(\nu_{ij})}\left(\frac{\|\bm h\|}{a}\right)^{\nu_{ij}} {\cal{K}}_{\nu_{ij}}\left( {\frac{\|\bm h\|}{a}}\right),
    \label{eq:para}
\end{equation}
for $i,j=1,2$. Here $\boldsymbol{\theta}$ consists of the following parameters: marginal smoothnesses $\nu_{ii}>0$, a common range $a>0$, marginal variances $\sigma^2_{ii}>0$, and the collocated correlation $\rho_{ij}$, $i \ne j$. The cross-smoothness 
$\nu_{ij}=\frac{1}{2}(\nu_{ii}+\nu_{jj})$ and $\rho_{ij}$ are related: $$\rho_{ij}=\beta_{ij}{\frac{\Gamma(\nu_{ii}+\frac{d}{2})^{\frac{1}{2}}}{\Gamma(\nu_{ii})^{\frac{1}{2}}} }
{\frac{\Gamma(\nu_{jj}+\frac{d}{2})^{\frac{1}{2}}}{\Gamma(\nu_{jj})^{\frac{1}{2}}} }
{\frac{\Gamma\{\frac{1}{2}(\nu_{ii}+\nu_{jj})\}}{\Gamma\{\frac{1}{2}(\nu_{ii}+\nu_{jj})+\frac{d}{2}\}}},$$
where $d = 2 $ and  $(\beta_{ij})^{p}_{i,j=1}$ is a symmetric and positive definite correlation matrix. More descriptions can be found in \cite{salvana2021high}. 

The flexible Mat\'ern cross-covariance function has the following form:
 \begin{equation}
    C_{ij}(\bm h;\bm \theta)= \frac{\rho_{ij} \sigma_{ii}\sigma_{jj}}{2^{\nu_{ij}-1}\Gamma(\nu_{ij})}\left(\frac{\|\bm h\|}{a_{ij}}\right)^{\nu_{ij}} {\cal{K}}_{\nu_{ij}}\left( {\frac{\|\bm h\|}{a_{ij}}}\right), 
    \label{eq:flex}
\end{equation}   
where $a^2_{ij}=(a^2_{ii}+a^2_{jj})/2+\bar \tau(a_{ii}-a_{jj})^2, 0 \leq\bar \tau < \infty$. In this case, the marginal ranges $a_{ii}$ are no longer constant across variables. 

Table~\ref{tab:3} shows two different settings to generate a $50$K (3a-1) and a $500$K (3b-1) datasets using the parsimonious bivariate model, with $\beta_{12}= 0.9$, $a=a_{11}=a_{22}=0.03$, $\nu_{11}= 0.6$ and $\nu_{22}= 1.4$. Table~\ref{tab:3} also shows two different settings of the flexible bivariate Mat\'ern model with $\beta_{12}= 0.9$ and $\bar\tau=0$:
   \begin{enumerate}[label=(\alph*)]
        \item[$\bullet$] Same smoothness parameters $\nu_{11}=\nu_{22}= 0.9$ and different range parameters with  
        $a_{11}=0.02$ indicating weak dependence and $a_{22}=0.3$ indicating strong dependence (3a-2 and 3b-2);
        \item[$\bullet$] Mixed parameters setting: smoothness parameters  $\nu_{11}=0.6$ and $\nu_{22}= 1.4$, and range parameters $a_{11}=0.03$ and $a_{22}=0.1$ (3a-3 and 3b-3).
    \end{enumerate}


\begin{table}[h!]
  \centering
  \caption{The settings to generate datasets in Sub-competitions 3a and 3b using a parsimonious/flexible Mat\'ern covariance model.}
    \scalebox{0.9}{
\begin{tabular}{ccccccccccccc}
		\hline
		Dataset & Size &  $\sigma_{11}^2$ & $\sigma_{22}^2$ & $\beta_{12}$ & $\nu_{11}$& $\nu_{22}$& $a_{11}$ &$a_{22}$&   \makecell{ EffRange\\{of $C_{11}$}} &  \makecell{ EffRange\\{of $C_{22}$}}&  $\rho_{12}$ & \makecell{ Mat\'ern\\{Model}} \\
		 	\hline
    3a-1 & 50K&    0.9  & 0.9 & 0.9 & 0.6  &1.4 & 0.03 & 0.03 &  0.097& 0.138& 0.824 & Parsimonious   \\
    3b-1  & 500K&    0.9   &0.9 & 0.9& 0.6 &1.4& 0.03&0.03 & 0.097& 0.138& 0.824 & Parsimonious  \\
  	\hline
   3a-2  & 50K&    0.9  & 0.9 & 0.9 & 0.9  & 0.9   &  0.02& 0.3&0.077 &1.15 & 0.9 & Flexible  \\
    
    3a-3 & 50K&    0.9  & 0.9 & 0.9 & 0.6  &1.4 &0.03&   0.1&0.097 & 0.46 & 0.824 & Flexible  \\

    3b-2 & 500K&    0.9   &0.9 & 0.9& 0.9 &0.9 &  0.02& 0.3 &0.077 &1.15 & 0.9 & Flexible  \\
    
    3b-3 & 500K&    0.9  & 0.9 & 0.9 & 0.6  & 1.4  &0.03&   0.1& 0.097 & 0.46 & 0.824 & Flexible  \\

		\hline
    \end{tabular}%
    }
  \label{tab:3}%
\end{table}%

\begin{figure}[h!]
\captionsetup[subfigure]{labelformat=empty}
  \begin{minipage}[t]{\linewidth}
\captionsetup[subfigure]{labelformat=empty}
\centering
  \vspace{3mm}
 \subcaption{Parsimonious bivariate Mat\'ern}
 \vspace{-2mm}
\begin{subfigure}[b]{0.2\textwidth}
\centering
\includegraphics[scale=0.2]{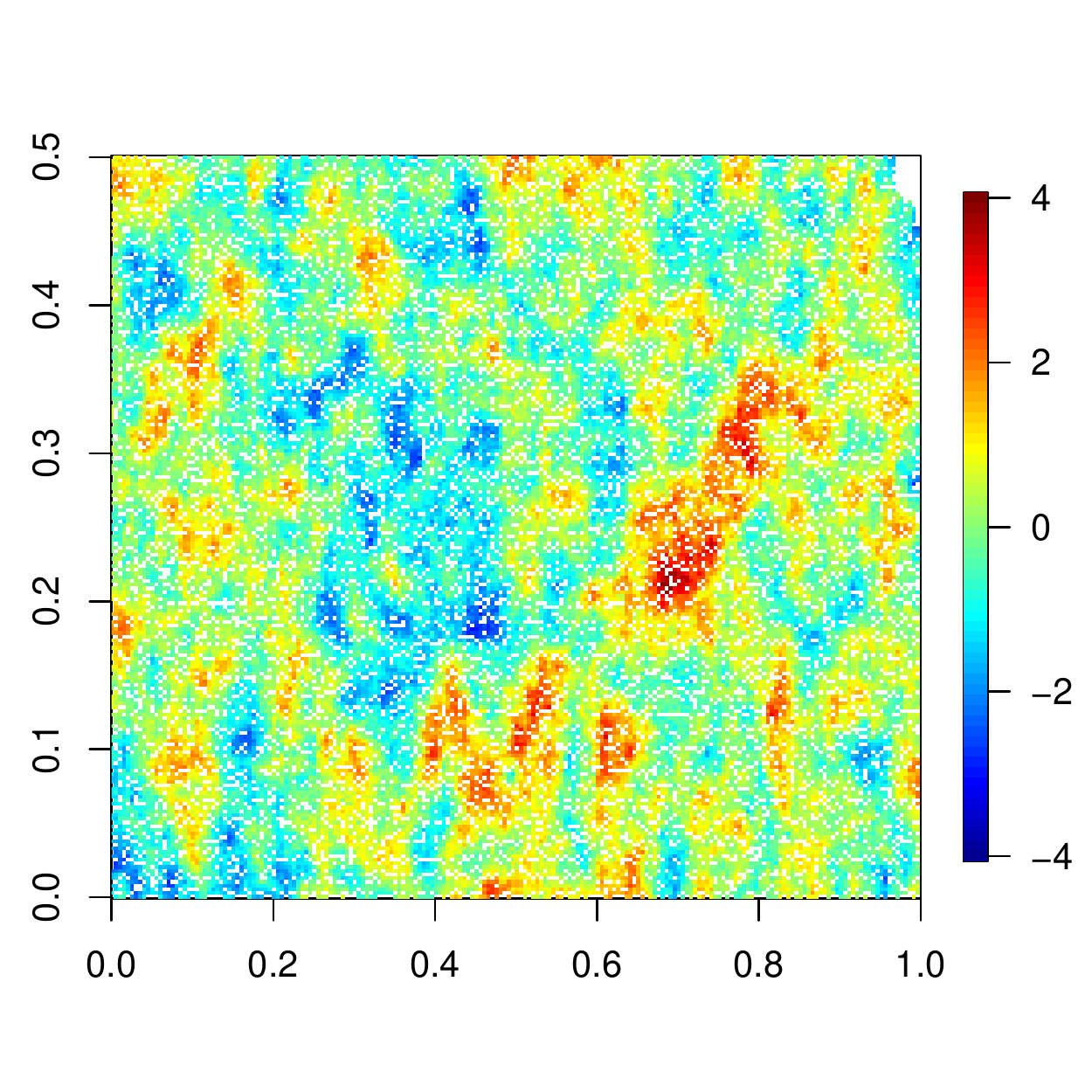}
\caption{3a-1 ($Z_1$)}
 \end{subfigure}
 \begin{subfigure}[b]{0.2\textwidth}
\centering
\includegraphics[scale=0.2]{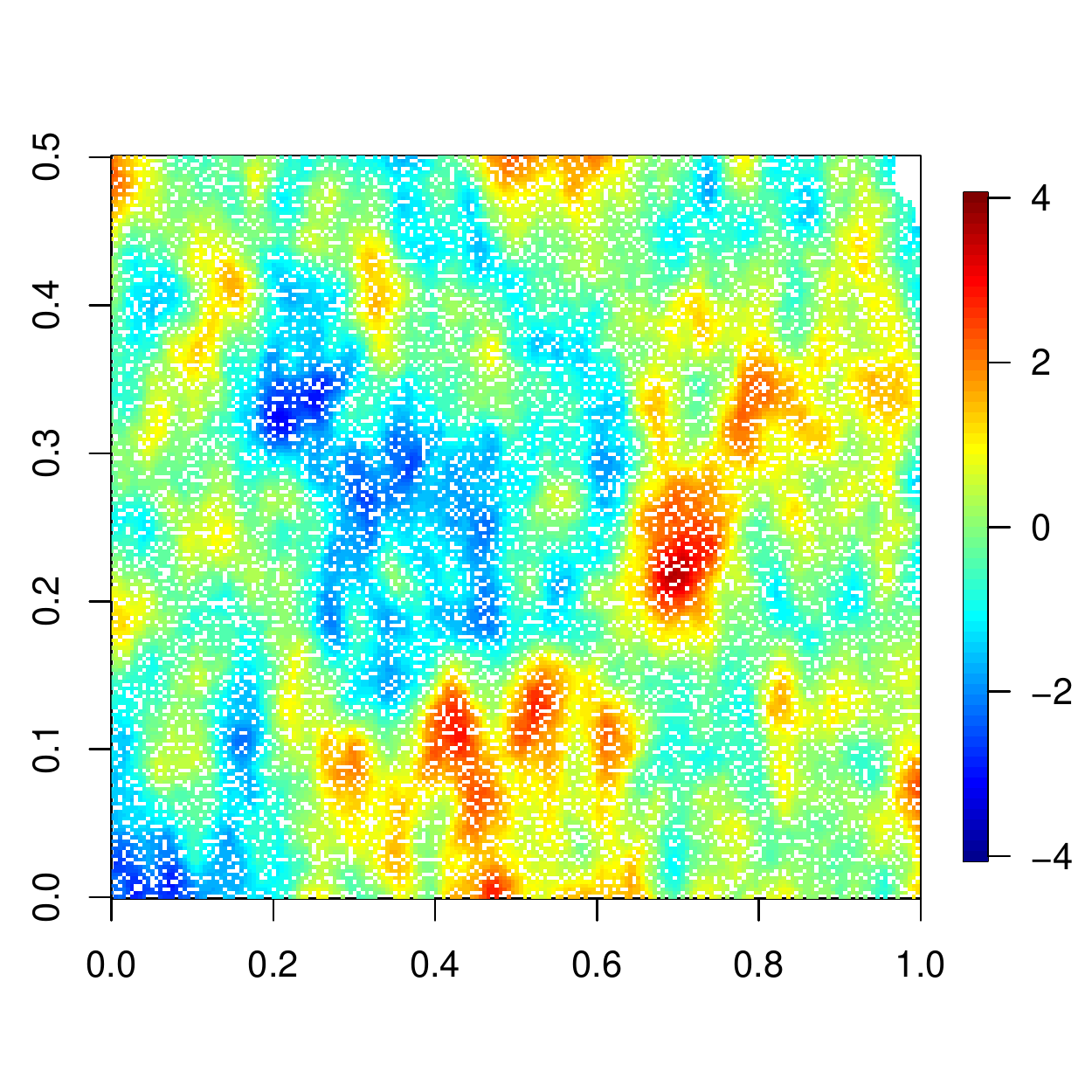}
\caption{3a-1 ($Z_2$)}
 \end{subfigure}
 \hspace{10mm}
 \begin{subfigure}[b]{0.2\textwidth}
  \centering
\includegraphics[scale=0.2]{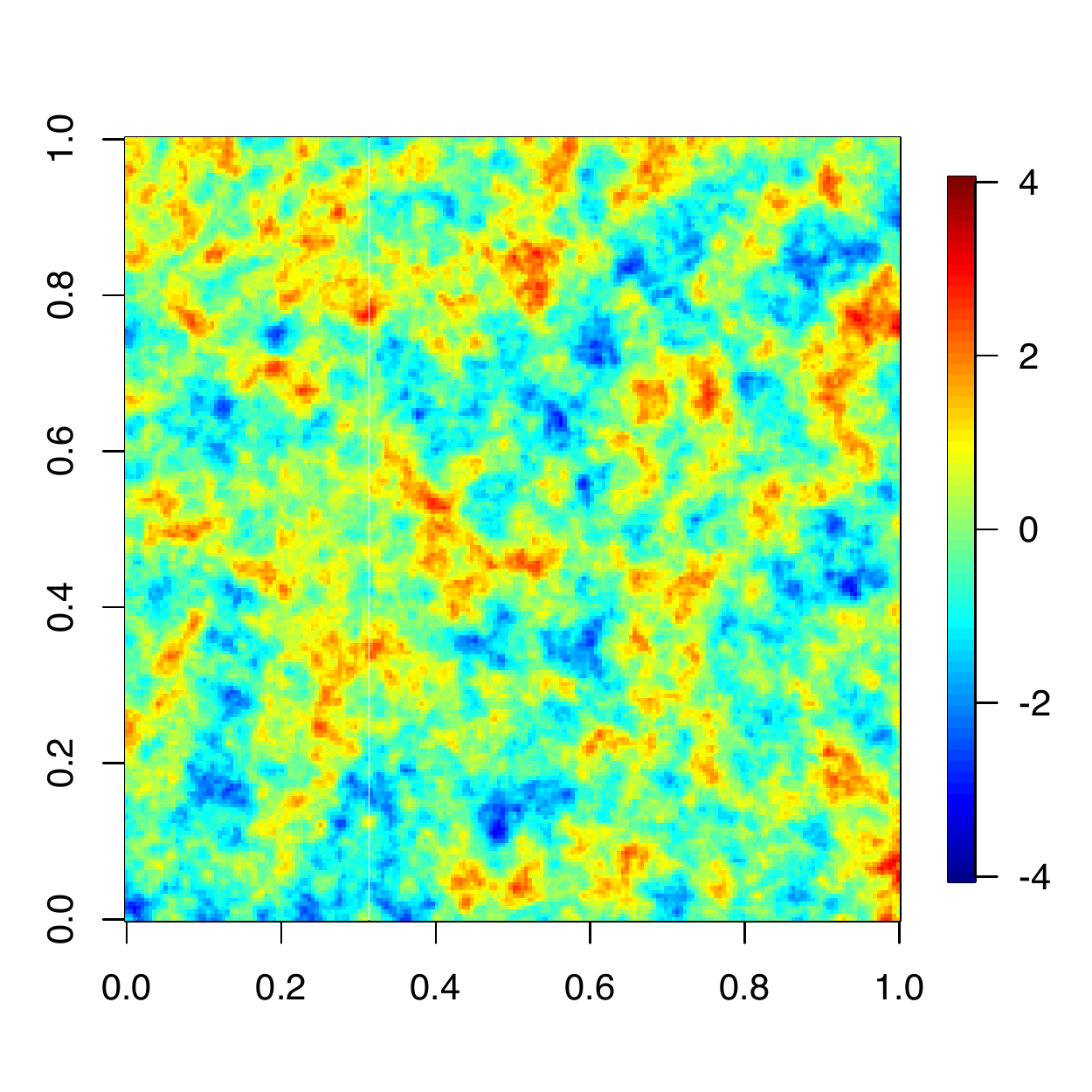}
\caption{3b-1 ($Z_1$)}
\label{3b-1-z1}
 \end{subfigure}
  \begin{subfigure}[b]{0.2\textwidth}
  \centering
\includegraphics[scale=0.2]{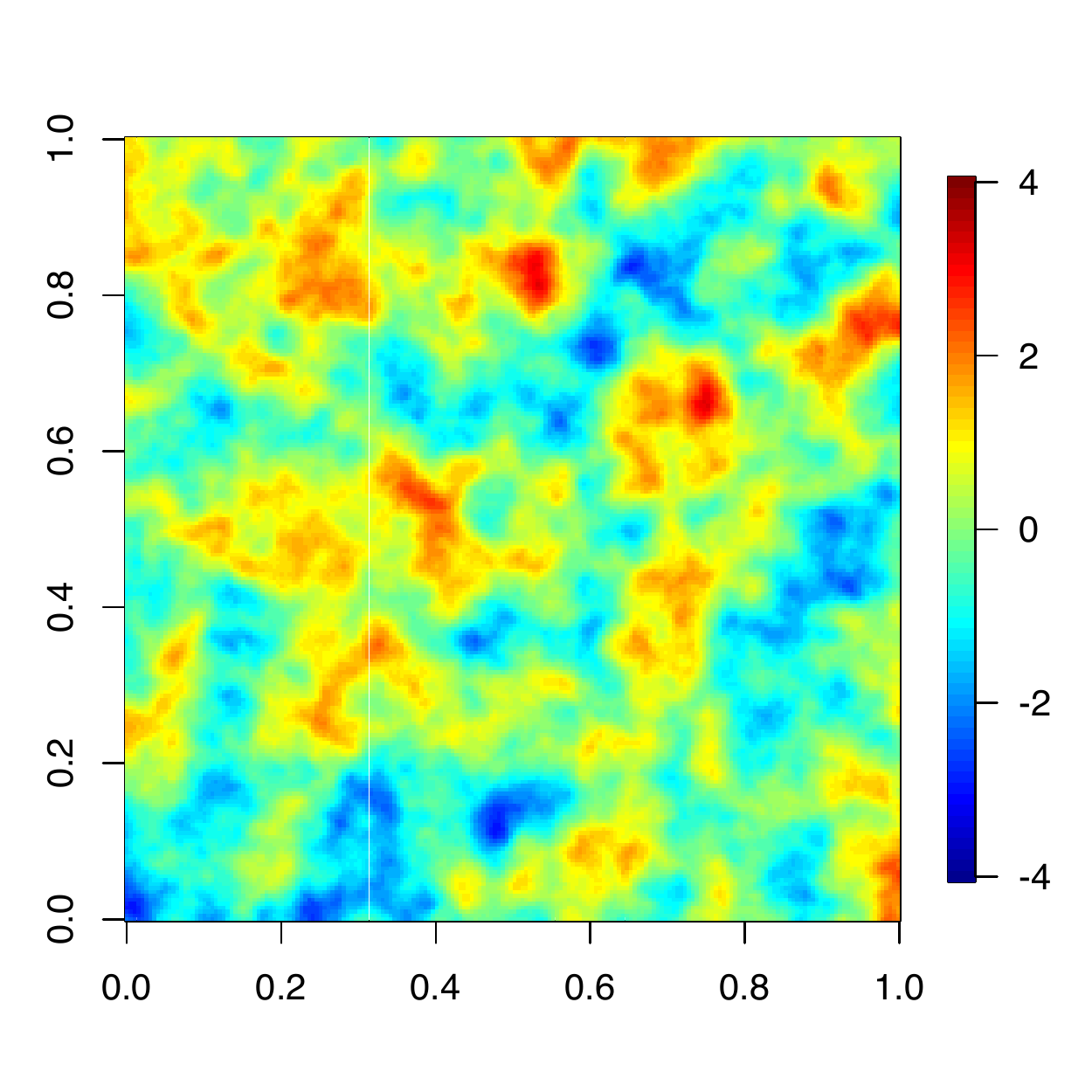}
\caption{3b-1 ($Z_2$)}
\label{3b-1-z2}
 \end{subfigure}

 \end{minipage}

\begin{minipage}[t]{\linewidth}
\centering
  \vspace{4mm}
  \subcaption{Flexible bivariate Mat\'ern}
    \vspace{-2mm}
\captionsetup[subfigure]{labelformat=empty}
 \begin{subfigure}[b]{0.2\textwidth}
\centering
\includegraphics[scale=0.2]{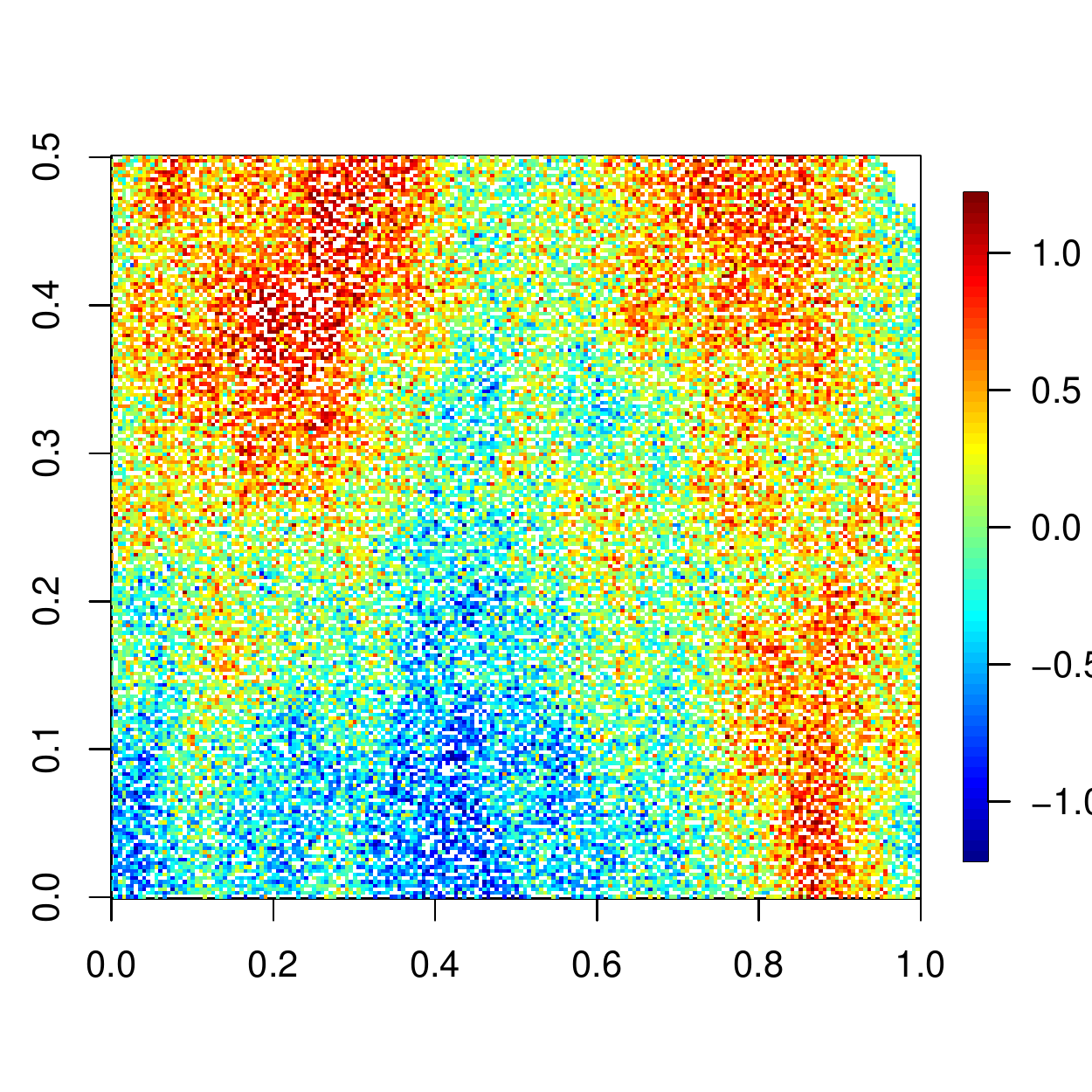}
\caption{3a-2 ($Z_1$)}
 \end{subfigure}
 \begin{subfigure}[b]{0.2\textwidth}
\centering
\includegraphics[scale=0.2]{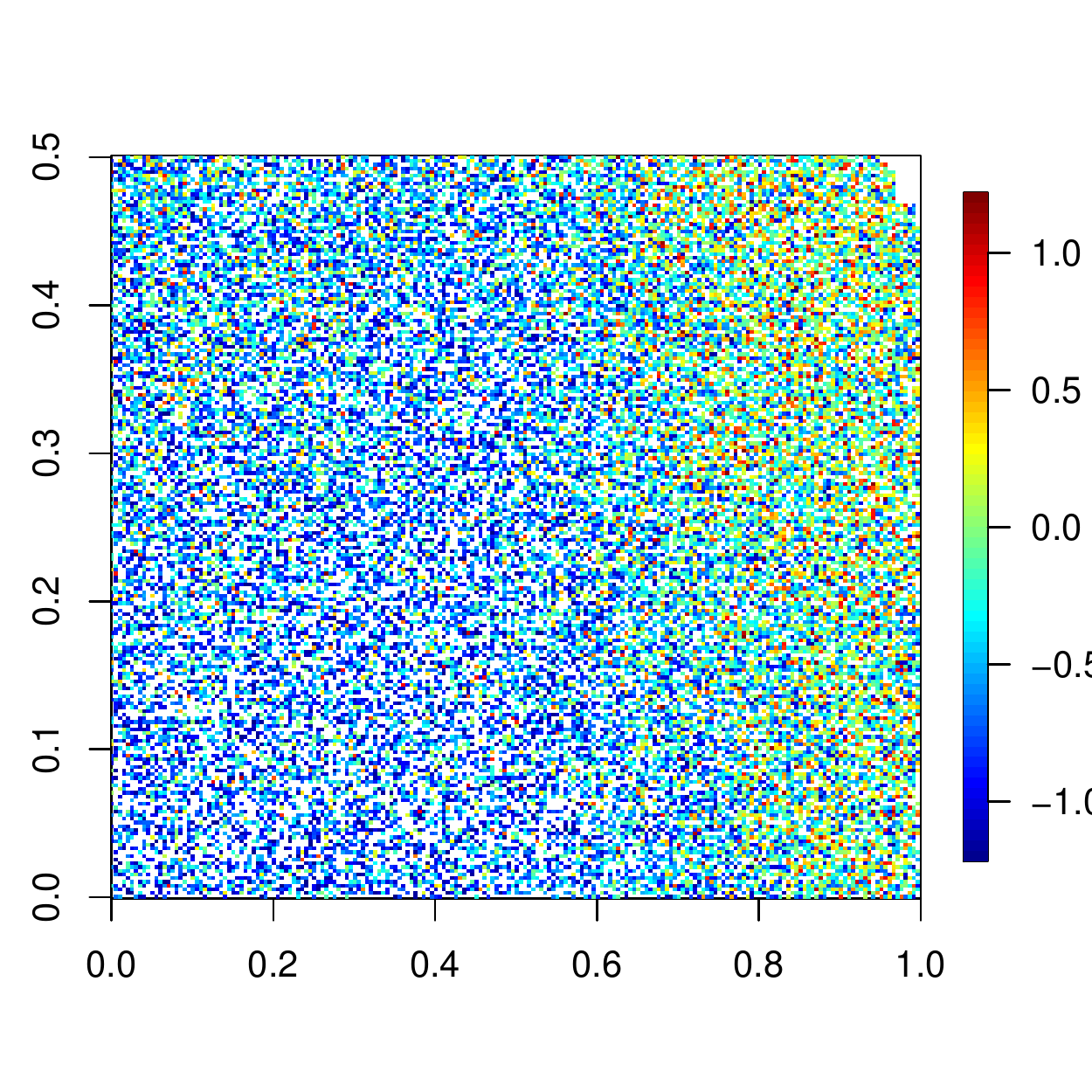}
\caption{3a-2 ($Z_2$)}
 \end{subfigure}
 \hspace{10mm}
 \begin{subfigure}[b]{0.2\textwidth}
  \centering
\includegraphics[scale=0.2]{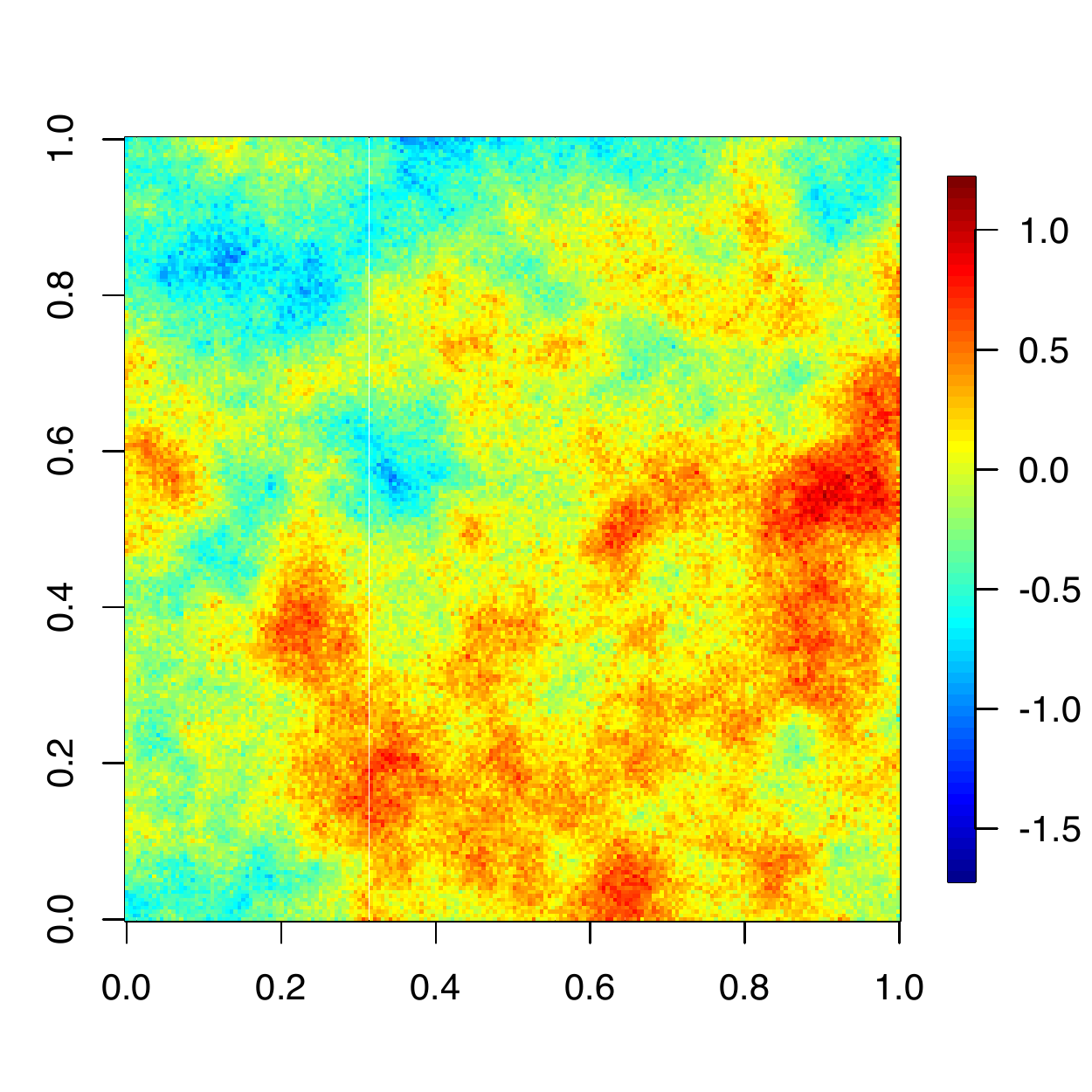}
\caption{3b-2 ($Z_1$)}
\label{3b-2-z1}
 \end{subfigure}
  \begin{subfigure}[b]{0.2\textwidth}
  \centering
\includegraphics[scale=0.2]{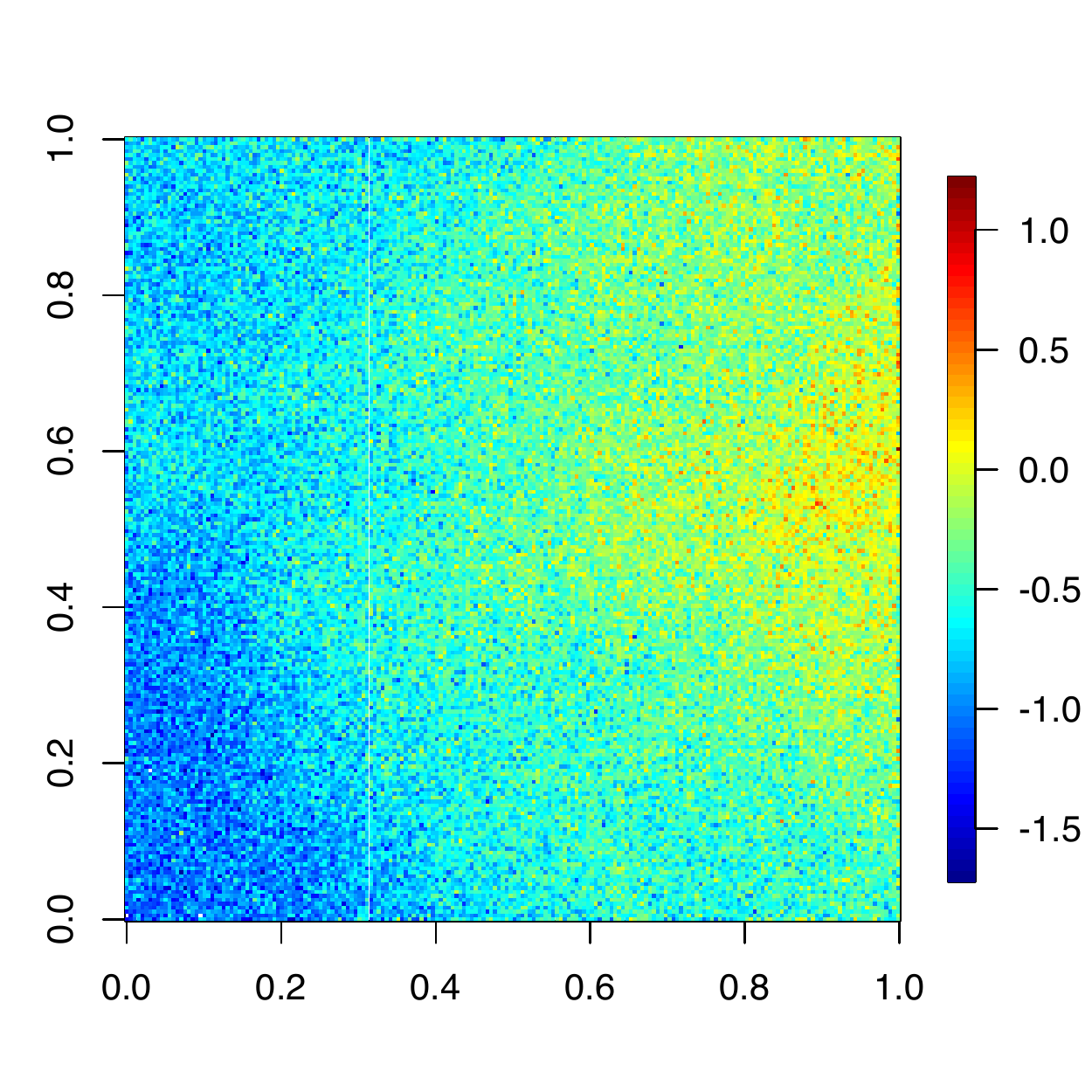}
\caption{3b-2 ($Z_2$)}
\label{3b-2-z2}
 \end{subfigure}
\begin{subfigure}[b]{0.2\textwidth}
\centering
\includegraphics[scale=0.2]{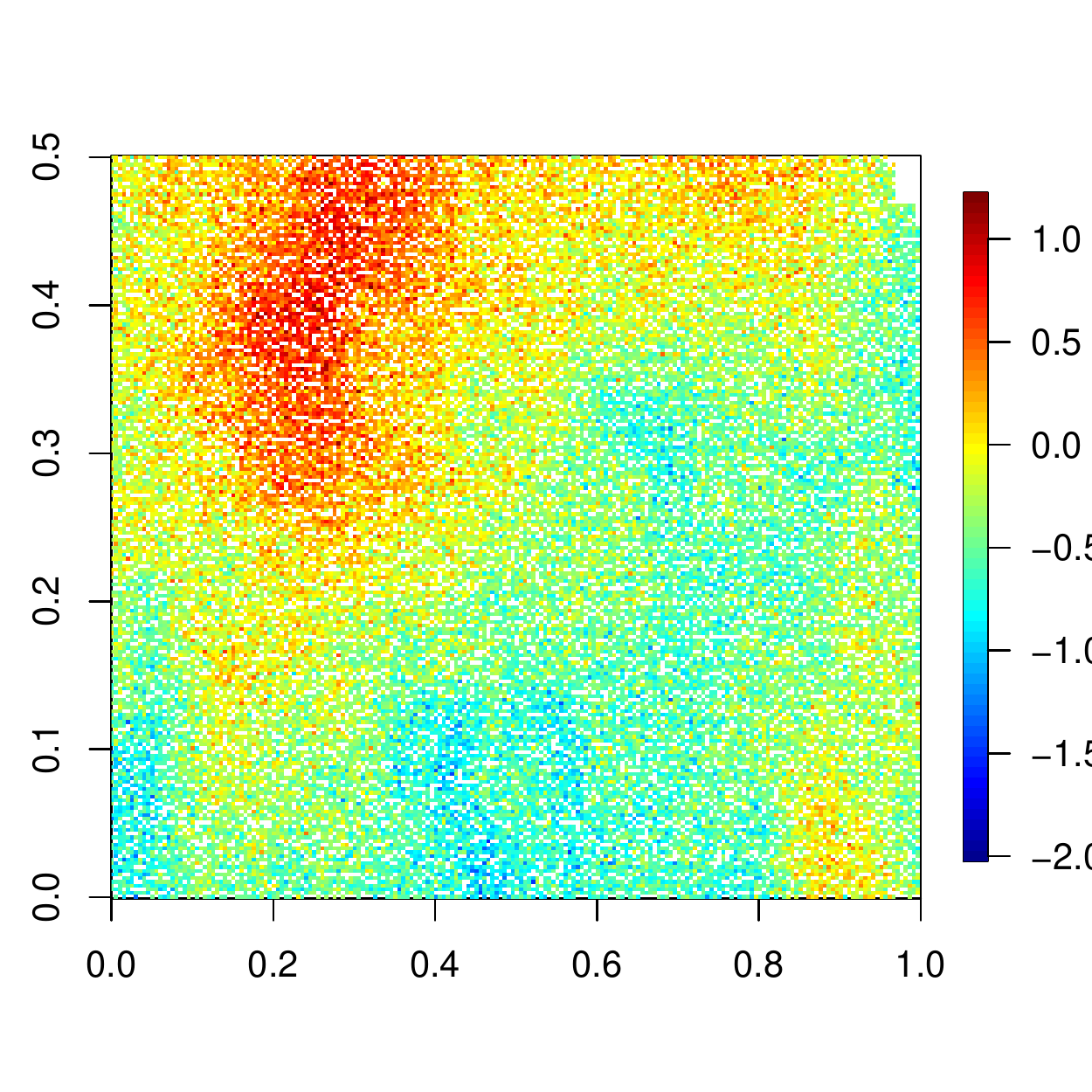}
\caption{3a-3 ($Z_1$)}
 \end{subfigure}
 \begin{subfigure}[b]{0.2\textwidth}
\centering
\includegraphics[scale=0.2]{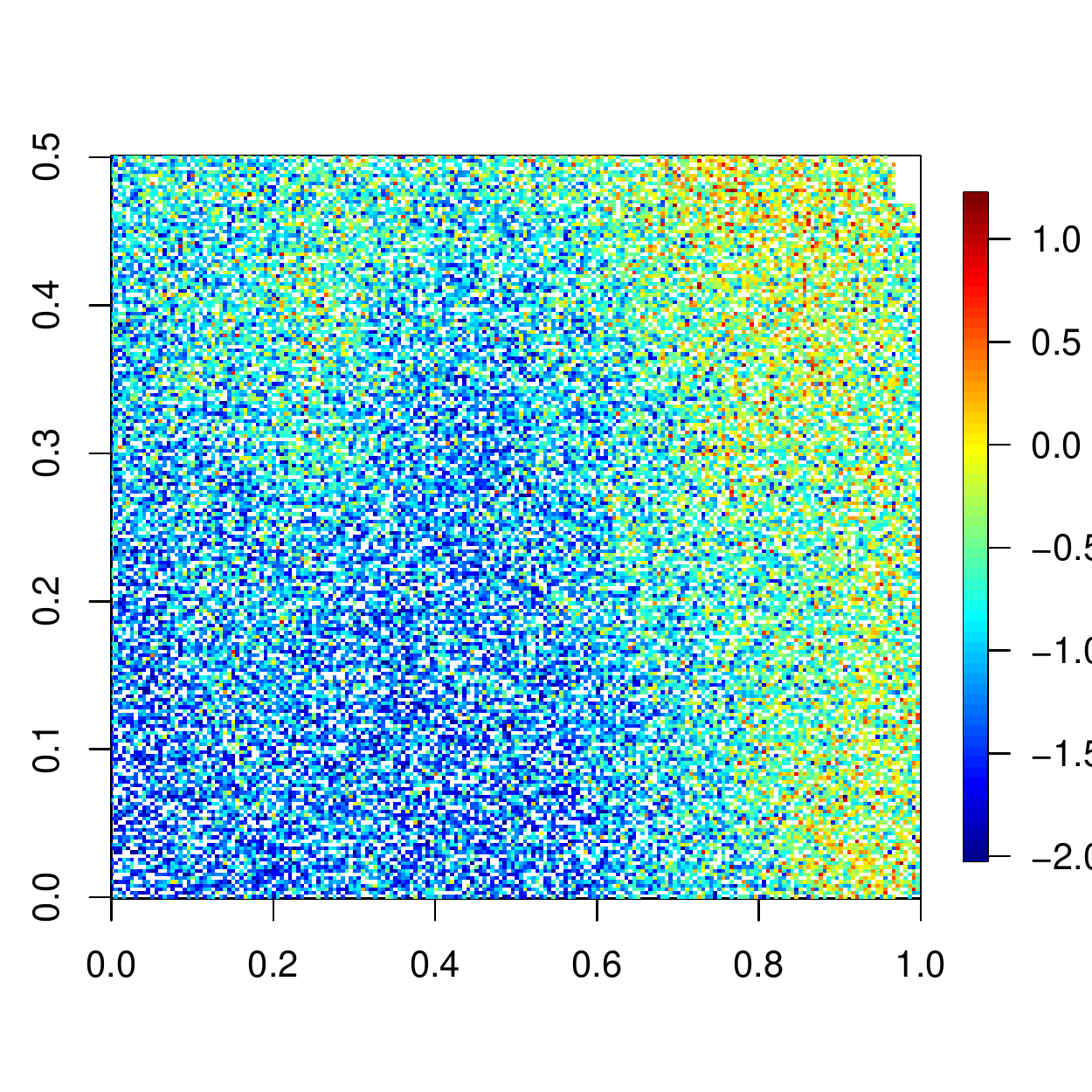}
\caption{3a-3 ($Z_2$)}
 \end{subfigure}
 \hspace{10mm}
 \begin{subfigure}[b]{0.2\textwidth}
  \centering
\includegraphics[scale=0.2]{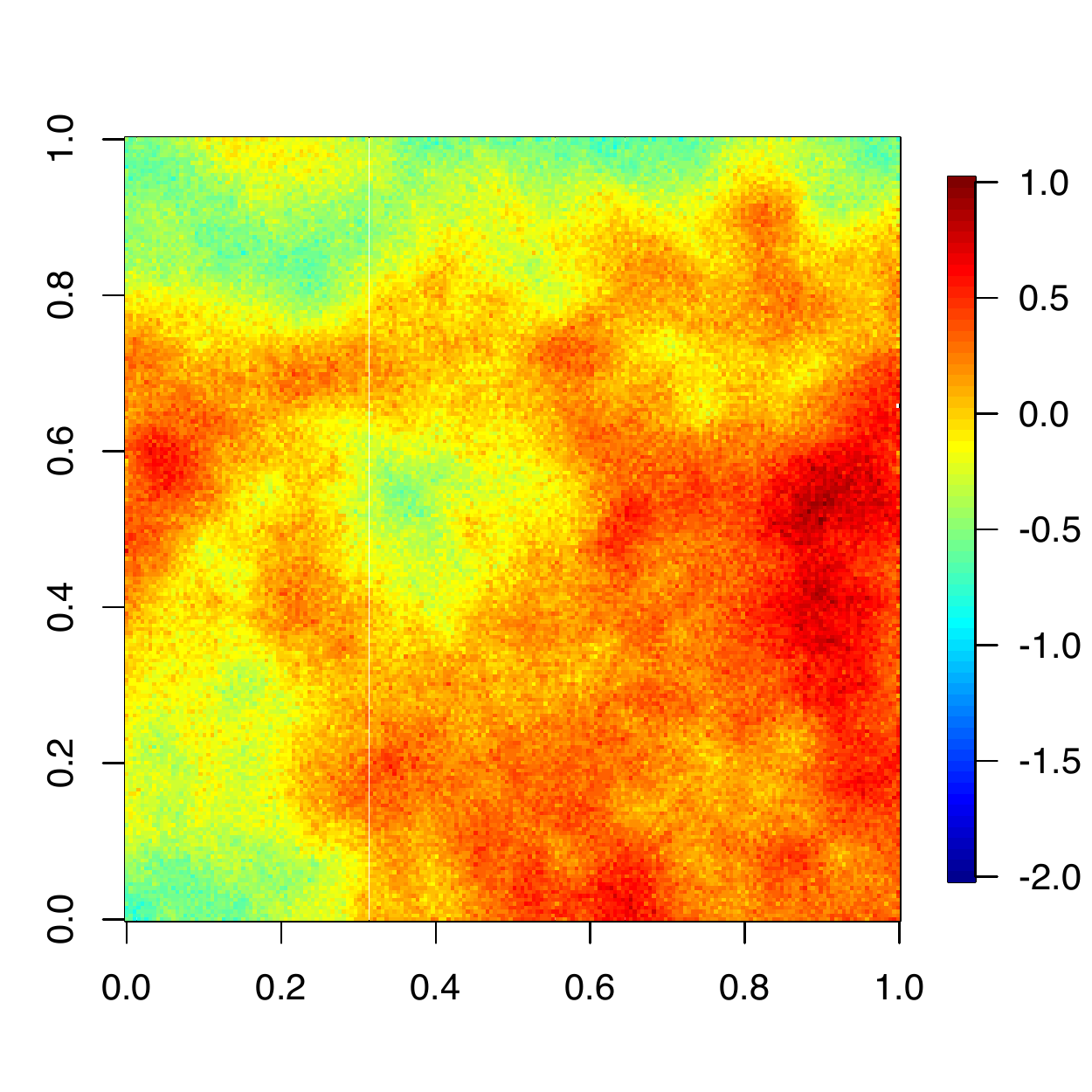}
\caption{3b-3 ($Z_1$)}
\label{3b-3-z1}
 \end{subfigure}
  \begin{subfigure}[b]{0.2\textwidth}
  \centering
\includegraphics[scale=0.2]{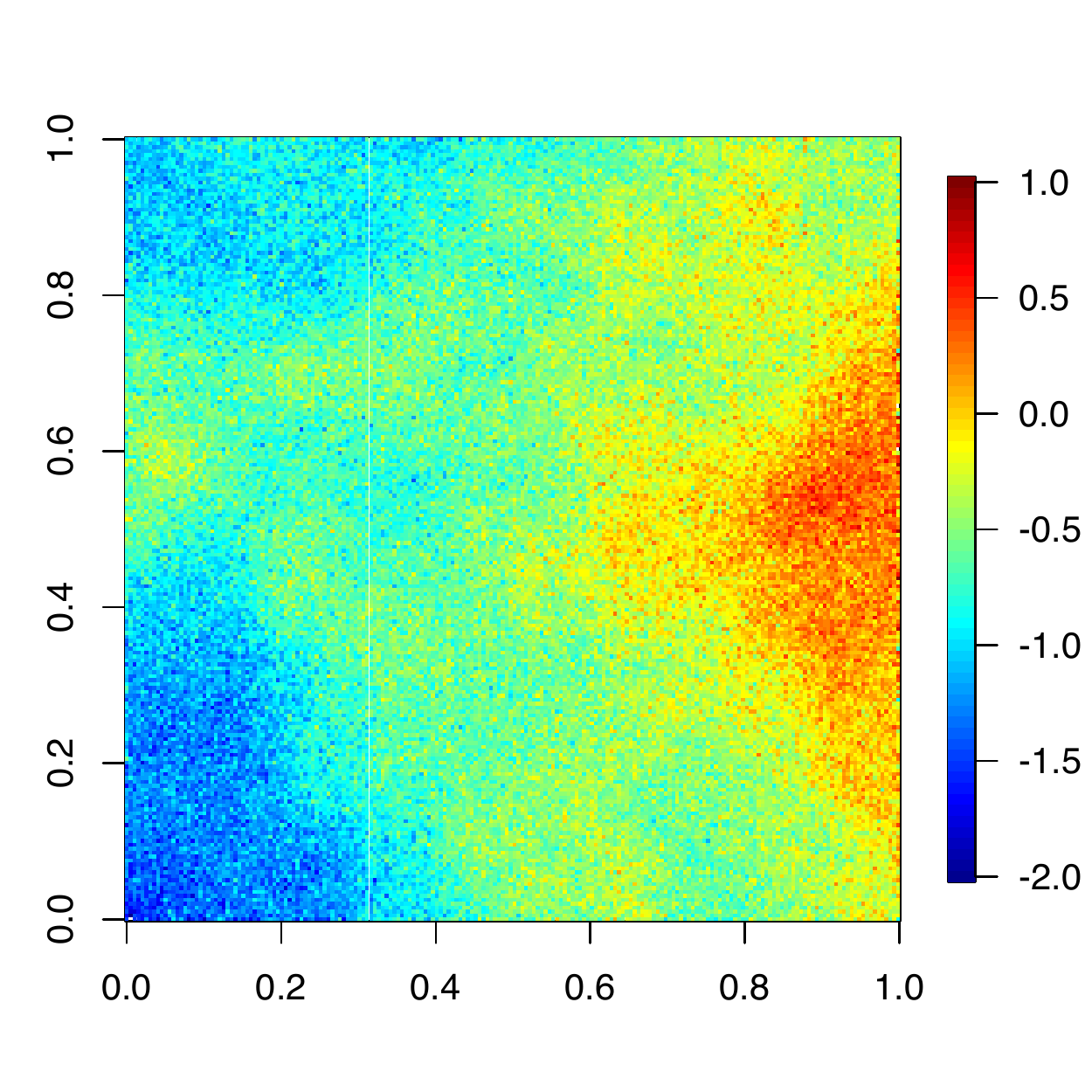}
\caption{3b-3 ($Z_2$)}
\label{3b-3-z2}
 \end{subfigure}

 \end{minipage}
 
 

  \caption{Bivariate datasets $(Z_1(\bm s),Z_2(\bm s))^\top$ using the parsimonious and flexible Mat\'ern models with two sizes: $50$K (3a-1, 3a-2, and 3a-3) and $500$K (3b-1, 3b-2, and 3b-3).}
   \label{fig:bivariate-flex}
\end{figure}

Figure~\ref{fig:bivariate-flex} shows visual images of datasets generated from the parsimonious and the flexible Mat\'ern models, respectively.  For prediction, 90\% of each dataset were provided as training data, while 10\% were kept as testing data with both variables left out simultaneously.

\subsection{Assessment Metric}

We hosted the competition this year on the Kaggle machine learning and data science platform. Six different Kaggle competitions were created to represent the six sub-competitions as follows:
\begin{itemize}
  \item Sub-competition 1a: two $100$K univariate nonstationary spatial datasets: \\
  \url{https://www.kaggle.com/competitions/2022-kaust-ss-competition-1a}
  \item Sub-competition 1b: two $1$M univariate nonstationary spatial datasets: \\
  \url{https://www.kaggle.com/competitions/2022-kaust-ss-competition-1b}
  \item Sub-competition 2a: nine $1$K spatial locations at $100$ time points (space-time) datasets:\\ \url{https://www.kaggle.com/competitions/2022-kaust-ss-competition-2a}
  \item Sub-competition 2b: nine $10$K spatial locations at $100$ time points (space-time) datasets:\\ \url{https://www.kaggle.com/competitions/2022-kaust-ss-competition-2b}
  \item Sub-competition 3a: three $50$K bivariate spatial datasets:\\ 
  \url{https://www.kaggle.com/competitions/2022-kaust-ss-competition-3a}
  \item Sub-competition 3b: three $500$K bivariate spatial datasets: \\
  \url{https://www.kaggle.com/competitions/2022-kaust-ss-competition-3b}
\end{itemize}
\vspace{.4cm}

The participating teams were ranked independently for each sub-competition. We used the Root Mean Square Error (RMSE) criterion to evaluate the prediction accuracy for each sub-competition:
\[
\hbox{RMSE} = \sqrt{\dfrac{1}{N_\textrm{test}}\sum^{N_\textrm{test}}_{i=1}\big(\hat Z_i- Z_i\big)^2},
\]
where $\hat Z_i$ and $Z_i$ are respectively the predicted and true realization values in the testing dataset, and $N_{\textrm{test}}$ is the total number of data points in the testing dataset. Depending on the sub-competitions, $\hat Z_i$ and $Z_i$ are either spatial only or spatio-temporal variables. The final score of each team for a given sub-competition is calculated in Kaggle using the Mean Columnwise Root Mean Squared Error (MCRMSE), i.e., the averaged RMSE over datasets of each sub-competition.
\vspace{.5cm}

{\small
\begin{longtable}{cccccc}
  \caption{Summary of the six sub-competitions.}\\
\hline
Sub-competition    &  Setting & True &   \# of   & Training  & Testing        \\ 
 & & Data Model & Datasets & Data Size & Data Size \\ \hline
1a   &  Univariate & GP with & 2   & 90K &  10K    \\
& Nonstationary & Nonstationary \\
& Spatial & Mean or Cov \\
\hline
1b   &  Univariate & GP with  & 2  & 900K &  100K     \\
& Nonstationary & Nonstationary \\
& Spatial & Mean or Cov \\
\hline
2a   &  Univariate & GP with  & 9   & 90K &  10K      \\
& Stationary Space-Time& Non-Separable Cov & & \\
\hline
2b   &  Univariate & GP with   & 9   & 900K &  100K      \\
& Stationary Space-Time& Non-Separable Cov & & \\
\hline
3a   &  Bivariate & GP with  & 3    & 45K &   5K  \\
& Stationary & Parsimonious or Flexible\\
& Spatial & Mat\'ern Cross-Cov\\
\hline
3b   &  Bivariate & GP with  & 3     & 450K &   50K \\
& Stationary & Parsimonious or Flexible\\
& Spatial & Mat\'ern Cross-Cov\\
\hline
\label{tbl:subcomp}
\end{longtable}
}

\section{Analysis of the Submitted Results}

In this section, we summarize the results submitted from different teams (list provided in Table~S1 of the Supplementary Material) and compare their performances. Specially, we have designed visual graphs to highlight the teams' performance related to each dataset and its impact on the teams' final rank in each sub-competition. The RMSE values for different teams on all datasets are provided in the Supplementary Material (Tables S2 to S11) as well as those obtained with {\it ExaGeoStat} for reference purpose.

\subsection{Results of Sub-competitions 1a and 1b (univariate nonstationary spatial)}

\begin{figure}[t!]
   \begin{minipage}[t]{0.45\linewidth}
      \centering
    \includegraphics[width=\textwidth]{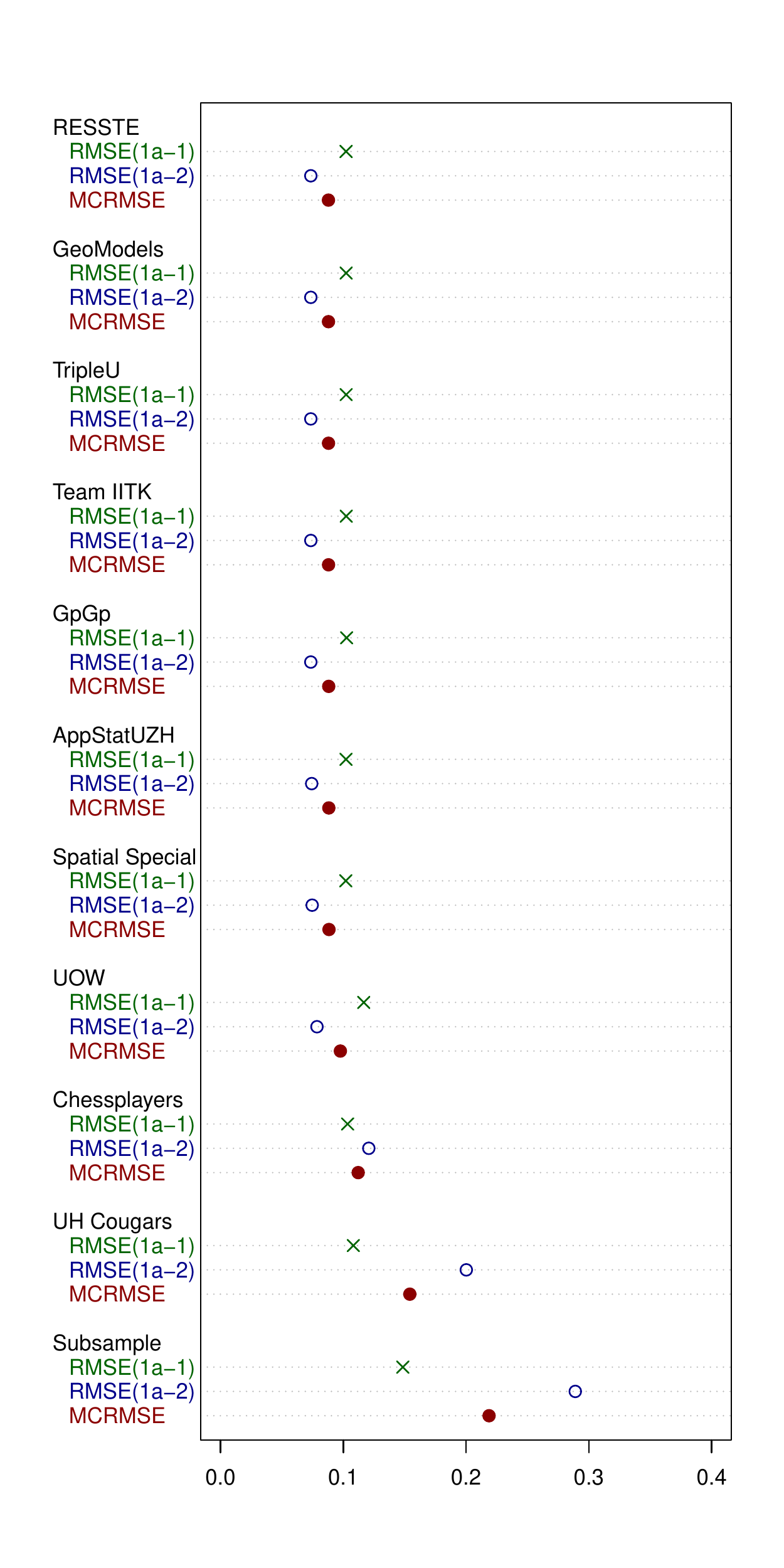}
        \subcaption{1a}
           \label{fig:R1a}
    \end{minipage}
    \begin{minipage}[t]{0.45\linewidth}
    \centering
\includegraphics[width=\textwidth]{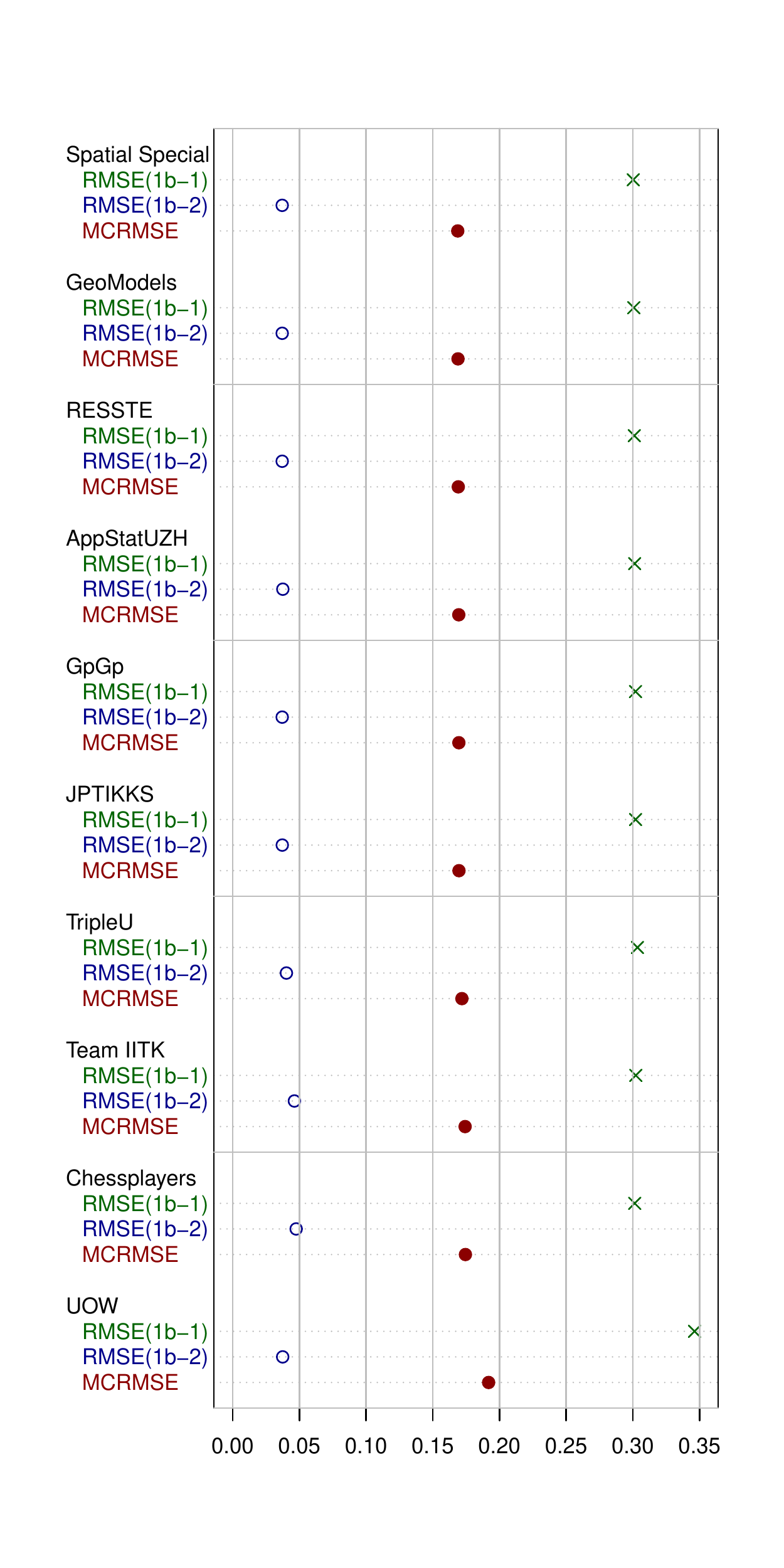}
    \subcaption{1b}
               \label{fig:R1b}
\end{minipage}
    \caption{Sub-competitions 1a and 1b leaderboard.}
    \label{fig:R1ab}
\end{figure}

In Sub-competition 1a, there are two $100$K datasets, 1a-1 and 1a-2, where 1a-1 has been generated using the deterministic mean function in~(\ref{eq:e1}), while 1a-2 has been generated using the nonstationary covariance function in~(\ref{eq:non_stat_cov_fun}) with specific settings as mentioned in Section~\ref{sec3.1}. Figure~\ref{fig:R1a} shows the competition results of the participating teams in Sub-competition~1a. All the teams except the \team{SubSample} team got close RMSE values for all the datasets. The \team{SubSample} team utilized a subsampling technique to fit the given data. However, their approach seems inappropriate for the given nonstationary data. In dataset 1a-2, the top eight teams, i.e., \team {RESSTE}, \team{GeoModels}, \team{TripleU}, \team{Team IITK}, \team{GpGp}, \team{AppStatUZH}, \team{Spatial Special}, and \team{UOW} obtained very close RMSEs with a variation of around $7\times10^{-6}$. The next three teams, i.e., \team{Chessplayers}, \team{UH Cougars}, and \team{SubSample} have larger RMSEs than the top eight teams. More details are needed to understand how to improve the performance of these three teams.

In Sub-competition 1b, we have two $1$M datasets, 1b-1 and 1b-2, where dataset 1b-1 has been generated using the deterministic mean function in~(\ref{eq:e2}), while dataset 1b-2 has been generated using the nonstationary covariance function in~(\ref{eq:non_stat_cov_fun}) with the settings mentioned in Section~\ref{sec3.1}. Figure~\ref{fig:R1b} summarizes the results from different teams. As shown by the figure, all the teams except the \team{UOW} team were able to obtain very close RMSEs in datasets 1b-1 and 1b-2 with variation around $10^{-5}$. In dataset 1b-2, the \team{UOW} team performed better than \team{TripleU}, \team{Team IITK}, and \team{Chessplayers} teams with RMSE equal to $0.0375$ compared to $0.0403$, $0.0463$, and $0.0476$ for the three teams, respectively. However, the overall MCRMSEs are better for the three teams than the \team{UOW} team, as shown in the figure, because of the performance of the \team{UOW} team in dataset 1b-1.

\subsection{Results of Sub-competitions 2a and 2b (univariate stationary space-time)}

\begin{figure}[H]
   \begin{minipage}[t]{0.47\linewidth}
      \centering
    \includegraphics[width=\textwidth]{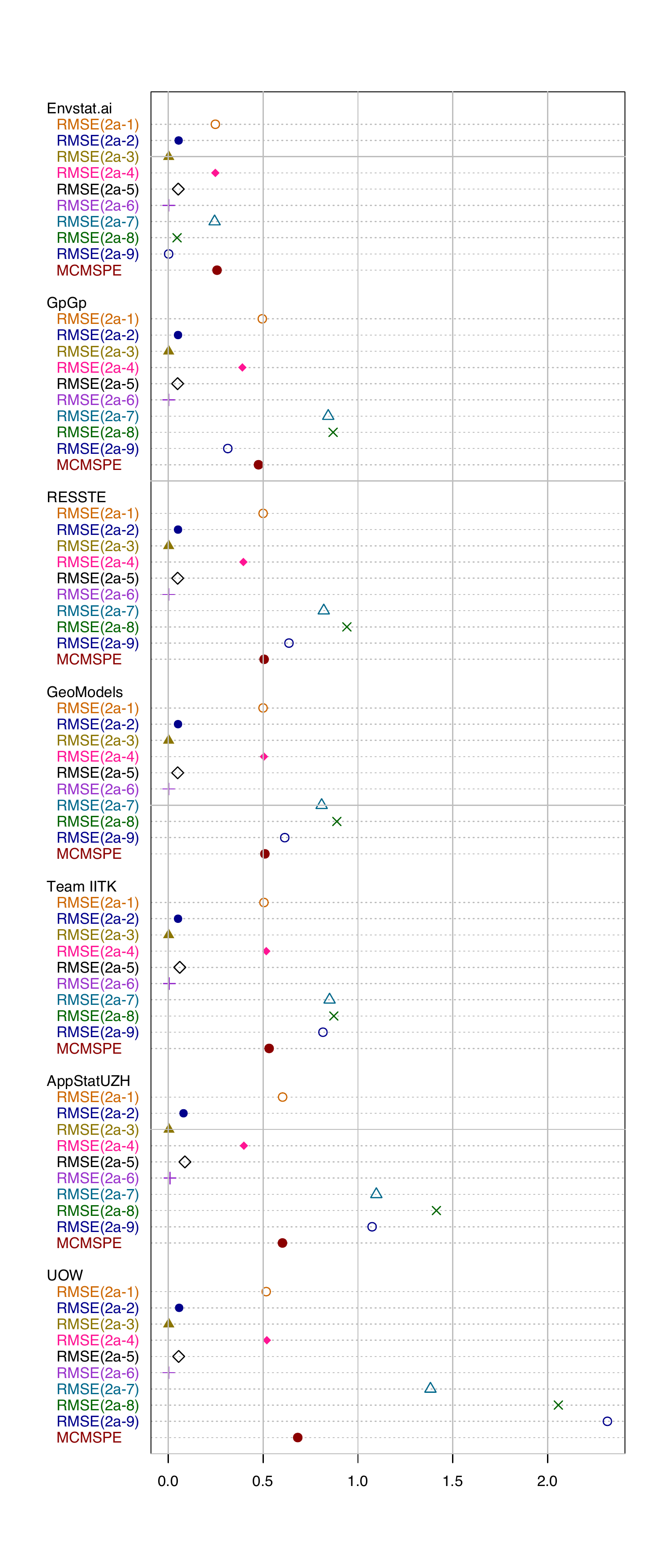}
        \subcaption{2a}
           \label{fig:R2a}
    \end{minipage}
    \hspace{0.5cm}
    \begin{minipage}[t]{0.47\linewidth}
    \centering
\includegraphics[width=\textwidth]{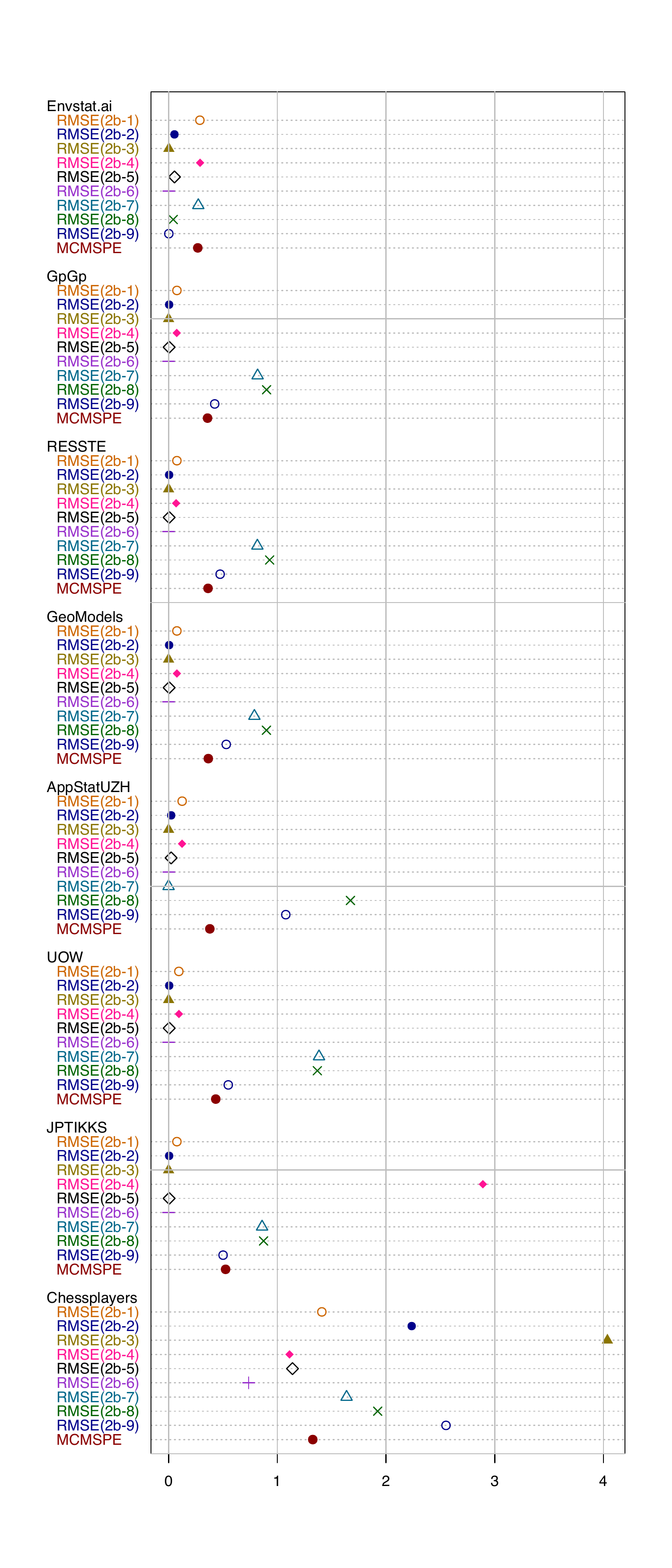}
    \subcaption{2b}
               \label{fig:R2b}
\end{minipage}
    \caption{Sub-competitions 2a and 2b leaderboard.}
    \label{fig:R2ab}
\end{figure}

In Sub-competition 2a, nine space-time datasets were generated using the non-separable stationary space-time covariance function in (\ref{eq:st-cov}) with $1$K locations and $100$ time-slots. The generated datasets were divided into a training dataset (90\%) and a testing dataset (10\%). The testing datasets have been chosen with three settings, i.e., RS, RST, and T10, as described in Section~\ref{sec:spacetime}. We have seven participants in this sub-competition and the results are presented in Figure \ref{fig:R2ab}. The \team{Envstat.ai} team was able to obtain the best MCRMSE, i.e., $0.2573$, for all the nine datasets in Sub-competition 2a. This result is $1.8$X better than the second-ranked team, i.e., \team{GpGp}. With a closer examination of the performance of the \team{Envstat.ai} team in different datasets, we observe that the team was able to obtain the best RMSEs in datasets 2a-7, 2a-8, and 2a-9, where all the spatial locations at the last 10 time-slots were missing (i.e., forecasting case). The improvements compared to the \team{GpGp} team are $1.36$X, $2.08$X, and	$3.47$X in datasets 2a-7, 2a-8, and 2a-9, respectively, which also shows that the \team{Envstat.ai} team performed better with strong space correlation and weak time correlation. The RMSEs were close to each other for different datasets for all other teams.

In Sub-competition 2b, nine space-time datasets were generated using the non-separable space-time covariance function in (\ref{eq:st-cov}) with $10$K locations and $100$ time-slots. Eight teams have successfully submitted their results to this sub-competition. The ranks of the first four teams were the same as in Sub-competition 2a. The \team{Envstat.ai} team obtained the first rank with a total MCRMSE value equals to $0.2687$, which is $1.33$X better than the second-ranked team, i.e., \team{GpGp}. The \team{Envstat.ai} team outperformed the \team{GpGp} team in only three datasets, 2b-7, 2b-8, and 2b-9, by $1.32$X, $2.14$X, and $3.71$X. Despite the fact that the \team{GpGp} team was able to outperform the \team{Envstat.ai} team in six datasets 2b-1, 2b-2, 2b-3, 2b-4, 2b-5, and 2b-6 by $1.39$X, $1.80$X, $1.90$X, $1.40$X, $1.81$X, $1.89$X, the differences in RMSE values were not enough to have an MCRMSE value lower than the \team{Envstat.ai} team.

\begin{figure}[t!]
   \begin{minipage}[t]{0.32\linewidth}
      \centering
    \includegraphics[width=\textwidth]{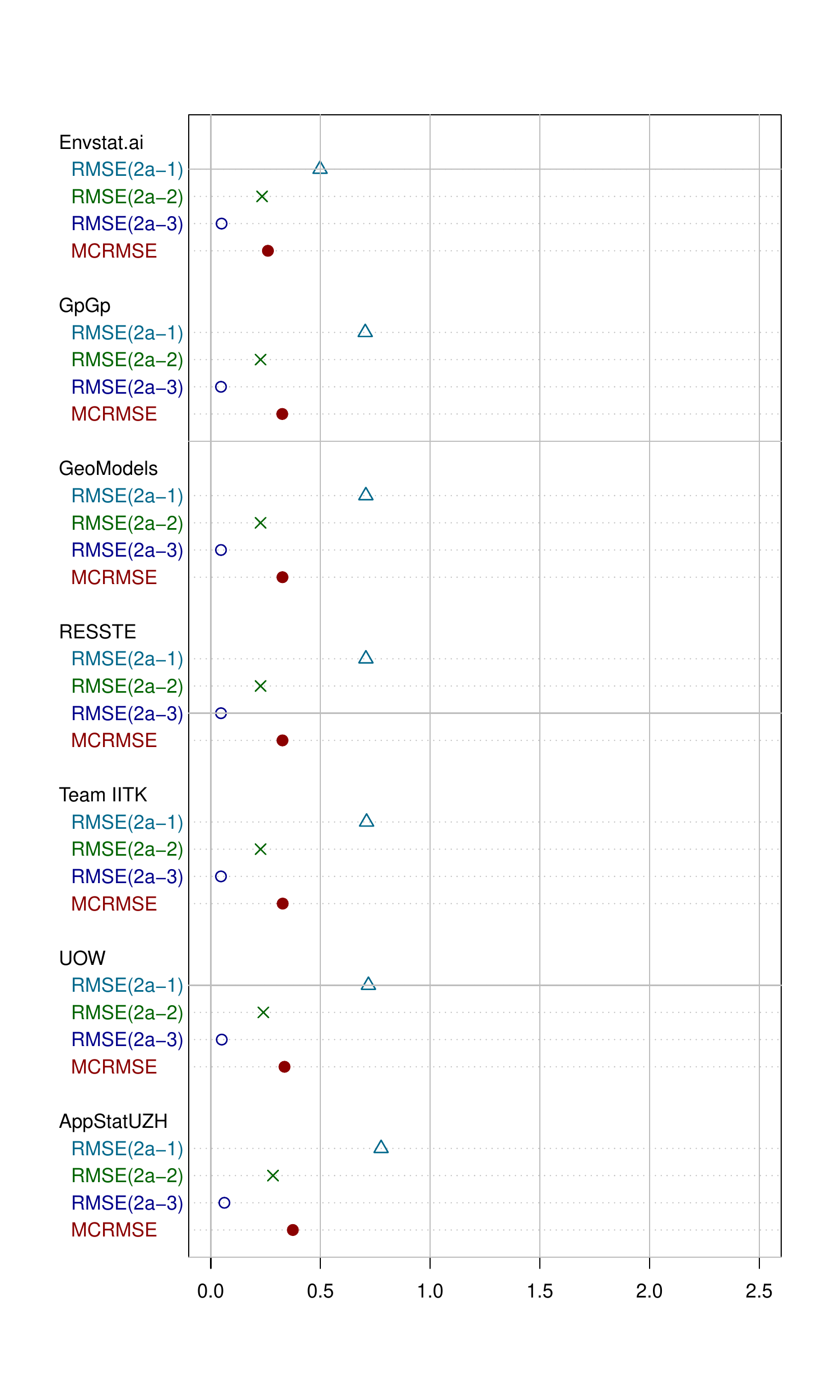}
        \subcaption{2a-1, 2a-2, and 2a-3 (RS)}
           \label{fig:R2a-123}
    \end{minipage}
    \begin{minipage}[t]{0.32\linewidth}
    \centering
\includegraphics[width=\textwidth]{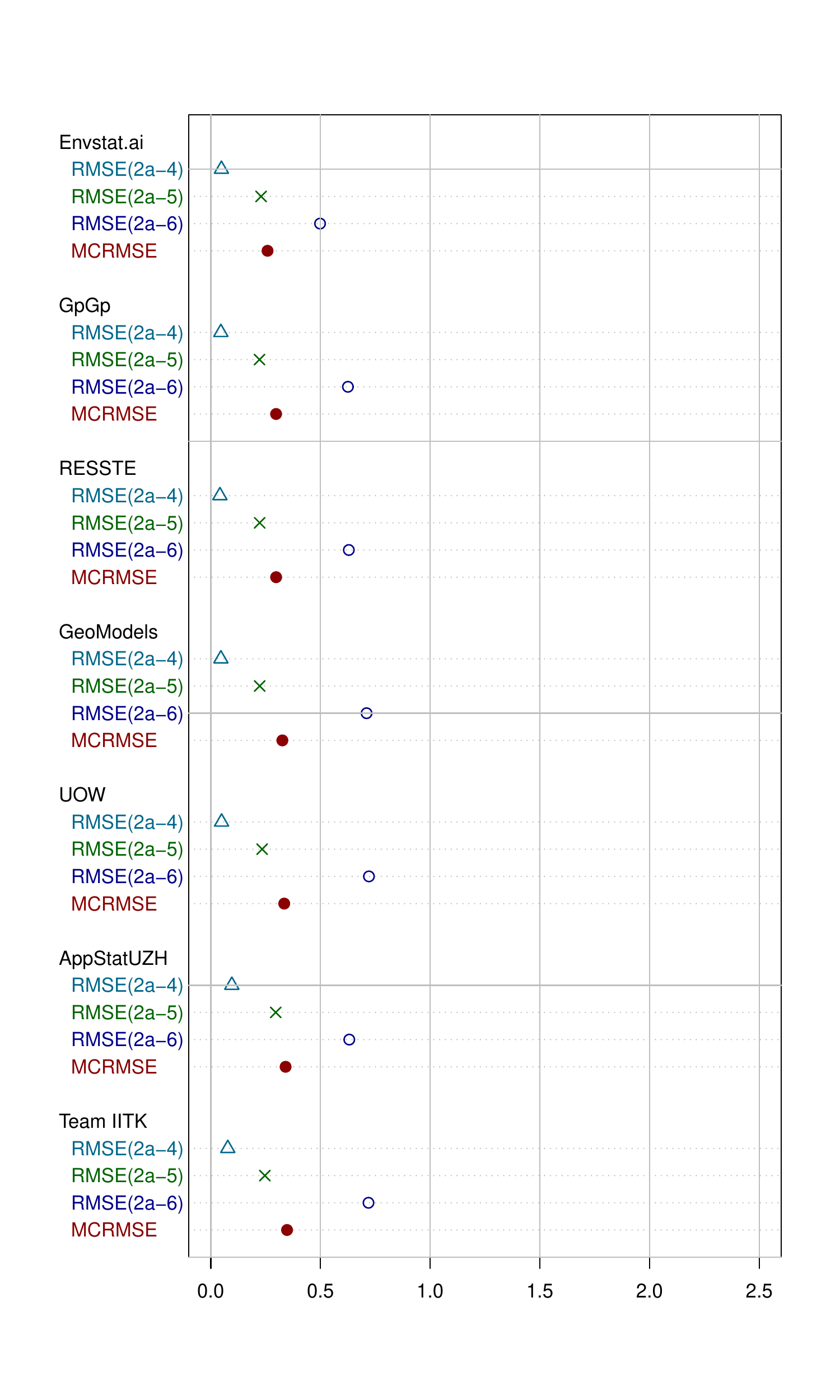}
        \subcaption{2a-4, 2a-5, and 2a-6 (RST)}
               \label{fig:R2a-456}
\end{minipage}
    \begin{minipage}[t]{0.32\linewidth}
    \centering
\includegraphics[width=\textwidth]{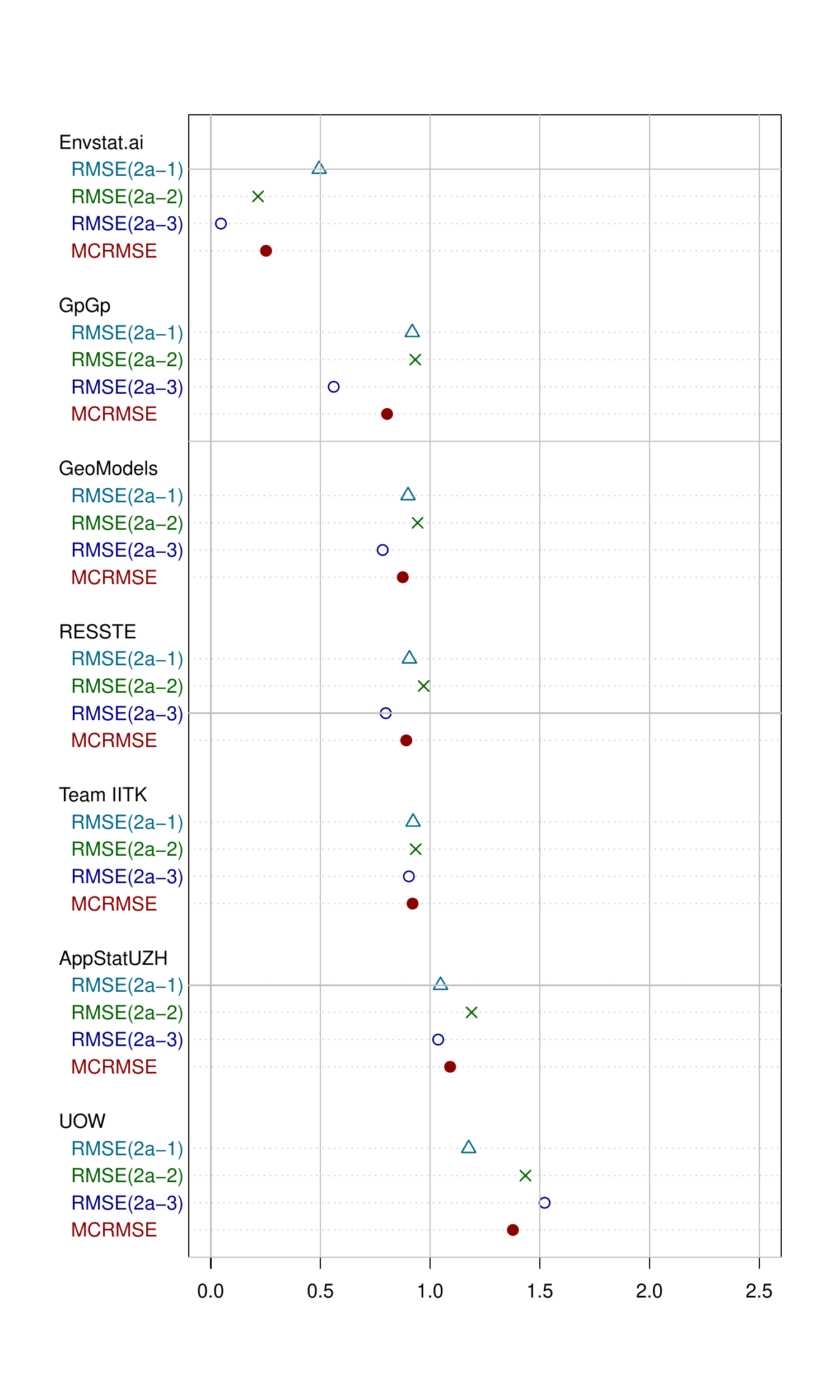}
        \subcaption{2a-7, 2a-8, and 2a-9 (T10)}
               \label{fig:R2a-789}
\end{minipage}

   \begin{minipage}[t]{0.32\linewidth}
      \centering
    \includegraphics[width=\textwidth]{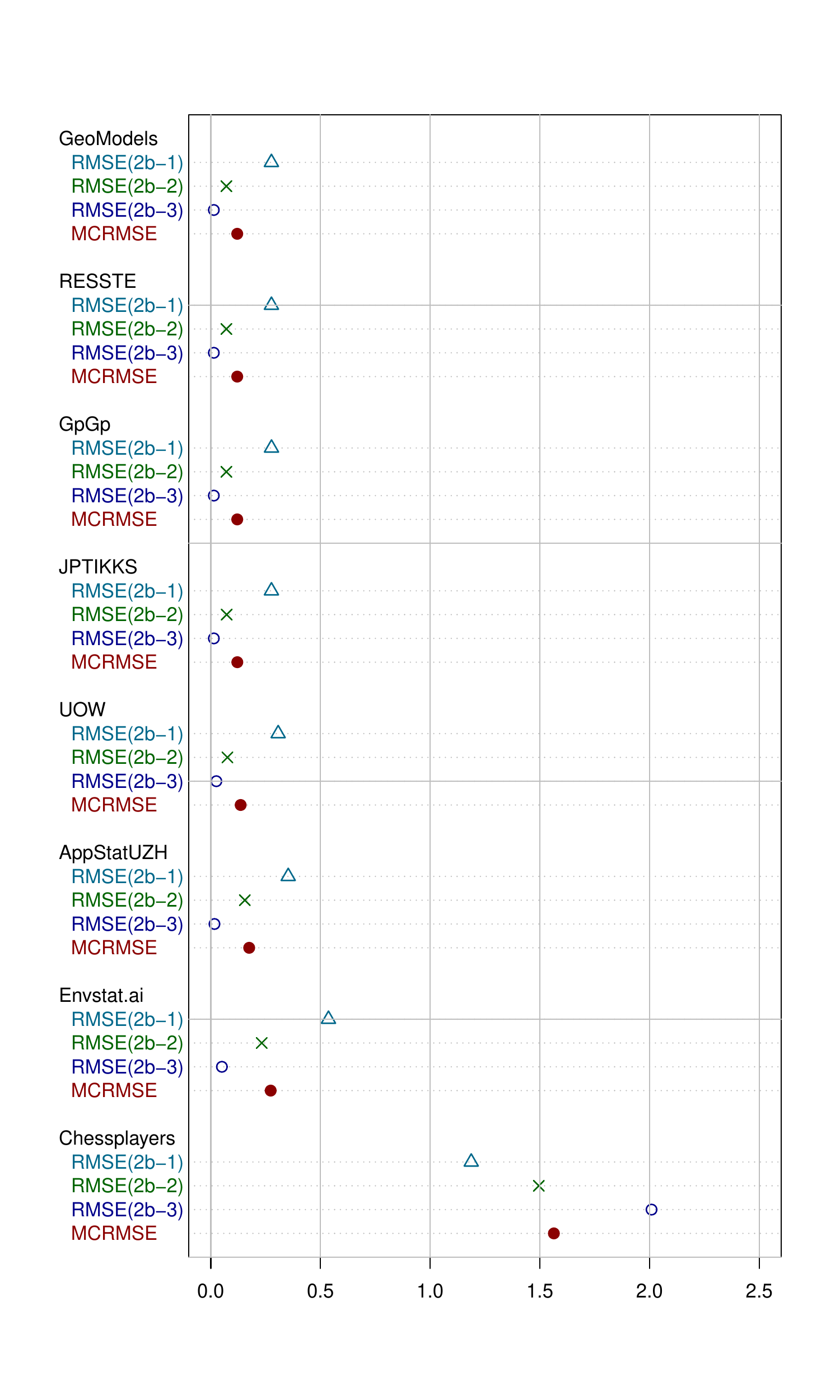}
        \subcaption{2b-1, 2b-2, and 2b-3 (RS)}
           \label{fig:R2b-123}
    \end{minipage}
    \begin{minipage}[t]{0.32\linewidth}
    \centering
\includegraphics[width=\textwidth]{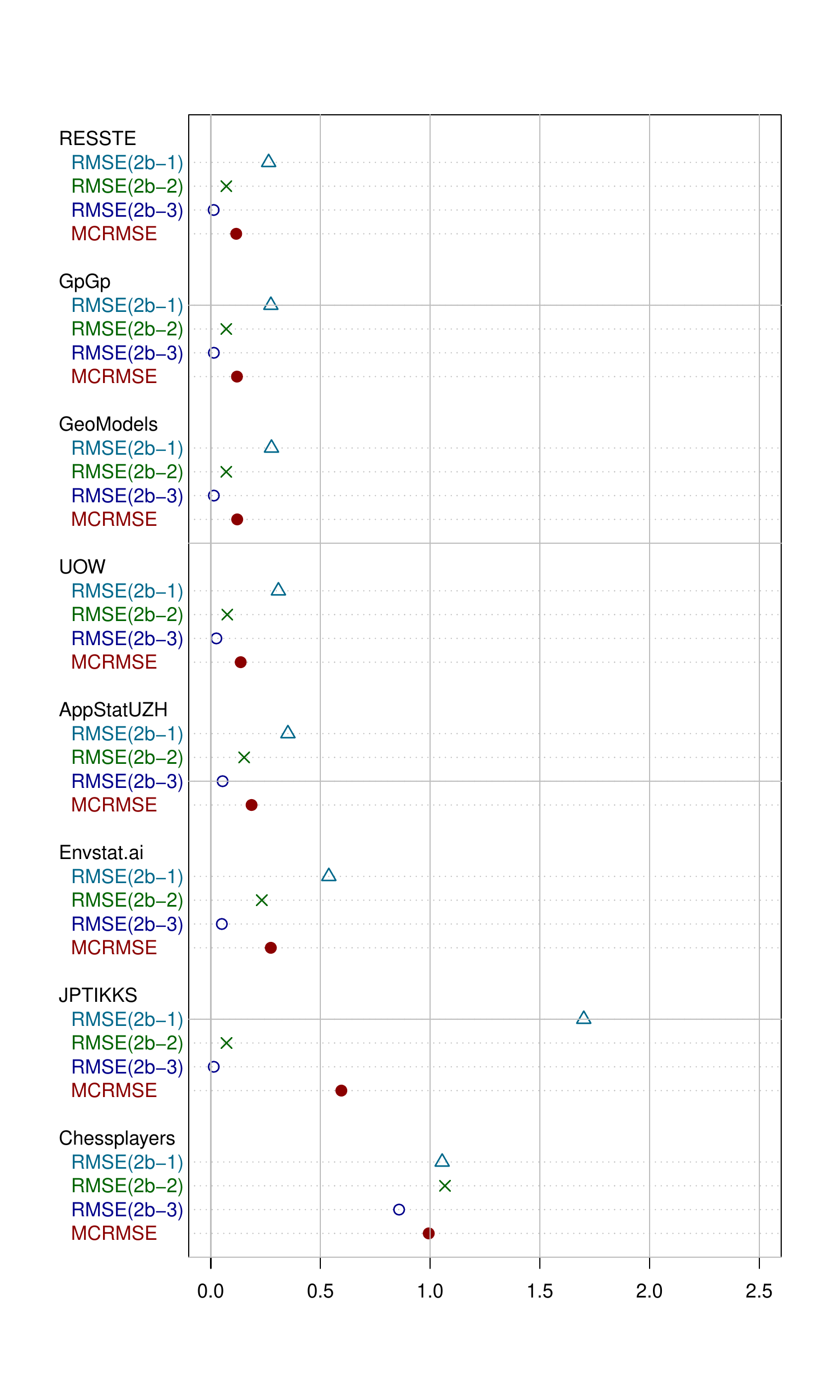}
        \subcaption{2b-4, 2b-5, and 2b-6 (RST)}
               \label{fig:R2b-456}
\end{minipage}
    \begin{minipage}[t]{0.32\linewidth}
    \centering
\includegraphics[width=\textwidth]{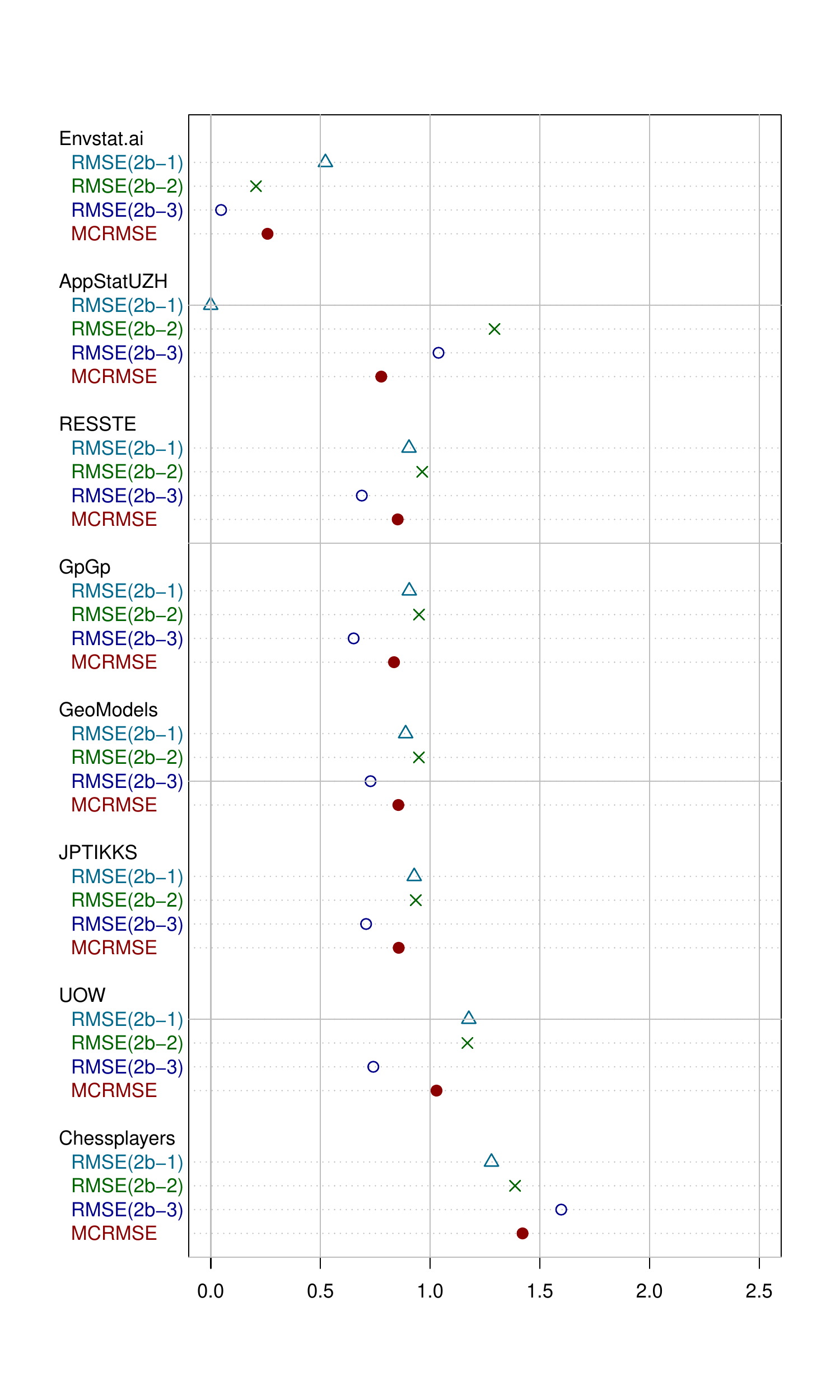}
        \subcaption{2b-7, 2b-8, and 2b-9 (T10)}
               \label{fig:R2b-789}
\end{minipage}
    \caption{Sub-competitions 2a and 2b individual leaderboards where we combine datasets with the same scenario of leaving out missing points for prediction (i.e., RS, RST, and T10).}
    \label{fig:R2ab-individual}
\end{figure}

Figure \ref{fig:R2ab-individual} shows the individual leaderboards in Sub-competition 2a and 2b for all the participating teams. It clearly shows that the \team{Envstat.ai} team was able to obtain the lowest RMSE values in the T10 datasets in both sub-competitions, i.e., forecasting case, with a noticeable improvement compared to the other teams. However, it did not perform well in 3 out of the 4 remaining datasets. The figure also shows that the \team{GpGp} team performed quite well in all the datasets compared to the  \team{Envstat.ai} team. The performance of the \team{Envstat.ai} team in 2a-7, 2a-8, 2a-9, 2b-7, 2b-8, and 2b-9 was good enough to win both sub-competitions.

\subsection{Results of Sub-competitions 3a and 3b (bivariate stationary spatial)}

\begin{figure}[t!]
   \begin{minipage}[t]{0.5\linewidth}
      \centering
    \includegraphics[width=\textwidth]{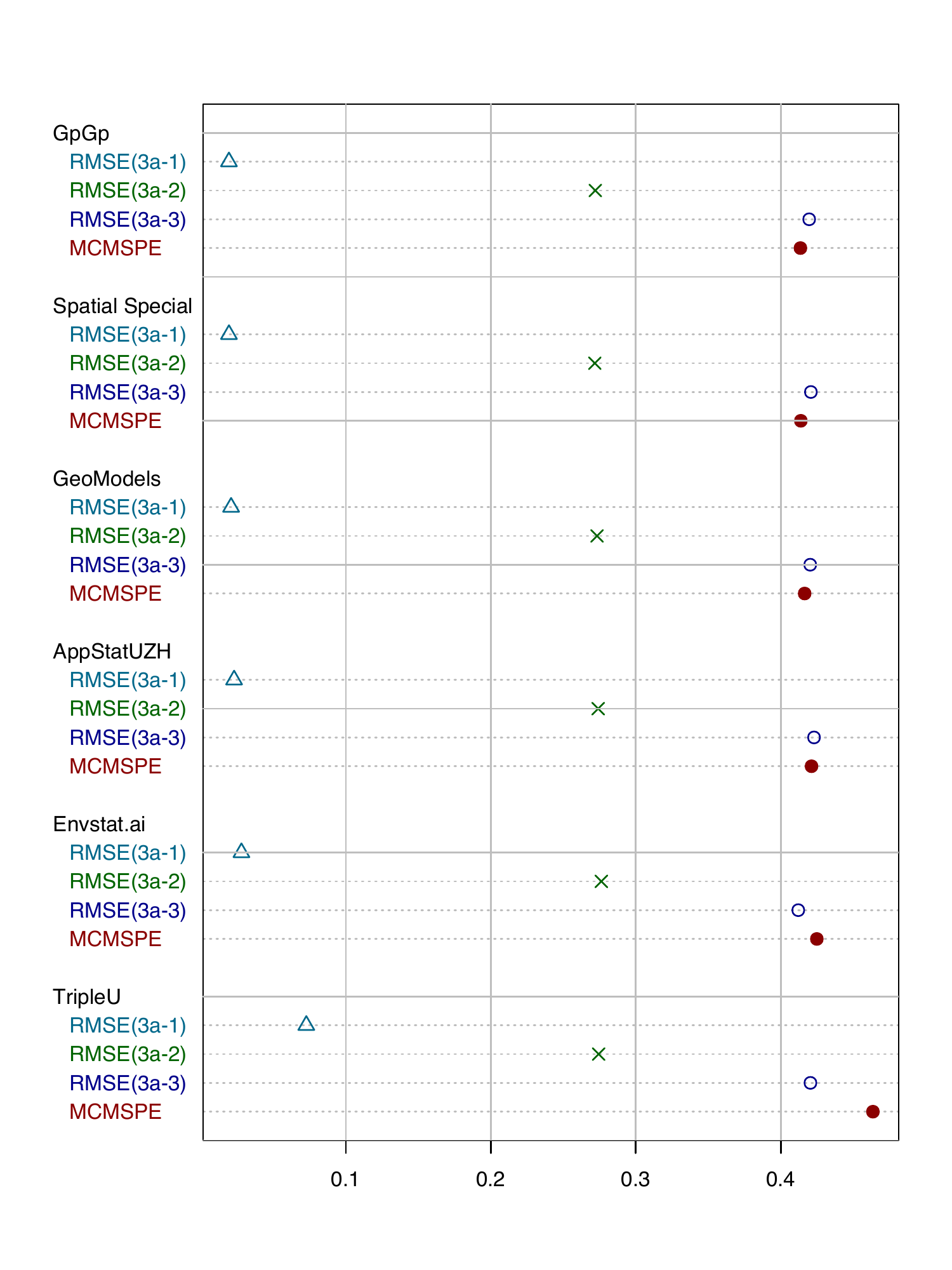}
        \subcaption{3a}
           \label{fig:R3a}
    \end{minipage}
    \begin{minipage}[t]{0.5\linewidth}
    \centering
\includegraphics[width=\textwidth]{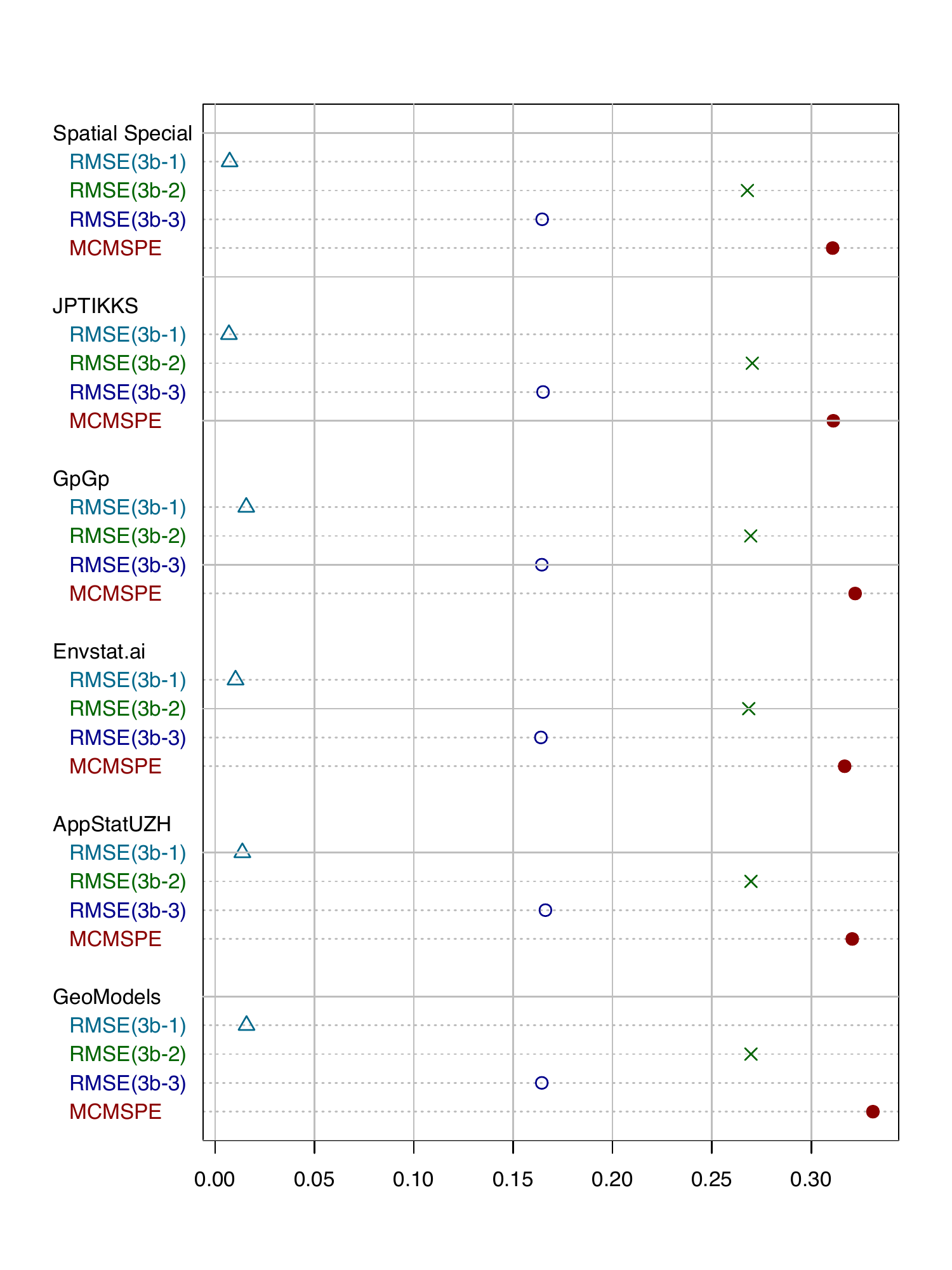}
    \subcaption{3b}
               \label{fig:R3b}
\end{minipage}
    \caption{Sub-competitions 3a and 3b leaderboard.}
    \label{fig:R3ab}
\end{figure}

In Sub-competition 3a, we generated one $50$K bivariate datasets using the parsimonious Mat\'ern covariance function in (\ref{eq:para}), i.e., 3a-1, and two $50$K bivariate datasets using the flexible Mat\'ern covariance function in~(\ref{eq:flex}), i.e., 3a-2 and 3a-3. Six teams submitted their results to this sub-competition. Figure~\ref{fig:R3ab} shows the two leaderboard lists of Sub-competitions 3a and 3b. As shown, the \team{GpGp} team outperformed all the other teams, slightly improving the final MCRMSE. The difference between teams was minimal. The only observation is that the \team{TripleU} team has a high RMSE in 3a-2, representing the flexible Mat\'ern model under $\nu_{11}=0.6$ and $\nu_{11}=1.4$, compared to the other teams.

In Sub-competition 3b, we generated one $500$K bivariate dataset using the parsimonious Mat\'ern covariance function in~(\ref{eq:para}), i.e., 3b-1, and two 500K bivariate datasets using the flexible Mat\'ern covariance function in~(\ref{eq:flex}), i.e., 3b-2 and 3b-3. Six teams submitted their results to this sub-competition, i.e., \team{Spatial Special}, \team{JPTIKKS}, \team{GpGp}, \team{Envstat.ai}, \team{AppStatZH}, and \team{GeoModels}. Although the main change in 3b compared to 3a is the sizes of the datasets, the ranks of the teams changed in 3b vs 3a. The \team{Spatial Special} team ranked first and the \team{GpGp} team ranked second in 3b, compared to third and first in 3a, respectively. The difference in the MCRMSE was very small, and there is no special observation from the final results for individual datasets.

\section{Methods Used in the Competition: A Closer Look}

The participants have relied on different methods to tackle the various sub-competitions. In this section, we comment on the performance of the top-ranked teams and the strategies they adopted based on the description they submitted to us.

\subsection{Sub-competition 1a (univariate nonstationary spatial, 100K)}

Sub-competition 1a included two $100$K nonstationary datasets: one was generated using a predefined deterministic mean function and the other using a nonstationary covariance function under certain settings. The \team {RESSTE} team ranked first in this sub-competition. The team applied the Vecchia composite likelihood approximation method~\citep{vecchia1988estimation,katzfuss2020vecchia,katzfuss2021general} to model the datasets in 1a. The Vecchia approximation method generally depends on a sparse covariance matrix instead of the usual dense matrix to allow faster computation of the matrix inverse operation. The \team {RESSTE} team used the {\em GpGp} R package~\citep{guinness2018gpgp} to implement their model for datasets 1a-1 and 1a-2. For 1a-1, the team realized the nonstationarity of the given data and decided to split the region into two parts, i.e., upper and lower, by the line $s_x+s_y=0.85$. The intercept $0.85$ was chosen by experimenting with several options and selecting the one with the lowest mean square prediction error  values on one part of the training dataset. They used an anisotropic Mat\'ern covariance function in the upper area, while an isotropic Mat\'ern function was chosen for the lower area. To apply the Vecchia approximation to 1a-1, the team set the number of neighbors to 10 for  estimation and to 500 for prediction. For 1a-2, an isotropic Mat\'ern covariance function was  used for the whole region without splitting. Herein, the team set the number of neighbors to 50 for estimation and to 70 for predicition.

The \team{GeoModels} team ranked second in Sub-competition 1a. The team applied the nearest neighbor weighted composite likelihood method to model both datasets in 1a. The analysis and the results were obtained using the {\em GeoModels} package~\citep{GM2022}. In 1a, the team considered Gaussian random fields with a geometrically anisotropic Mat\'ern correlation model and Tukey non-Gaussian random fields.

\subsection{Sub-competition 1b (univariate nonstationary spatial, 1M)}
\label{sec:1b}

Sub-competition 1b had two 1M datasets: 1b-1 was generated using a deterministic mean function, while 1b-2 was generated using a nonstationary covariance function in~(\ref{eq:e2}).

The \team{Spatial Special} team ranked first in this sub-competition. The team was able to find that dataset 1b-1 is nonstationary, and they adopted a deformation technique to map the locations from the original space to a latent space as follows: $r= \sqrt{s_x+s_y}$ and $\theta = \tan^{-1}(\sqrt{s_y/s_x})$.
This deformation allowed the team to assume the data are stationary in the latent space. 

The prediction was performed using a deep feed-forward neural network model with two inputs (i.e., $r$ and $\theta$), four hidden layers (i.e., 50, 50, 20, and 20 neurons, respectively), and one output (i.e., MSPE). The objective function was constructed as follows:
$$
MSPE=\frac{1}{N} {\sum}_{i=1}^{N}\{\hat{Z}(r_i,\theta_i)-Z(r_i, \theta_i)\}^2, 
$$
where $\hat{Z}(r_i,\theta_i)$ is the predicted value, $Z(r_i, \theta_i)$ is the true value at the location $(r_i, \theta_i)$, the activation function was the hyperbolic tangent function, and the optimizer was the Adam optimizer~\citep{kingma2014adam}. All the experiments were performed using TensorFlow 1.15.

In 1b-2, the \team{Spatial Special} team followed a different strategy since it  assumed that the data are stationary. The team fit the data with a Gaussian process, taking a constant mean $\mu$ and a Mat\'ern covariance function with nugget $\tau^2$. Parameter estimation was done using the {\em GpGp} package with 10-neighbors approximation, then the 30-neighbors approximation was maximized using the 10-neighbors estimates as starting values. For the prediction, 200-neighbors have been used to obtain predicted values. The parameter estimates are $\hat\mu=-0.8148$, $\hat \sigma^2=3.1046$, $\hat\beta=0.0031$, $\hat\nu=0.7250$, and $\hat\tau=8.01\times 10^{-5}$.

The \team{GeoModels} team was ranked second in Sub-competition 2b. They also applied the nearest neighbor weighted composite likelihood-based methods to model both datasets in 1b  using the {\em GeoModels} R package.
The strategy of modeling and prediction was the same as in 1a.

The \team{RESSTE} team was ranked third in this sub-competition. They also applied the Vecchia approximation methods as in Sub-competition 1a. For dataset 1b-1, the data region was divided using the line $s_x+s_y=0.85$ as in 1a-1. However, an anisotropic Mat\'ern model was chosen for the lower region, and an isotropic Mat\'ern model was chosen for the upper region. For Vecchia approximation, the number of neighbors was set to 10 and the estimation step to 500. For 1b-2, an anisotropic Mat\'ern covariance function was used for the whole region without splitting. The team set the number of neighbors to 50 and the estimation step to 70.


\subsection{Sub-competitions 2a and 2b (univariate stationary space-time)}
\label{sec:2a-2b}

We combine two sub-competitions in this subsection since they have the same top-ranked teams adopting the same methods to deal with the datasets. Sub-competition 2a has nine $1$K datasets in 100 time-lots, using three different settings. Sub-competition 2b has a larger size, $10$K datasets in 100 time-lots. The description of the datasets in both sub-competitions is shown in Section~\ref{sec:spacetime}.

The \team{Envstat.ai} was ranked first in Sub-competitions 2a and 2b. The \team{Envstat.ai} relied on a Deep Neural Network (DNN) for spatio-temporal predictions, an extension of the spatial version in \cite{2022deepkrig}. They used basis functions to capture the spatio-temporal dependence. For interpolation, they relied on regression, whereas they used a 2-stage modeling approach for forecasting at new locations. First, they interpolated at the new locations for the existing time points. Then they trained Long-Short Term Memory Network (LSTM) on these points to get the forecast for future time pointss. The results obtained by the \team{Envstat.ai}  team show a considerable improvement compared to the following ranked teams in this sub-competition.

The \team{GpGp} team was ranked second in Sub-competitions 2a and 2b. The team initially fit the Mat\'ern  space-time covariance function in the {\em GpGp} package, i.e., {\em matern\_spacetime(args)} to estimate the model parameters. Then, they used the two estimated range parameters to rescale the coordinates to select the ordering and neighbors for Vecchia's approximation. They relied on 30-neighbors to fit the model for prediction purpose. The team used the prediction function in the {\em GpGp} package, i.e., {\em predictions(args)}, where the number of selected neighbors was $m=60$.

The \team{RESSTE} team was ranked third in both sub-competitions. The team performed an exploratory data analysis step on the given data using a space-time covariance function and found that the data were generated from a positively non-separable space-time kernel. Based on this finding, the team chose to rely on a covariance function from the Gneiting class, with a Mat\'ern spatial covariance and a Cauchy temporal covariance. Due to the size of the given datasets, the team used  a block-composite likelihood to estimate the spatial and temporal parameters separately. Then, the team used the estimated parameters to fit the space-time model and estimate the full set of parameters. To predict the missing values, the \team{RESSTE} team applied ordinary kriging conditioned on the 1/9 nearest data points in space and 3 preceding, current, and 3 following time-slots in time for 2a, and 1/100 nearest data points in space and 2 preceding, and 3 following time-slots in time for 2b. The team applied slight changes to datasets 7, 8, and 9 in 2a and 2b by using 20 time-slots and 10 time-slots for 2a and 2b, respectively.

\subsection{Sub-competition 3a (bivariate stationary spatial, 50K)}

Sub-competition 3a included three bivariate spatial datasets, i.e., 3a-1, 3a-2, and 3a-3. Dataset 3a-1 was generated using the parsimonious Mat\'ern cross-covariance function in (\ref{eq:para}). Datasets 3a-2 and 3a-3 were generated using the flexible Mat\'ern cross-covariance function in (\ref{eq:flex}). The size of each dataset was $50$K.

The \team{GpGp} team was ranked first in Sub-competition 3a. The team relied on the flexible bivariate Mat\'ern model in~\cite{apanasovich2012valid} to fit the data in 3a-1, 3a-2, and 3a-3.
It applied the Fisher scoring algorithm~\citep{guinness2021gaussian} to perform the optimization of the likelihood function through Vecchia's approximation method. The team used the {\em GpGp} package to perform the modeling and the prediction tasks using 30-neighbors and 100-neighbors, respectively. 

The \team{Spatial Special} team was ranked second in this sub-competition. The team considered $Z_1$ and $Z_2$ as individual variables and modeled them separately and not jointly. The team used the Mat\'ern covariance function and its variogram to fit all the datasets in 3a (except for the $Z_2$ variable in 3a-1) by weighted least squares for parameter estimation. It defined the weights $n_h/h^2$, where $n_h$ is the number of point pairs and $h$ is the distance. The team used the {\em gstat} package \citep{gstat2004} to implement their method. For 3a-1 ($Z_2)$, the team relied on the same method it used in 1b-2, as described in Section~\ref{sec:2a-2b}. The only change is the number of considered neighbors when performing the prediction, for which the team used 100-neighbors instead of 200.

The \team{GeoModels} team was ranked third in this sub-competition. They applied the methods of the weighted composite likelihood based on pairs, where the weight function was based on the spatial nearest neighbors. The covariance function was assumed to be the bivariate isotropic Mat\'ern covariance function. The prediction was performed using local kriging.




\subsection{Sub-competition 3b (bivariate stationary spatial, 500K)}

Sub-competition 3b included three bivariate spatial datasets, i.e., 3b-1, 3b-2, and 3b-3. Dataset 3b-1 was generated using the parsimonious Mat\'ern cross-covariance function in (\ref{eq:para}). Datasets 3b-2 and 3b-3 were generated using the flexible Mat\'ern cross-covariance function in (\ref{eq:flex}). The size of each dataset was 500K.

The \team{Spatial Special} team was ranked first. For 3b-1 ($Z_1$) and 3b-1 ($Z_2$), the team used the same strategy as in 1b, with the number of nearest neighbors being 100 for prediction. For 3b-2 ($Z_1$), 3b-2 ($Z_2$), 3b-3 ($Z_1$), and 3b-3 ($Z_2$), the team divided the prediction region into 100 smaller prediction subregions. The team also used the ordinary kriging method to predict the missing data in each subregion independently. The missing values at the intersection between subregions were obtained by averaging the predicted values. The team fit the data using the Mat\'ern covariance function and its variogram. The weighted least squares method was used for parameter estimation where the weights were again $n_h/h^2$. The modeling and the prediction were implemented using the {\em gstat} R package. 

The \team{JPTIKKS} team was ranked second in this sub-competition. The team used covariance tapering with a bivariate Mat\'ern covariance function to estimate the parameters for the first dataset, 3b-1. For datasets 3b-2 and 3b-3, they adopted the sparse version of the scalable Geographically Weighted Regression (GWR) method \citep{murakami2020scalable} with the help of {\em scgwr}  R package to model the data. The sparsity helped in dealing with large datasets and reduced the computation complexity. Local regression was used to estimate spatially varying intercepts. The kernel was considered as a linear combination of known exponential sub-kernels. Each sub-kernel used 100-neighbors to estimate the parameter vector. The team used the leave-one-out cross-validation method for estimation. 

The \team{GpGp} team was ranked third in this sub-competition. The team fit the flexible bivariate Mat\'ern model to the data and used Vecchia’s approximation through the {\em GpGp} package. They chose 30-neighbors when fitting the model and 100-neighbors to perform predictions.

\section{Discussion}

This work proposed a framework for designing and assessing a competition for predicting missing values in large spatial and spatio-temporal datasets. Thanks to the {\em ExaGeoStat} software, very large datasets with various settings were generated from popular spatial and spatio-temporal models with exact computations. We have made all the datasets publicly available online  (\href{http://dx.doi.org/10.25781/KAUST-4ADYZ}{http://dx.doi.org/10.25781/KAUST-4ADYZ}) for future assessments of other existing or new methods as needed. This work also serves as a reference for comparing new results with the results obtained in this study. The competition covered three types of data: nonstationary spatial data, stationary space-time data, and bivariate stationary spatial data. We have reviewed the methods used by each participating team and ranked their performances based on prediction accuracy. The ranks provide a fair comparison among these teams, which may shed light on developing even better prediction methods.

In the process of evaluating the performance of different teams, we noticed the two top performers 
in Sub-competitions 2a and 2b, outperforming other teams by a large margin. After contacting them, we learned that both teams had combined several datasets from the same sub-competition to perform predictions. Since the same model actually generated different datasets in each sub-competition, the prediction accuracy was significantly improved by combing datasets. Therefore, we decided to exclude both teams from the leaderboard of Sub-competition 2 for fair comparisons. We believe this was a unintended flaw in the competition and such data-generating problems should be avoided in future competitions. 

For the second year, we had another successful competition with a wonderful response from the community. We believe that conducting such competitions provides opportunities for researchers to assess their methods on the same large datasets that are generated by exact computations. These datasets are valuable to the community, and we have made them publicly available. We also believe it is worthwhile to build benchmarking tools to assess the performance of existing approximation methods, which will significantly help to better understand the advantages and disadvantages of these methods. Although we focused on point forecasts, there is value in probabilistic forecasting for uncertainty quantification purpose and we plan to explore this topic in a future competition.

\section*{Supplementary Material}
In the Supplementary Material, we list the members of all the teams participating in this competition in Table S1.  Moreover, Tables S2 to S11 summarize the RMSE values obtained by different teams in each dataset of different sub-competitions, as well as those obtained with
{\it ExaGeoStat} for reference purpose.

\section*{Acknowledgement}
We want to thank the Supercomputing Laboratory (KSL) 
for providing computational resources to this project on the Shaheen-II Cray XC40 Supercomputer. Finally, the authors would
like to thank the KSL team for their valuable help in running the experiments in this publication.

\section*{Funding}
The research in this manuscript was funded by the King Abdullah University of Science and Technology (KAUST) in Thuwal, Saudi Arabia. We want to thank the Supercomputing Laboratory (KSL) at KAUST \footnote{https://www.hpc.kaust.edu.sa/} for supporting this research by providing the hardware resources, including the Shaheen-II Cray XC40 supercomputer used to generate the datasets in this competition.

\bibliographystyle{jds}
\bibliography{main}

\end{document}